\pdfoutput=1

\newcommand*{\ATLASLATEXPATH}{latex/}

\documentclass[PAPER, atlasdraft=true, texlive=2016,cernpreprint, english]{\ATLASLATEXPATH atlasdoc}

\usepackage{\ATLASLATEXPATH atlaspackage}

\usepackage{\ATLASLATEXPATH atlasbiblatex}

\usepackage{\ATLASLATEXPATH atlasphysics}

\graphicspath{{logos/}{figures/}}

\usepackage{lq3_common_paper-defs}
\usepackage{multirow}

\AtlasTitle{Searches for third-generation scalar leptoquarks in \tttev $p\mkern -1.5mu p$ collisions with the ATLAS detector}

\AtlasAbstract{%
Limits are set on the pair production of scalar leptoquarks,  
where all possible decays of the leptoquark into a quark ($t$, $b$)
and a lepton ($\tau$, $\nu$) of the third generation are considered. 
The limits are presented as a function of the leptoquark mass   
  and the branching ratio into charged leptons for up-type (\lqthreeu\ $\rightarrow t \nu / b \tau$)
and down-type (\lqthreed\ $\rightarrow b \nu / t \tau$) leptoquarks.
Many results are reinterpretations of previously published ATLAS searches. In all cases, LHC proton--proton collision 
data at a centre-of-mass energy of \tttev recorded by the ATLAS detector in 2015 and 2016 are used, 
corresponding to an integrated luminosity of \ilumi. 
Masses below 800~\GeV\ are excluded for both 
\lqthreeu\ and \lqthreed\
independently of the branching ratio, with masses below about 1~\TeV\ being excluded for the limiting cases
of branching ratios equal to zero or unity.
}

\AtlasRefCode{EXOT-2017-30}

 \PreprintIdNumber{CERN-EP-2019-026}

\arXivId{1902.08103}

 \AtlasJournalRef{JHEP 06 (2019) 144}
 \AtlasDOI{10.1007/JHEP06(2019)144}

\AtlasCoverCommentsDeadline{11 January 2019}

\hypersetup{pdftitle={ATLAS document},pdfauthor={The ATLAS Collaboration}}

\begin{document}

\maketitle

\tableofcontents

\section{Introduction}
\label{sec:intro}

Leptoquarks (LQ) are predicted by many extensions \cite{dimop_suss,techni2, techni3, string, comp, pati_salam_colour,georgi_glashow_unification} of the Standard Model (SM) and provide a connection between the quark and lepton sectors, which
exhibit a similar structure in the SM.
Also, recent hints of a potential violation of lepton universality in measurements of $B$-meson decays can be attributed, if confirmed, 
to the exchange 
of LQs~\cite{Hiller:2014yaa, Gripaios:2014tna, Freytsis:2015qca, Bauer:2015knc, DiLuzio:2017chi, Buttazzo:2017ixm, Cline:2017aed}.
LQs are bosons carrying colour and fractional electrical charge. They possess non-zero baryon and lepton numbers and decay into a quark--lepton pair. 
Leptoquarks can be scalar or vector bosons and can be produced singly or in pairs in proton--proton collisions.

This analysis focuses on the pair-production of third-generation scalar leptoquarks, i.e.\ LQs that decay into third-generation 
SM particles.
The assumption that LQs can only interact with leptons and quarks of the same family follows the minimal Buchm\"uller--R\"uckl--Wyler model \cite{Buchmuller:1986zs},
which is the benchmark model used in this analysis. 
The LQs couple to the quark--lepton pair via a Yukawa interaction. The couplings are determined by two parameters: 
a model parameter $\beta$ and the coupling parameter $\lambda$. The coupling to the charged lepton ($\tau$) is given by $\sqrt{\beta} \lambda$, and
the coupling to the $\tau$-neutrino $\nu$ by $\sqrt{1-\beta}\lambda$. 
The search is carried out for an up-type (\lqthreeu\ $\rightarrow t \nu / b \tau$) 
and a down-type (\lqthreed\ $\rightarrow b \nu / t \tau$) leptoquark.

The LQ model
is identical to the one used for a recent ATLAS search for first- and second-generation scalar LQs
using the dataset from 2015 and 2016, consisting of \ifb{36.1} of data taken at \tttev \cite{Aaboud:2019jcc}. 
In the following, the third-generation results are presented for that same dataset, where all possible decays of the pair-produced
\lqthreeu\ and \lqthreed\ into a quark ($t$, $b$) and a lepton ($\tau$, $\nu$) of the third generation are considered.
The results are presented as a function of the leptoquark mass and the branching ratio ($B$) into charged leptons,
in contrast to using mass and $\beta$ as done for the first and second generations.
This is due to the fact that $\beta$ is not equal to the branching ratio for third-generation LQs due to the sizeable top-quark mass. 
Previous ATLAS results for third-generation LQs for the case $B = 0$ for both \lqthreeu\ and \lqthreed\
used the $\sqrt{s} = 8$~\TeV\ dataset from 2012 \cite{EXOT-2014-03}.
The most recent CMS results are based on the 2016 dataset targeting $B = 0$ for both \lqthreeu\ and \lqthreed\ \cite{CMS-SUS-18-001}
as well as $B = 1$ for \lqthreeu\ 
\cite{Sirunyan:2018vhk}
and \lqthreed\ \cite{Sirunyan:2018nkj}.

The results presented here are from a
dedicated LQ search based on a search for pair-produced Higgs bosons 
decaying into two $b$-jets and two $\tau$-leptons~\cite{HIGG-2016-16}, 
where the search 
is optimized for the \lqthreeu\ pair production 
with $B \approx 1$, and four 
reinterpretations of ATLAS searches for supersymmetric particles.   
Supersymmetric particles can have similar or even identical experimental signatures and very similar kinematics to pair-produced LQs. 
Pair production of the supersymmetric partner of the top (bottom) quark, known as the top (bottom) squark, 
has the same experimental signature of a \antibar{t}-pair (\antibar{b}-pair) and missing transverse momentum 
as \lqthreeu\ (\lqthreed) pair production with 
$B =0$ (see Figure~\ref{fig:feynman-diagrams}).
Hence, the ATLAS searches for top squarks in final states with one \cite{SUSY-2016-16} or zero \cite{stop0LMoriond2017} leptons 
and for bottom squarks \cite{SUSY-2016-28}
are optimal when searching for \lqthreeu\ and \lqthreed\ with $B = 0$, respectively.   
The final state of two $\tau$-leptons, $b$-jets, and missing transverse momentum
is targeted in another top-squark pair-production search~\cite{SUSY-2016-19}
and is expected to be sensitive to medium and high branching ratios into charged leptons. 
For all analyses, the results are presented as a function of $B$ and the leptoquark mass for both \lqthreeu\ and \lqthreed.

\begin{figure}
\begin{center}
\includegraphics[width=0.49\linewidth]{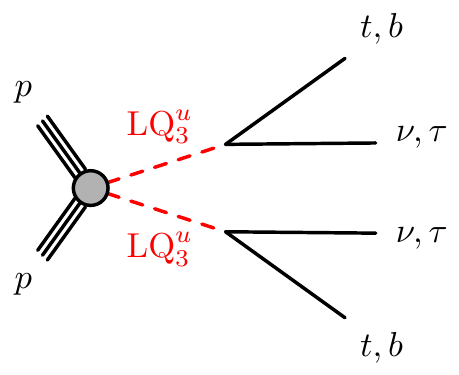}
\includegraphics[width=0.49\linewidth]{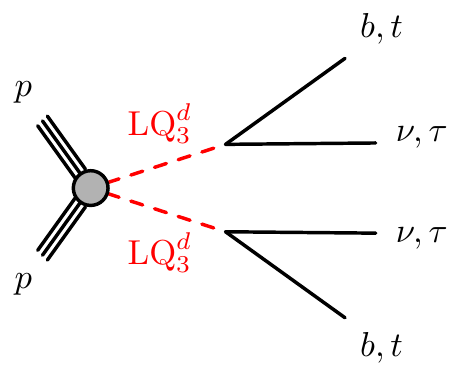}
\caption{
Pair production and decay of \lqthreeu\ and \lqthreed.
}
\label{fig:feynman-diagrams}
\end{center}
\end{figure}

The paper is structured as follows. After a brief description of the ATLAS detector, the Monte Carlo (MC) simulations 
of LQ pair production are discussed.
This is followed by a description of the five analyses, starting with the 
dedicated search for two $b$-jets and two $\tau$-leptons.
Four brief sections describe the reinterpretations
of searches for supersymmetric particles, 
as the published analyses are not modified for this reinterpretation. 
Each of the five analysis sections finish with cross-section and mass limits for a fixed value of $B$ to which 
the analysis is particularly sensitive.
Finally, the results of all analyses are presented as a function of $B$ 
and leptoquark mass.

\section{ATLAS detector}
\label{sec:detector}

The ATLAS detector~\cite{PERF-2007-01} is a multipurpose particle detector at the LHC with nearly $4\pi$ coverage around the collision point.\footnote{
	ATLAS uses a right-handed coordinate system with its origin at the nominal interaction point (IP) in the centre of the detector
	and the $z$-axis along the beam pipe. The $x$-axis points from the IP to the centre of the LHC ring, and the $y$-axis points upward. Cylindrical coordinates ($r$,$\phi$) are used in the transverse plane, $\phi$ being the azimuthal angle around the beam pipe.
	The pseudorapidity is defined in terms of the polar angle $\theta$ as $\eta= -\ln\tan(\theta/2)$. Angular distance is measured in units of $\Delta R = \sqrt{(\Delta \eta)^2+(\Delta \phi)^2}$.}
Closest to the beam is the inner detector, which provides charged-particle tracking in the range $|\eta| < 2.5$. During the LHC shutdown between Run 1 and Run 2, a new innermost layer of silicon pixels was added, which improves the track impact parameter resolution and vertex position resolution~\cite{ATLAS-TDR-19,ATL-PHYS-PUB-2015-051}. 
 The inner detector is surrounded by a superconducting solenoid providing a 2\,T axial magnetic field, 
followed by a lead/liquid-argon electromagnetic sampling calorimeter and a steel/scintillator-tile hadronic calorimeter. 
The endcap and forward regions are instrumented with liquid-argon calorimeters for both the electromagnetic and hadronic energy measurements up to $|\eta|=4.9$.
The outer part of the detector consists of a muon spectrometer with high-precision tracking chambers for coverage up to $|\eta|=2.7$,
fast detectors for triggering over $|\eta| < 2.4$, and three large superconducting toroidal magnets with eight coils each.
Events are selected by a two-level trigger system consisting of a hardware-based trigger for the first level and a software-based system for the second level~\cite{TRIG-2016-01}.

\section{Signal simulation}
\label{sec:signal_mc}

Signal samples were generated at next-to-leading order (NLO) in QCD with MadGraph5\_aMC@NLO 2.4.3 \cite{Alwall:2014hca},
using the LQ model of Ref.~\cite {Mandal:2015lca} that adds parton showers to previous fixed-order NLO OCD calculations~\cite{Kramer:2004df},
and the NNPDF 3.0 NLO \cite{Ball:2014uwa} parton distribution functions (PDF),
interfaced with \PYTHIA 8.212 \cite{Sjostrand:2014zea} using the A14 set of tuned parameters~\cite{ATL-PHYS-PUB-2014-021}
for the parton shower and hadronization.
The leptoquark signal production cross-sections are taken from calculations~\cite{xsref} of direct top-squark pair production,
as both are massive, coloured, scalar particles with the same production modes. The calculations are at NLO plus next-to-leading-logarithm accuracy, with uncertainties determined by variations of factorization and renormalization scales, $\alpha_\mathrm{s}$, and PDF variations.

Madspin \cite{Artoisenet:2012st} was used for the decay of the LQ. The parameter $\lambda$ was set to 0.3, resulting in 
a LQ width 
of about 
0.2\% of its mass~\cite{Buchmuller:1986zs,Belyaev:2005ew}.
The samples were produced for a model parameter of $\beta=0.5$, where 
desired branching ratios $B$ were obtained by reweighting the samples based on generator information.
Additional samples for $\beta=1$ are 
used in the analysis
of the final state with two $b$-jets and two $\tau$-leptons.

Due to the difference between the model parameter $\beta$ and $B$ for third-generation LQs, 
the branching ratio $\hat{B}$ in the simulated sample with $\beta=0.5$
can either be calculated from the given parameter $\lambda$ and the resulting decay width or be taken from the generator information of the MC sample. 
It is shown in Figure \ref{fig:lq3_br} as a function of the leptoquark mass.
\begin{figure}[h]
\centering
\includegraphics[width=0.7\linewidth]{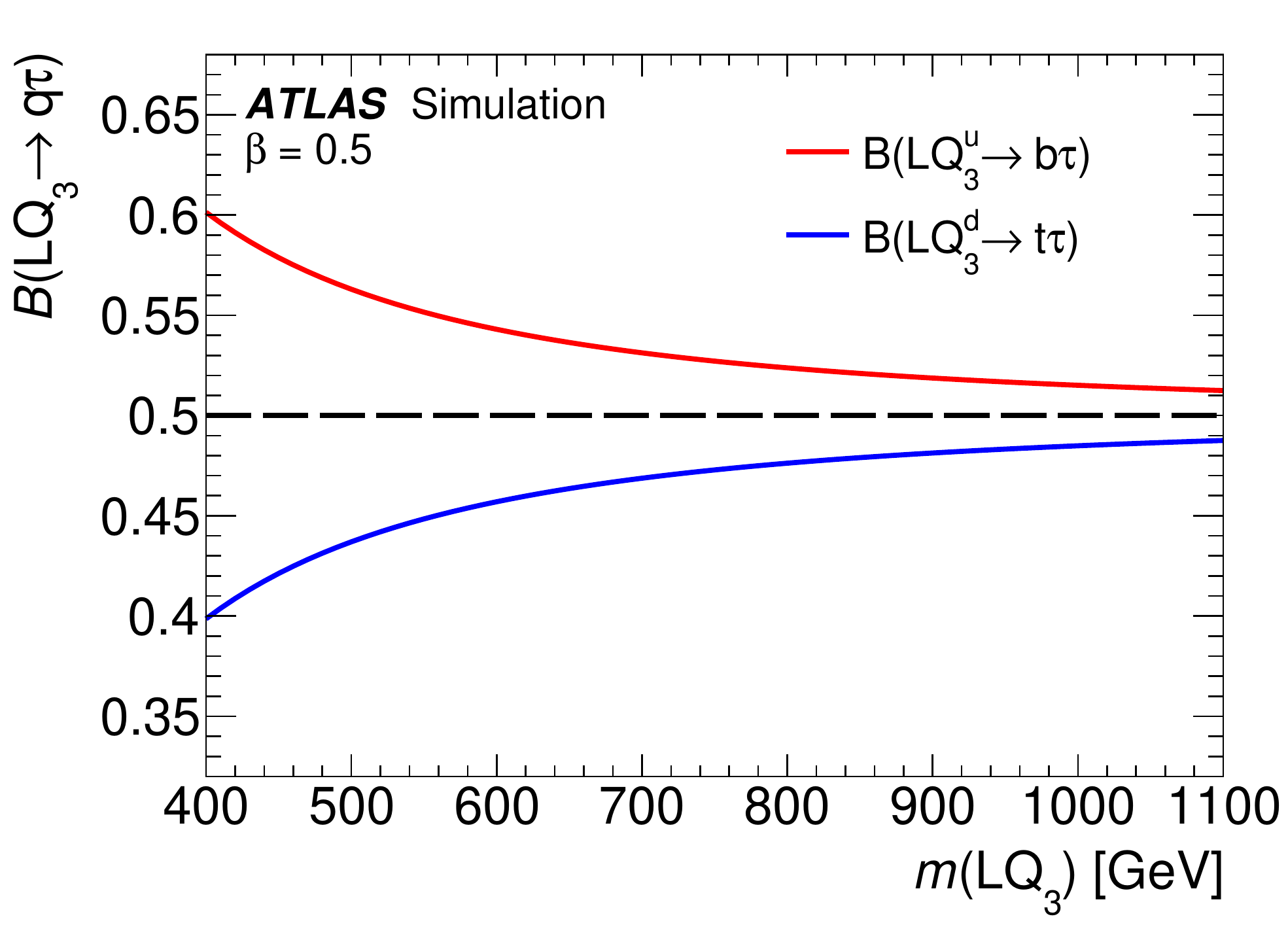}
\caption{Branching ratio into charged leptons for $\beta=0.5$ and different LQ masses for 
\lqthreeu\ $\rightarrow b \tau / t \nu$ and \lqthreed~$\rightarrow t \tau / b \nu$.
}
\label{fig:lq3_br}
\end{figure}
The reweighting is based on the number of charged leptons $n_{\mathrm{cl}}$ at generator level originating directly from the decaying leptoquarks for each event. The weight $w$ depends on the $\hat{B}$ of the produced MC sample and on the $B$ of the decay channel and 
is
\begin{equation}
\nonumber
w(B) = \left(\frac{B}{\hat{B}}\right)^{n_{\mathrm{cl}}} \times \left(\frac{1-B}{1-\hat{B}}\right)^{\left(2-n_{\mathrm{cl}}\right)}.
\end{equation}

\section{The $b \tau b \tau$ channel}

To search for pair-produced scalar leptoquarks decaying into \btaubtau,
final states in which one $\tau$-lepton decays leptonically and the other hadronically (\tlhad), as well as the case in which both $\tau$-leptons decay hadronically (\thadhad), are considered. This analysis utilizes the same analysis strategy employed by the ATLAS search for pair-produced Higgs bosons decaying into $bb \tau\tau$ final states \cite{HIGG-2016-16} but optimized for a leptoquark signal with decays into $b \tau b \tau$.

\begin{figure}
  \centering
  \small
  \includegraphics[width=0.75\textwidth]{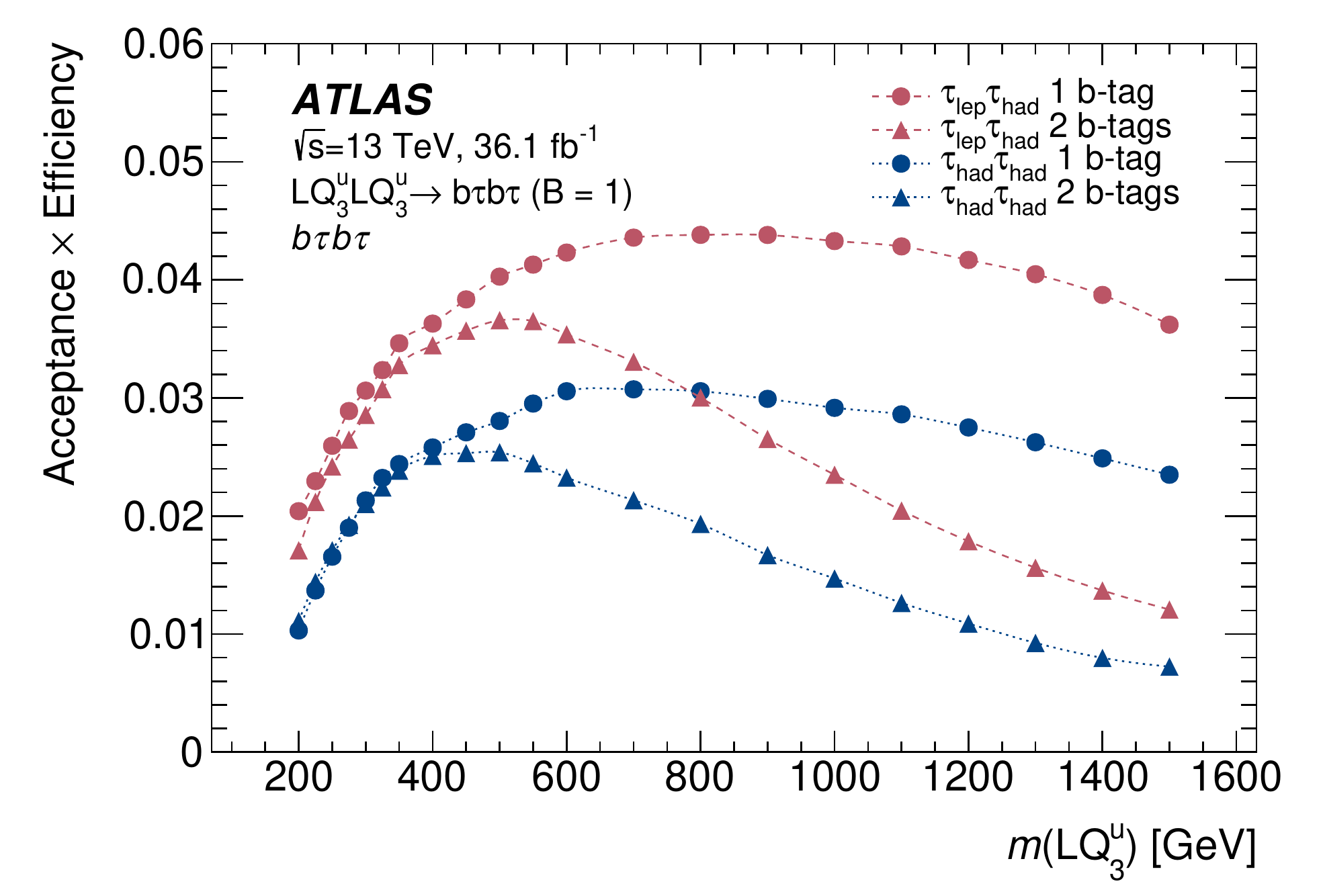}
  \caption{The acceptance times efficiency for the up-type leptoquark signal is shown for the \tlhad\ and \thadhad\ channels as a function of the leptoquark mass.  The offline selection is summarized in Table \ref{selection}. The results are given separately for selected events with one and two b-tagged jets.
  }
  \label{fig:acceptance}
\end{figure}

Object reconstruction in the detector (electrons, muons, $\tau$-leptons, jets, and $b$-jets) employed for \tlhad\ and \thadhad\ channels is the same as in Ref. \cite{HIGG-2016-16}.  The data were collected through three triggers, a single-lepton (electron or muon) trigger (SLT), a single-$\tau$-lepton trigger (STT), and a di-$\tau$-lepton trigger (DTT).  The offline selection is dependent on the trigger and is summarized in Table \ref{selection}. Events in each channel must pass the associated trigger or else the event is discarded.  The selection criteria are chosen to optimize the trigger efficiency for the associated data samples.  Events are split into categories according to the multiplicity of $b$-tagged jets.  Events with one or two $b$-tagged jets are considered signal-like events (1-tag and 2-tag events), which form two separate signal regions.  In events with one $b$-tag, the highest-\pt\ non-tagged jet is considered for leptoquark event reconstruction.  The acceptance times efficiency for the \lqthreeu\ signal is shown in Figure \ref{fig:acceptance} for \tlhad\ and \thadhad\ channels as a function of the leptoquark mass with $B = 1$.  The decrease in acceptance times efficiency with leptoquark mass is driven by a combination of reduced b-tagging effiency and the efficiency in pairing the bottom quark and $\tau$-lepton.  After applying the selection criteria, boosted decision trees (BDTs) are used to improve discrimination between signal and background.  The BDTs are only trained on the up-type leptoquark signal. The sensitivity to the down-type leptoquark decay channel is due to the final state $t\tau t\tau \rightarrow Wb\tau Wb\tau$, where the \Wboson bosons decay into jets.  Because this analysis does not veto additional jets, it is sensitive to this decay chain, although it is not optimal.

\begin{table}
\centering
\caption{Summary of applied event selection for the \tlhad\ and \thadhad\ channels.}
\label{selection}
\footnotesize
\begin{tabular}{ |l|m{0.85\textwidth}|}
\hline
\tlhad (SLT) & 

\begin{itemize} 
\item{Exactly one $e$ passing `tight' identification criteria or one $\mu$ passing `medium' identification criteria \cite{PERF-2013-03,PERF-2015-10}.  Events containing additional electrons or muons with `loose' identification criteria and \pt $>$ 7 ~\GeV\ are vetoed.} 
\item{Exactly one hadronically decaying $\tau$-lepton with transverse momentum (\pt) $>$~25~\GeV\ and $|\eta|$ < 2.3.} 
\item{Opposite-sign charge between the $\tau$-lepton and the
light lepton ($e/\mu$).} 
\item{At least two central jets in the event with \pt $>$ 60 (20)~\GeV\ for the
leading (subleading) jet.} 
\end{itemize}\\

\hline

\thadhad (STT) & 
\begin{itemize} 
\item{Events containing electrons or muons with `loose' identification criteria and \pt $>$ 7 ~\GeV\ are vetoed.} 
\item{Exactly two hadronically decaying $\tau$-leptons with $|\eta| < 2.3$. The leading $\tau$-lepton must have \pt $>$ 100, 140, or 180~\GeV\ for data periods where the trigger \pt~threshold was 80, 125, or 160~\GeV\, respectively.  The subleading $\tau$-lepton is required to have \pt\ $>$ 20~\GeV.} 
\item{The two $\tau$-leptons must have opposite-sign charge. } 
\item{At least two jets in the event with \pt $>$ 45 (20)~\GeV\ for the leading (subleading) jet.} 
\end{itemize} \\

\hline
\thadhad (DTT) & \begin{itemize}
    \item{Selected events from STT are vetoed as are events containing electrons or muons with `loose' identification criteria and \pt $>$ 7 ~\GeV.}
    \item{Exactly two hadronically decaying $\tau$-leptons with $|\eta| < 2.3$. The leading (subleading) $\tau$-lepton must have \pt $>$ 60 (30)~\GeV.} 
    \item{The two $\tau$-leptons must have opposite-sign charge. } 
    \item{At least two jets in the event with \pt $>$ 80 (20)~\GeV\ 
          for the leading (subleading) jet.} 
  \end{itemize}\\
  \hline
\end{tabular}
\end{table}

BDTs are trained to separate the signal from the expected backgrounds, and the BDT score distributions are used as the final discriminant to test for the presence of a signal. The BDTs utilize several input variables, shown in the list below, including those derived from a mass-pairing strategy which extracts the most likely $b\tau$ pairs by minimizing the mass difference between the leptoquark candidates:  

\begin{itemize} 
  \item{$s_\text{T}$: the scalar sum of missing transverse momentum (\met), the \pt\ of any reconstructed $\tau$-lepton(s), the \pt of the two highest-\pt jets, and the \pt of the lepton in \tlhad\ events} 
  \item{$m_{\tau, \mathrm{jet}}$: the invariant mass between the leading $\tau$-lepton and its mass-paired jet 
  \item{$m_{\ell, \mathrm{jet}}$: the invariant mass between the lepton and its matching jet from the mass-pairing strategy (\tlhad\ only)}
  \item{$\Delta R (\mathrm{lep}, \mathrm{jet})$: the $\Delta R$ between the electron or muon (leading $\tau$-lepton) and jet in \tlhad\ (\thadhad)} 
  \item{\MET $\phi$ centrality: quantifies the $\phi$ separation between the \met and $\tau$-lepton(s).  Full definition is in Ref.~\cite{HIGG-2016-16}} 
  \item{$p_\text{T}^{\tau}$}: the \pt of the leading $\tau$-lepton}
  \item{$\Delta\phi(\mathrm{lep},\MET)$: the opening angle between the lepton and \met (\tlhad\ only)} 
\end{itemize}

Kinematic distributions for \lephad\ and \hadhad\ signal regions in 2-tag events are shown in Figure \ref{fig:bdtinput}.  Separate BDTs are trained for \tlhad\ and \thadhad\ categories for each mass point of the \lqthreeu\ MC sample and for each $b$-tag signal region. 
The signal samples used in the training include events with a small range of leptoquark masses around the given mass point to ensure the BDT is sensitive to signals that have masses between the hypotheses simulated. In the \tlhad\ channel the training is performed against the dominant $t\bar{t}$ background only. BDTs for the \thadhad\ channel are trained against simulated $t\bar{t}$ and $Z\rightarrow \tau\tau$ events and multi-jet events from data.

\begin{figure}
  \centering
 {
    \includegraphics[width=0.43\textwidth]{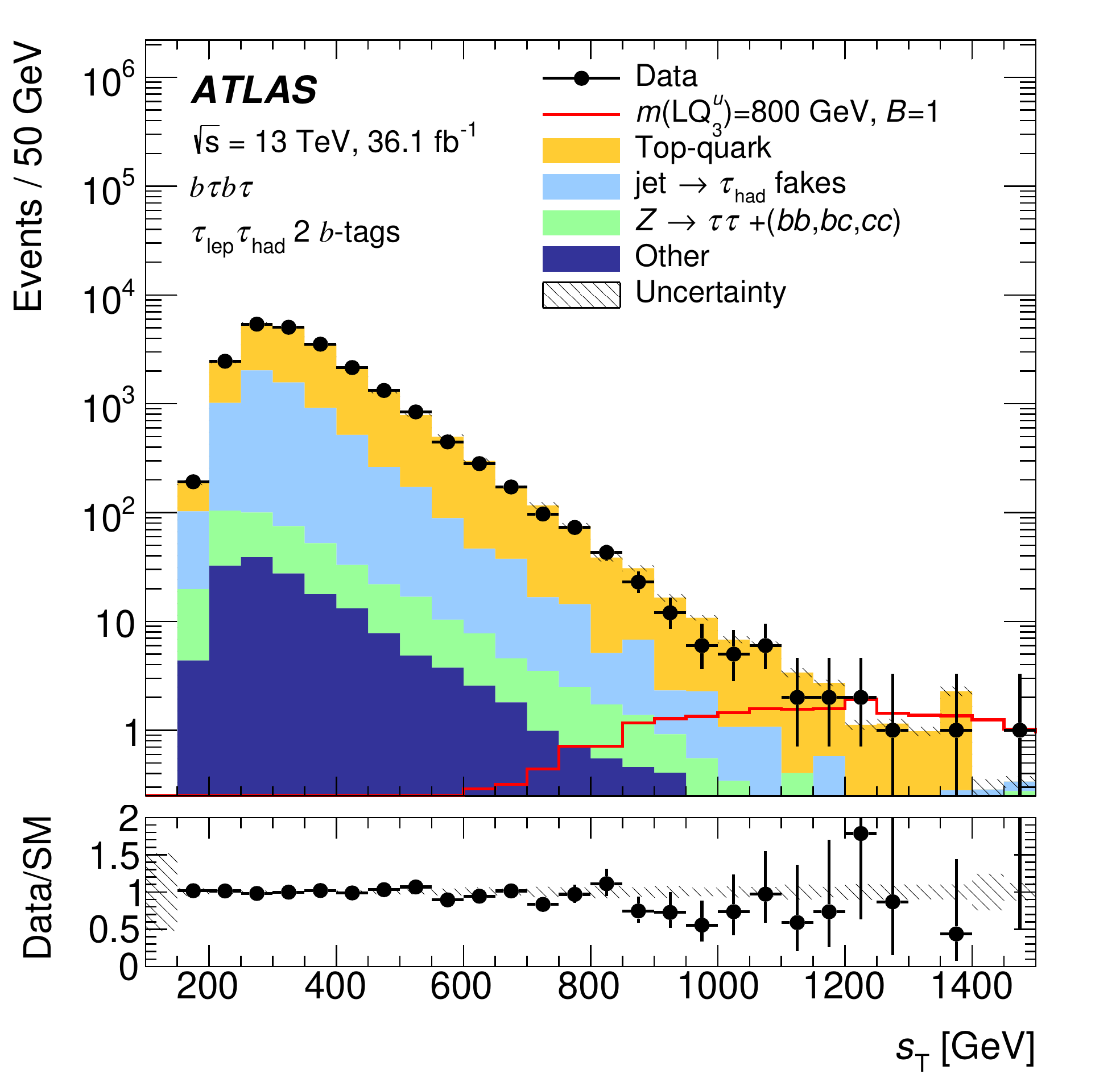}
  }
 {
    \includegraphics[width=0.43\textwidth]{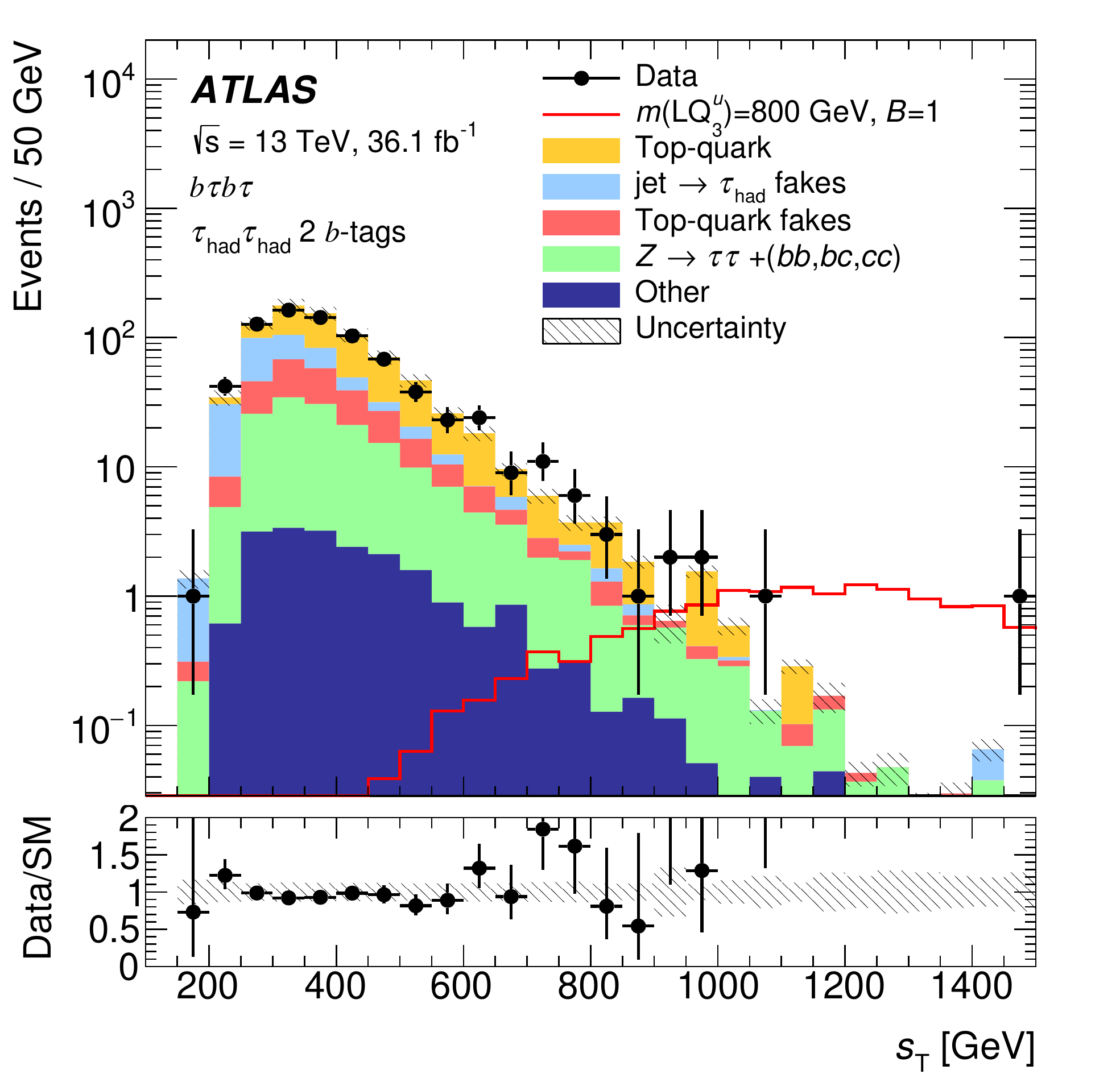}
  }\\
 {
    \includegraphics[width=0.43\textwidth]{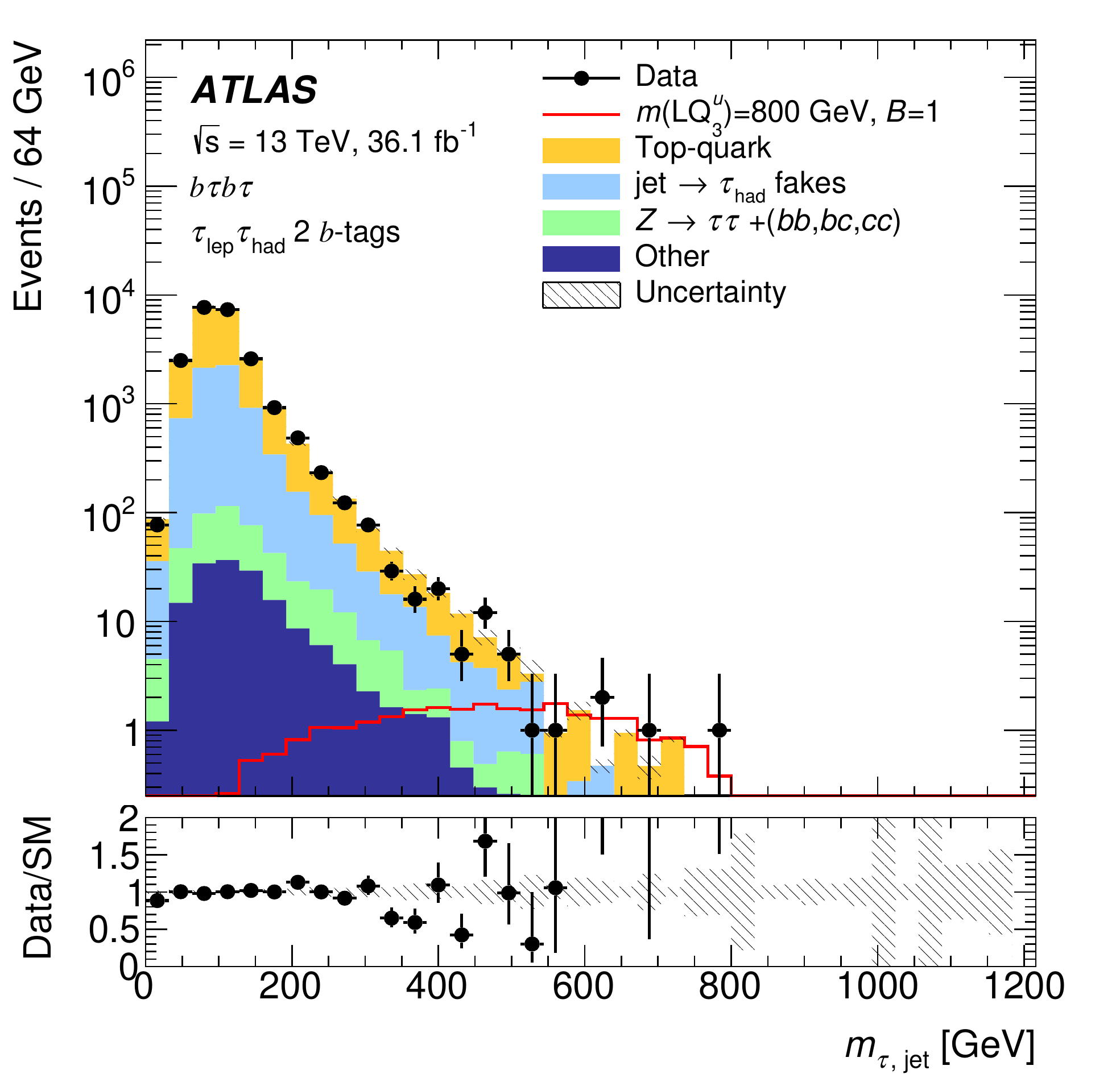}
  }
 {
    \includegraphics[width=0.43\textwidth]{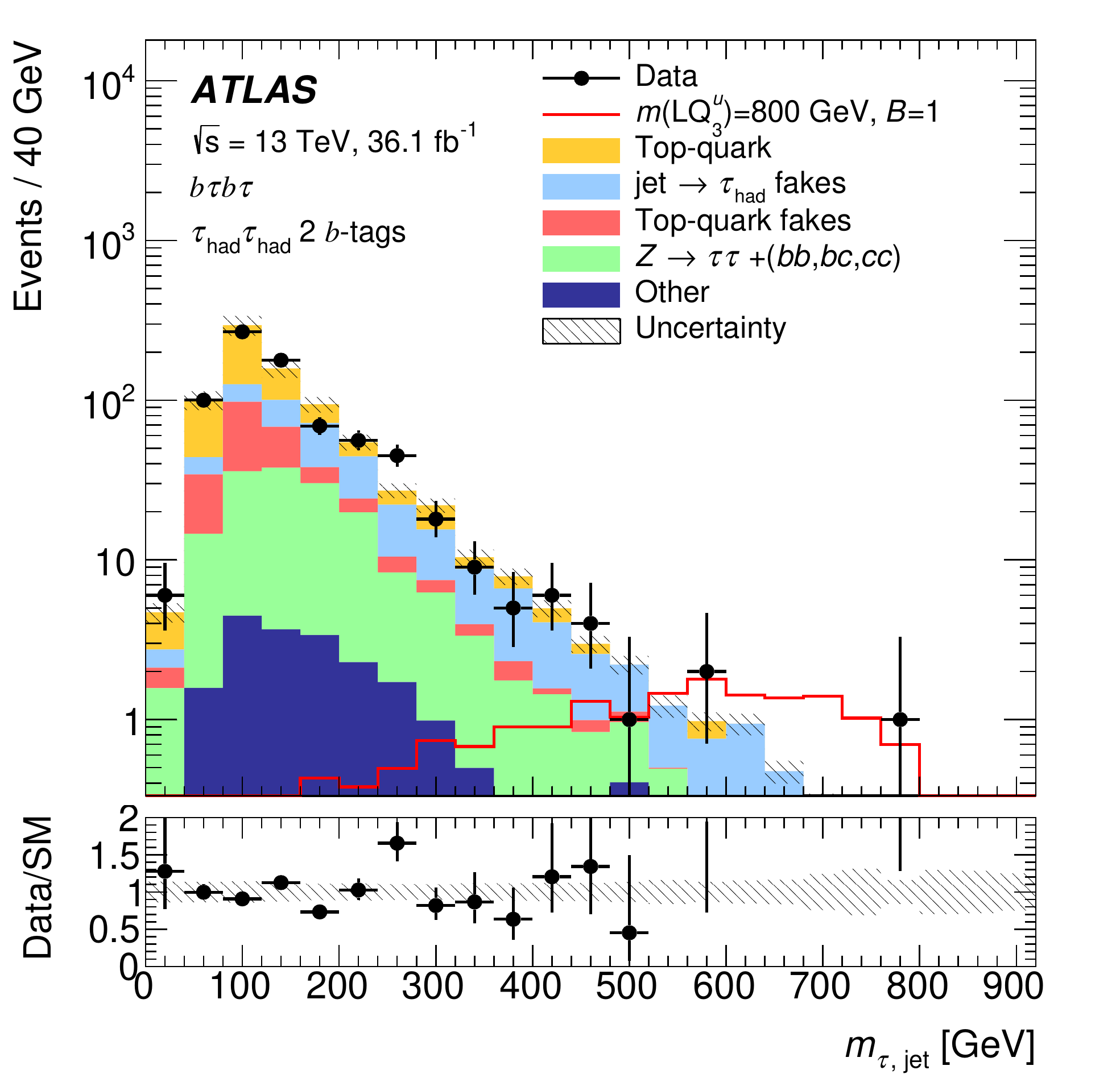}
  }\\
 {
    \includegraphics[width=0.43\textwidth]{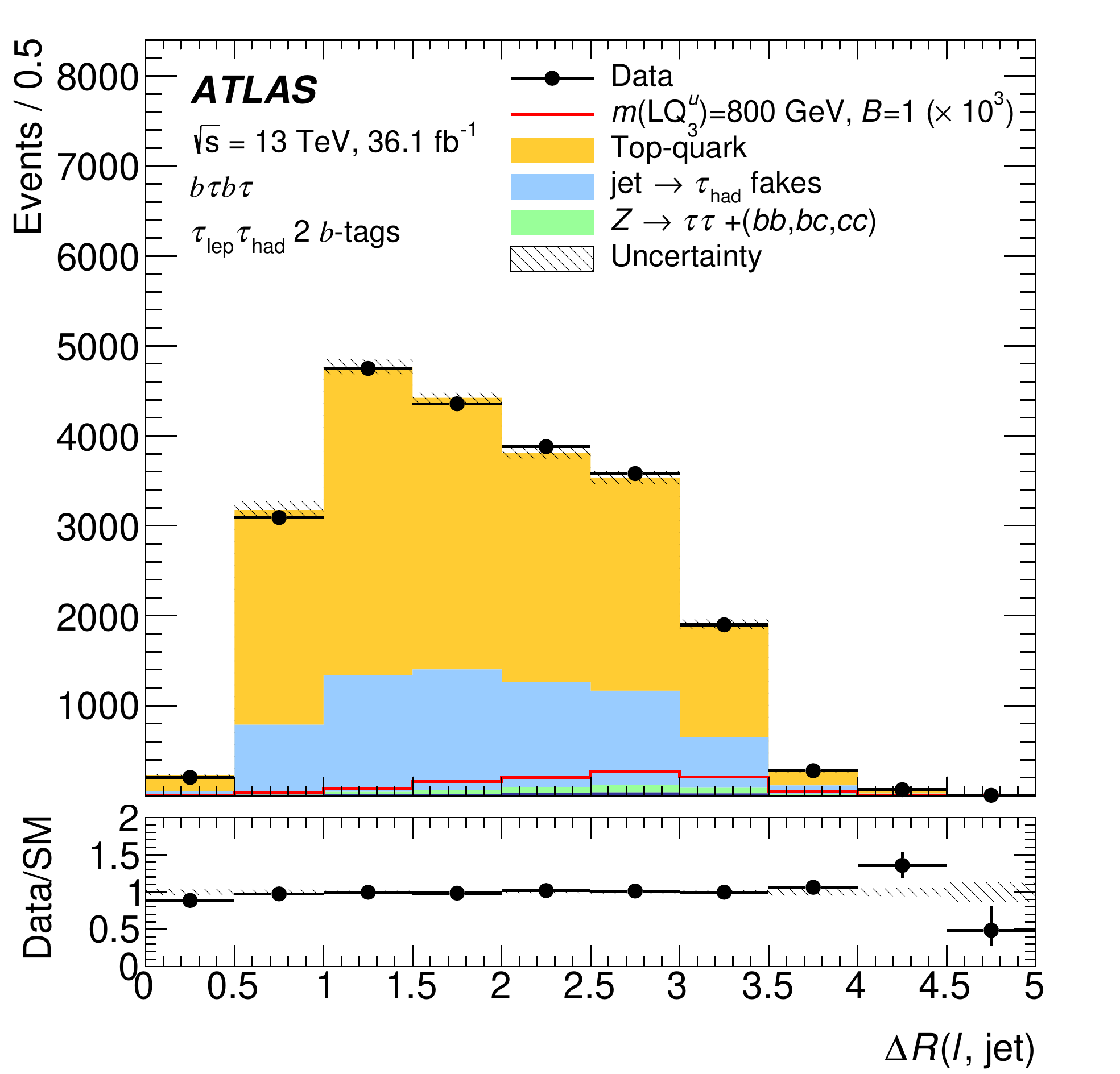}
  }
 {
    \includegraphics[width=0.43\textwidth]{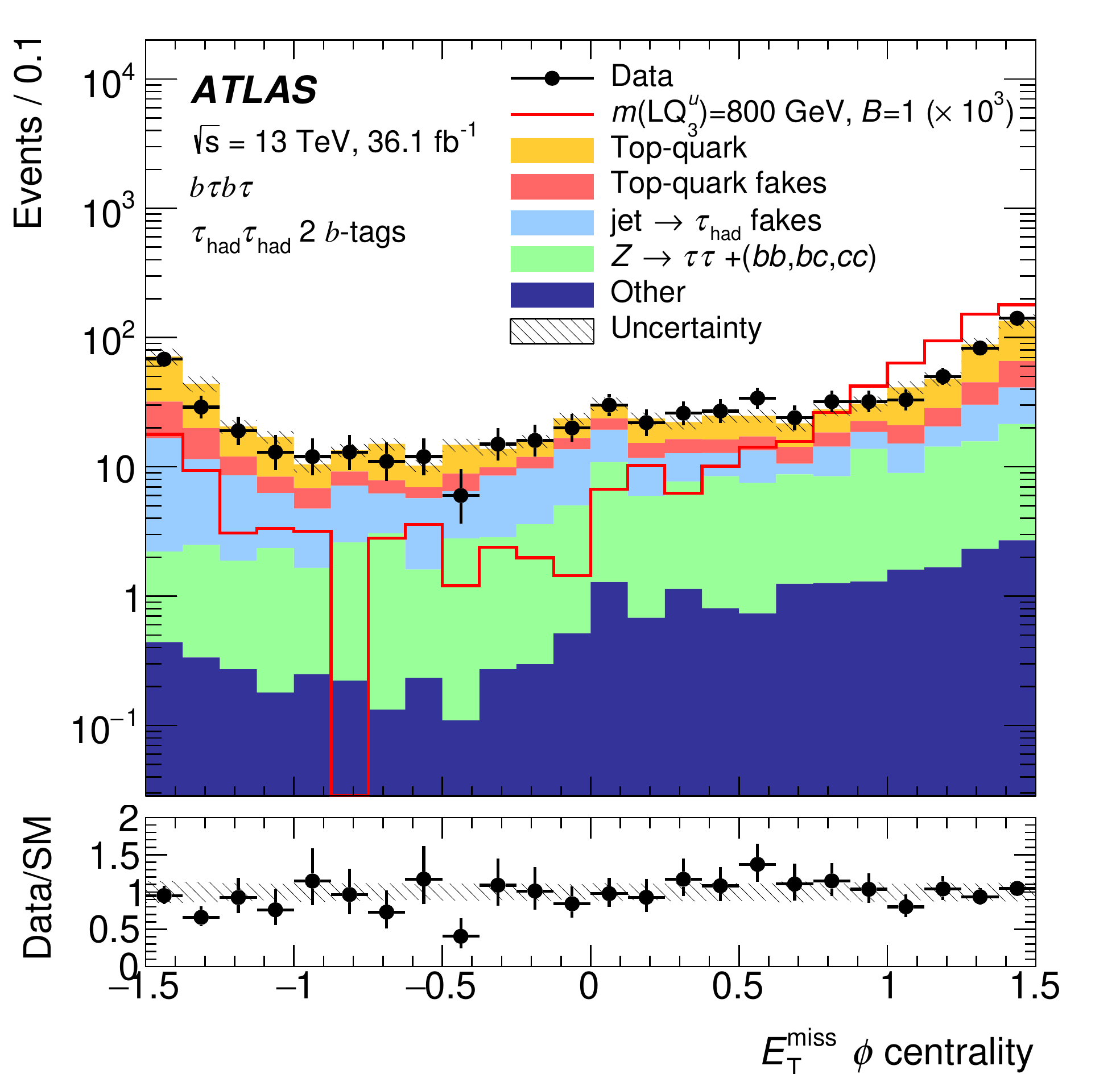}
  }
  \caption{Kinematic distributions for \lephad\ (left) and \hadhad\ (right) signal regions in 2-tag events after performing the combined channel fit.  The ratio of the data to the sum of the backgrounds is shown in the lower panel.  The hatched bands indicate the combined statistical and systematic uncertainties in the background.  Distributions include the scalar sum of transverse momentum of reconstructed objects in the event ($s_\text{T}$), the invariant mass between the leading $\tau$-lepton and its mass-paired jet ($m_{\tau, \mathrm{jet}}$), the invariant mass between the lepton and its matching jet from the mass-pairing strategy ($m_{\ell, \mathrm{jet}}$), the $\Delta R$ between the leading $\tau$-lepton and jet ($\Delta R (\mathrm{lep}, \mathrm{jet})$), and  the \MET $\phi$ centrality, which quantifies the $\phi$ separation between the \met and $\tau$-lepton(s).  }
  \label{fig:bdtinput}
\end{figure}

The background estimation techniques in this analysis are the same as used in Ref. \cite{HIGG-2016-16} and are summarized here. In both channels, background processes containing true $\tau$-leptons are taken from simulation.  The dominant background processes are $\ttbar$ and $\Ztautau$ produced in association with jets originating from heavy-flavour quarks ($bb,bc,cc$).  The $\ttbar$ events are normalized using events with low BDT output score in the \tlhad\ category.  Events from $\Ztautau$ plus heavy-flavour jets are normalized using a control region of events that include two muons with combined invariant mass consistent with that of a \Zboson\ boson.  Backgrounds in which quark- or gluon-initiated jets are misidentified as hadronically decaying $\tau$-leptons are estimated using data-driven methods.  In both channels, $\ttbar$ events are estimated separately from other background sources if one or more reconstructed $\tau$-lepton decays are mis-reconstructed jets (so-called `fake $\tau$-leptons').  In the \tlhad\ channel all fake-$\tau$-lepton contributions from $\ttbar$, $W$+jets, and multi-jet processes are estimated using an inclusive fake-factor method, described in Ref.~\cite{HIGG-2016-16}. Theory uncertainties in the modeling of the $\ttbar$ and $Z$+jets background containing true $\tau_\mathrm{had}$ are assessed by varying the matrix element generator and the parton shower model, and by adjusting the factorization and renormalization scales along with the amount of additional radiation.  The resulting variations in the BDT distributions are included as shape uncertainties in the final fit.  

In the \thadhad\ channel, the fake-$\tau$-lepton $\ttbar$ component is estimated as follows:  the probability for a jet from a hadronic $W$-boson decay to be reconstructed as a hadronically decaying $\tau$-lepton is measured in data.  This is used then to correct the MC prediction, after subtracting the predicted number of true $\tau$-leptons from the MC.  Three control regions are defined for both \tlhad\ and \thadhad: these include 1-$b$-tag and 2-$b$-tag same-sign lepton events that are mostly events with fake $\tau$-leptons and a $\ttbar$ control region as defined in Ref.~\cite{HIGG-2016-16}. The uncertainty in the modeling is estimated by varying the fake-factors and fake-rates within their statistical uncertainties and varying the amount of true $\tau_\mathrm{had}$ background subtracted.  Systematic uncertainties on the extrapolation of the fake $\tau_\mathrm{had}$ backgrounds from the control regions, where they are derived, to the signal regions are estimated by varying the control region definition; an uncertainty due to the difference in the flavour composition of the jet faking the $\tau_\mathrm{had}$ is also assigned based on simulation.  

The BDT responses in the 2-tag same-sign and top-quark control regions are shown in Figure \ref{fig:bdtoutput_CR}.  The background modeling was checked in these control regions, which validates the signal-sensitive region at high BDT score.  The 1-tag and 2-tag \lephad\ and \hadhad\ signal regions are shown in Figures \ref{fig:bdtoutput_low} and \ref{fig:bdtoutput_high} for up-type samples with a mass of 400~\GeV\ ($B=1$) and down-type leptoquark samples with a mass of 800~\GeV\ ($B=1$). The binning choice differs between the figures as this follows an algorithm depending on the number of signal (in the case of the signal region only) and background events in each bin.  Yield tables are shown for both 1-tag and 2-tag regions in Table \ref{tab:postfitYields}, where the numbers quoted are after performing the combined fit, described below.   

\begin{figure}
  \centering
{
    \includegraphics[width=0.49\textwidth]{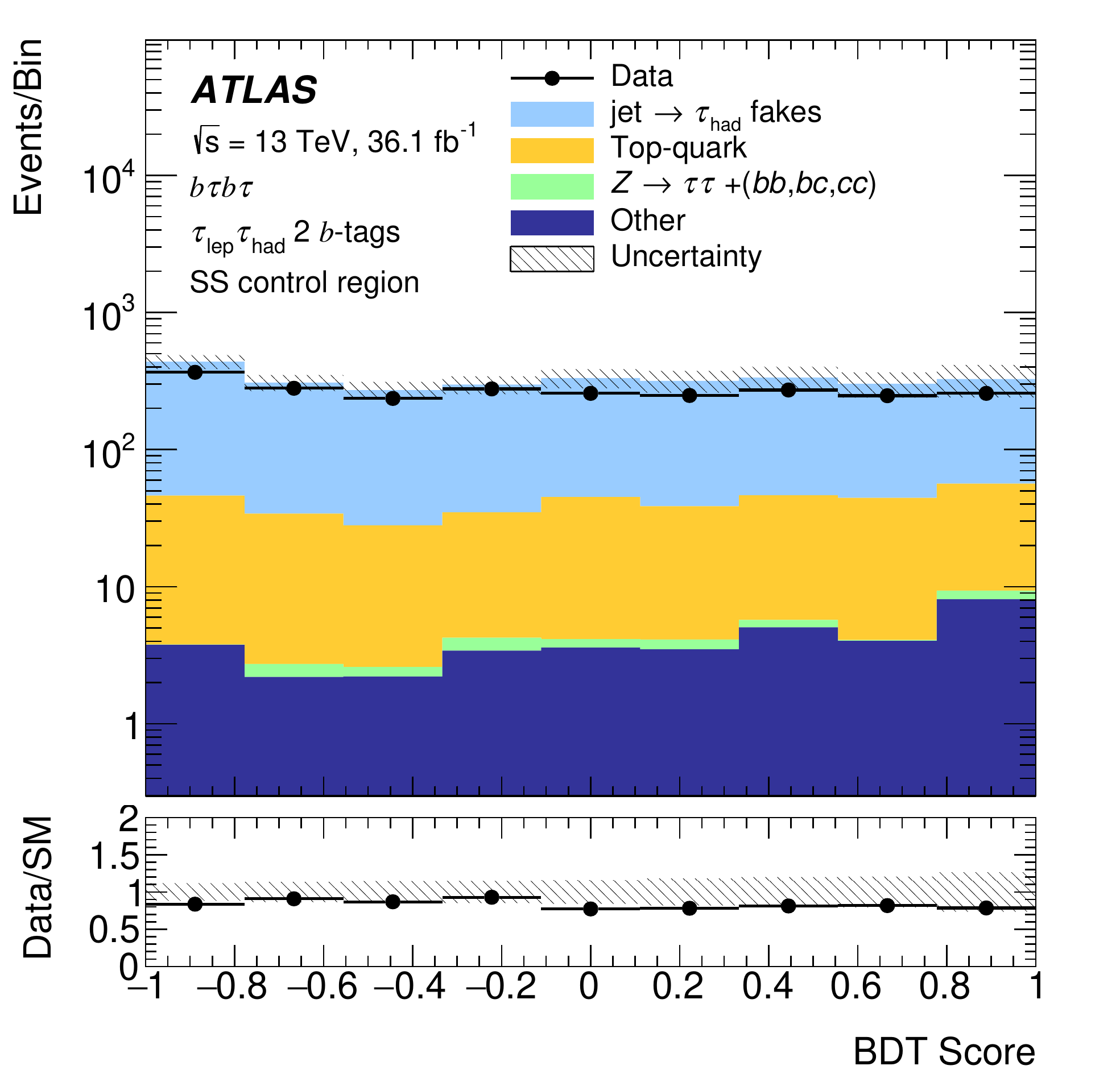}
  }
{
    \includegraphics[width=0.49\textwidth]{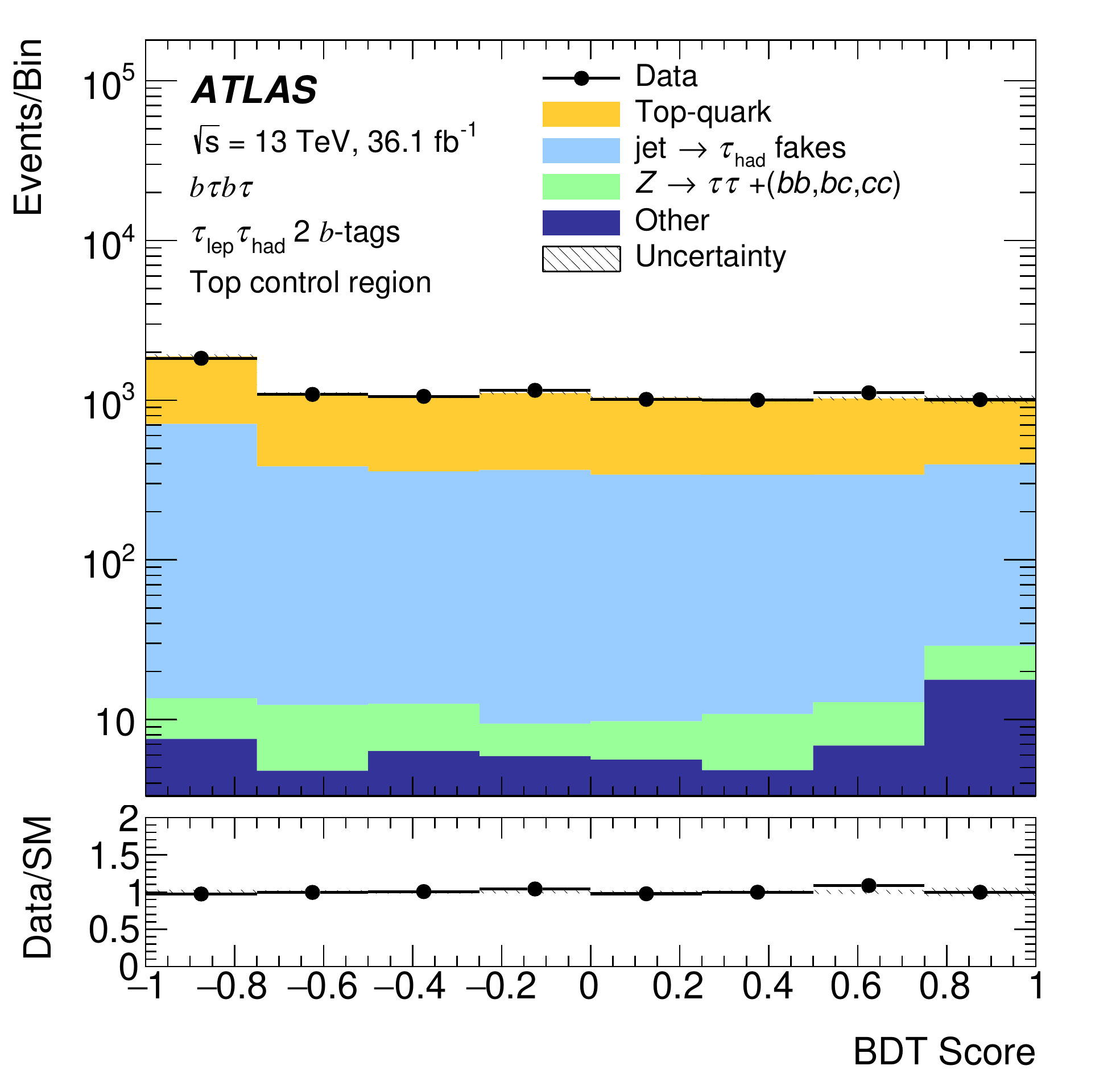}
  }\\
{
    \includegraphics[width=0.49\textwidth]{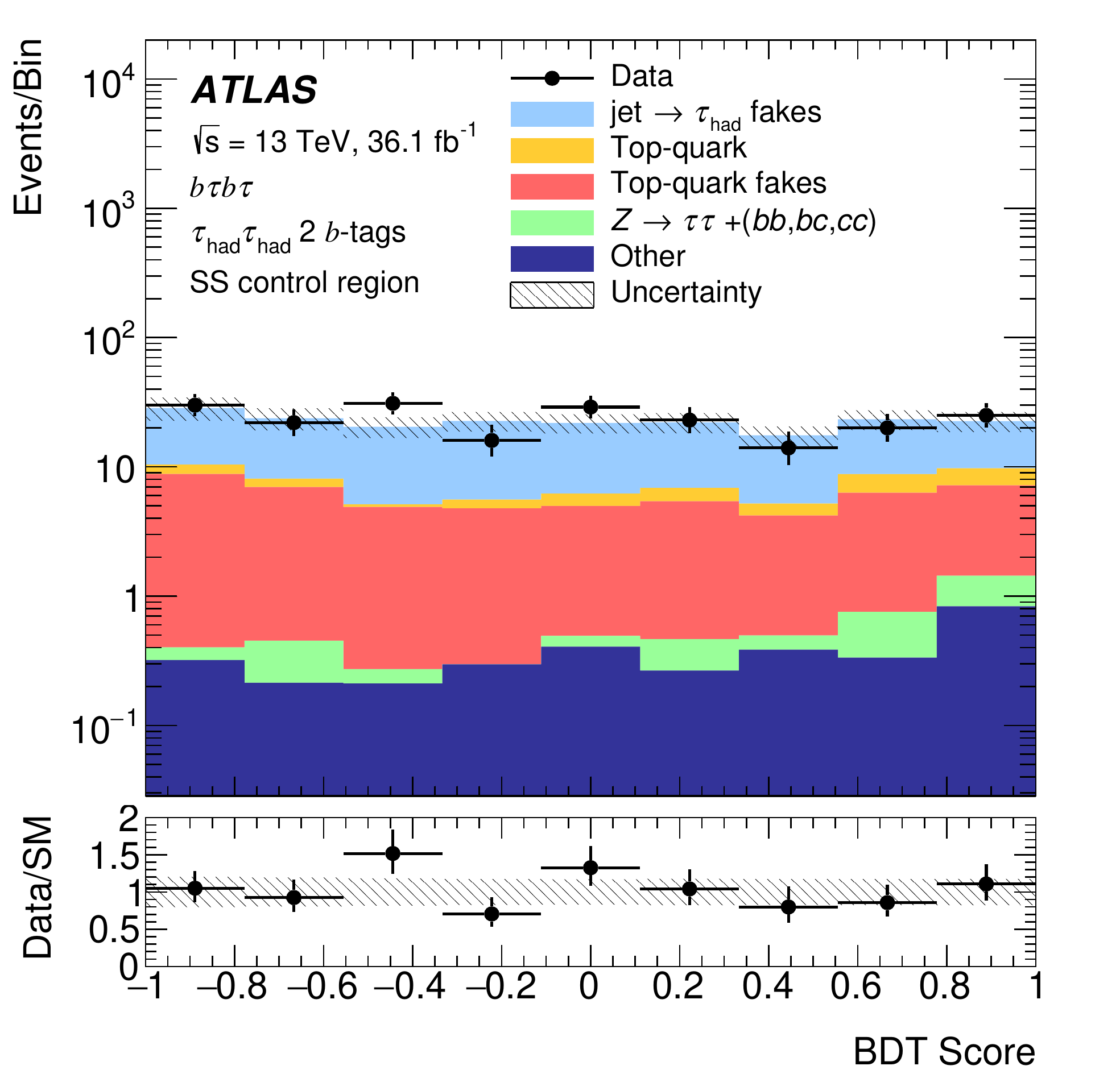}
  }
{
    \includegraphics[width=0.49\textwidth]{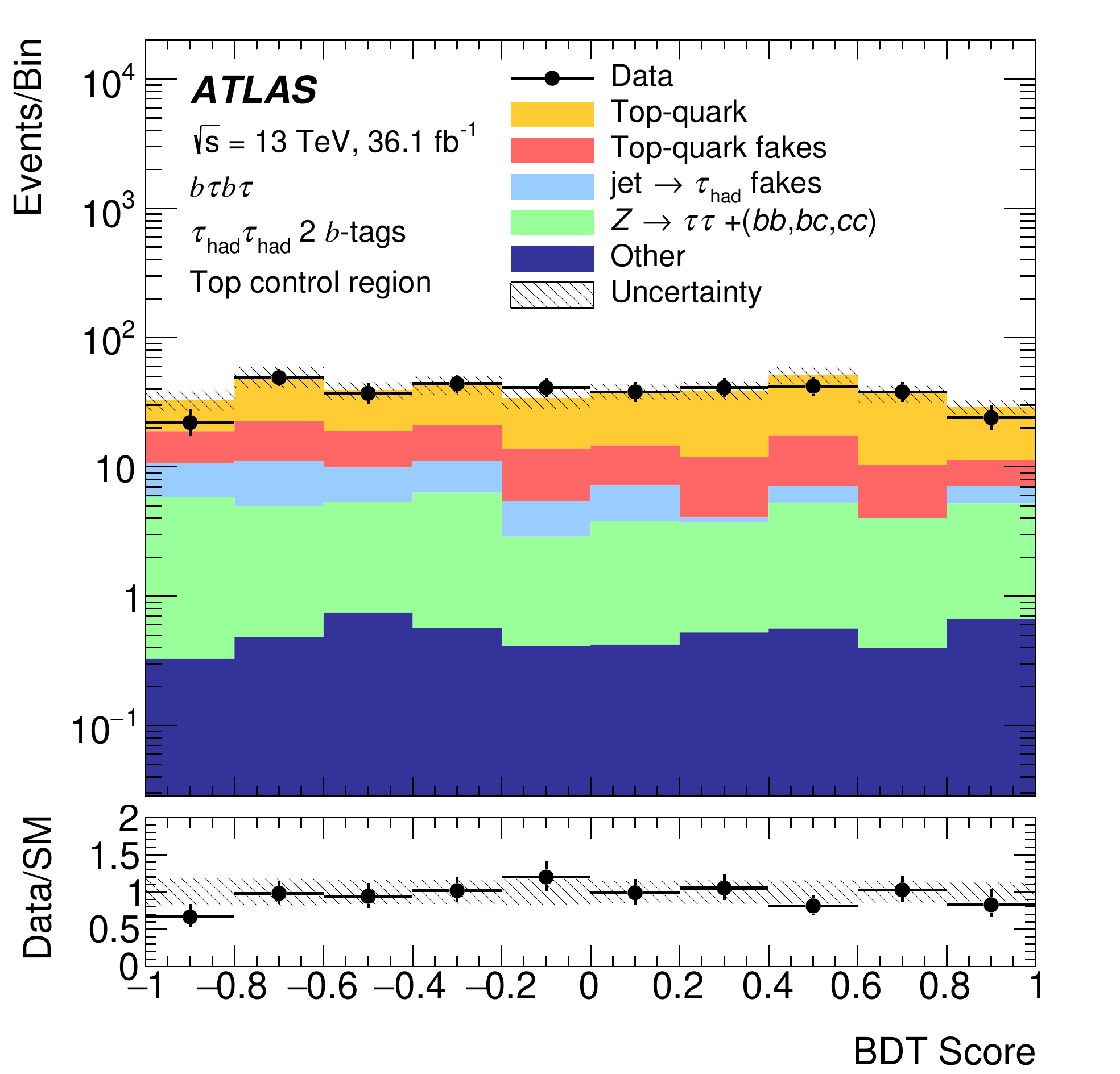}
  }
  \caption{BDT score distributions for \lephad\ (top) and \hadhad\ (bottom) 2-tag channels in the same-sign (left) and top-quark (right) control regions after performing the combined fit to all channels.  The ratio of the data to the sum of the backgrounds is shown in the lower panel.  The hatched bands indicate the combined statistical and systematic uncertainties in the background.}
  \label{fig:bdtoutput_CR}
\end{figure}

\begin{figure}
  \centering
{
    \includegraphics[width=0.49\textwidth]{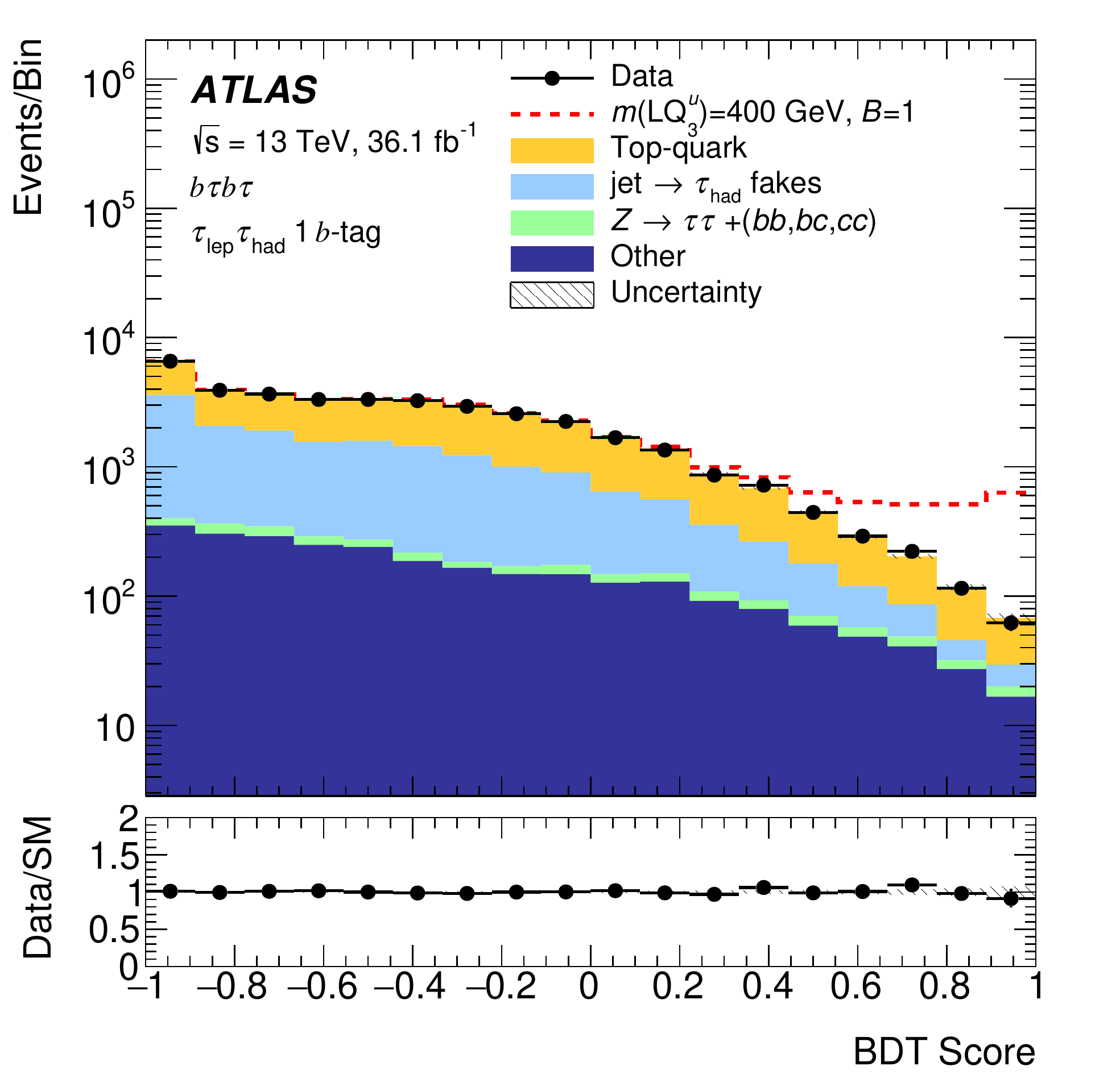}
  }
{
    \includegraphics[width=0.49\textwidth]{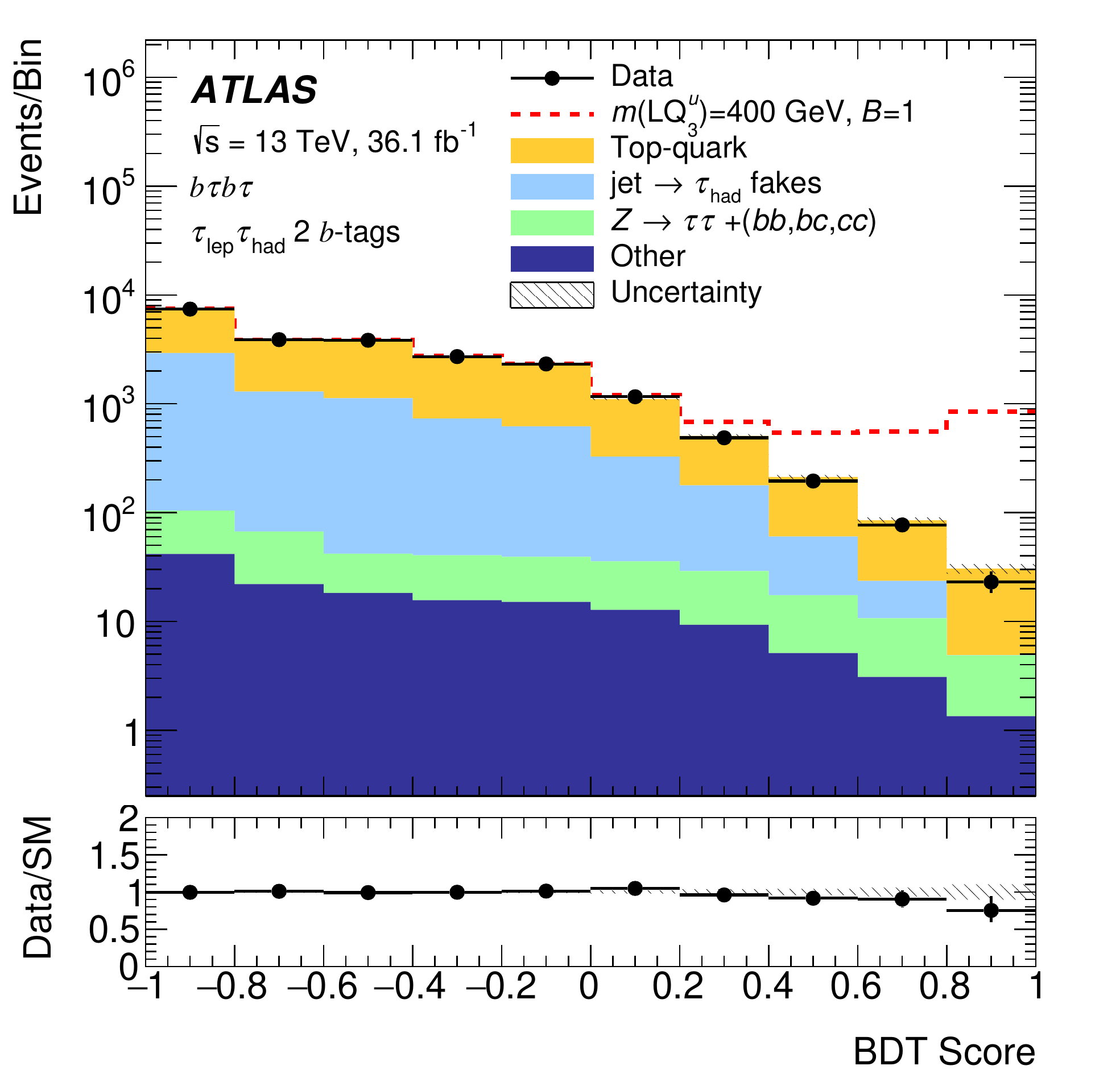}
  }\\
{
    \includegraphics[width=0.49\textwidth]{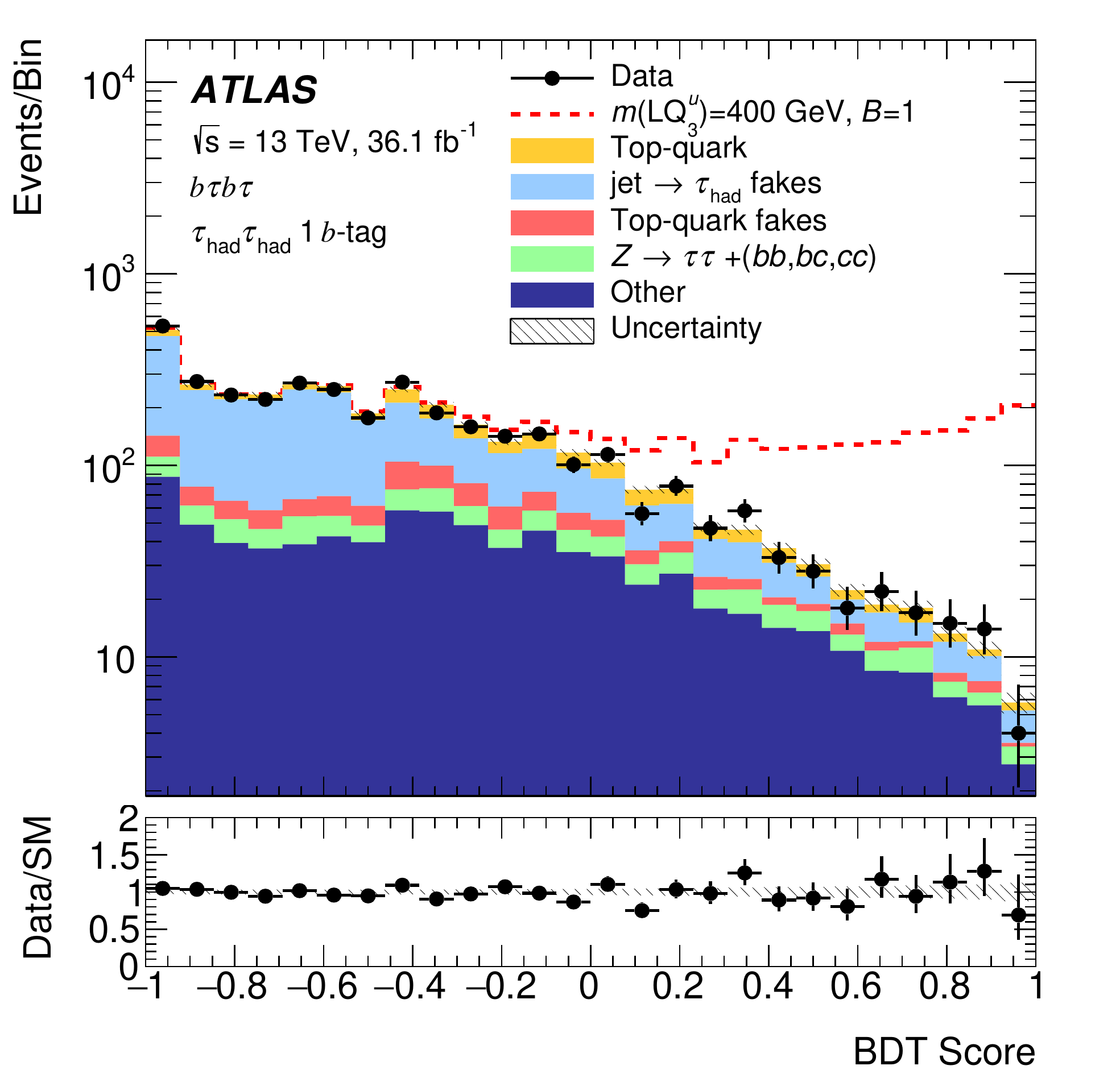}
  }
{
    \includegraphics[width=0.49\textwidth]{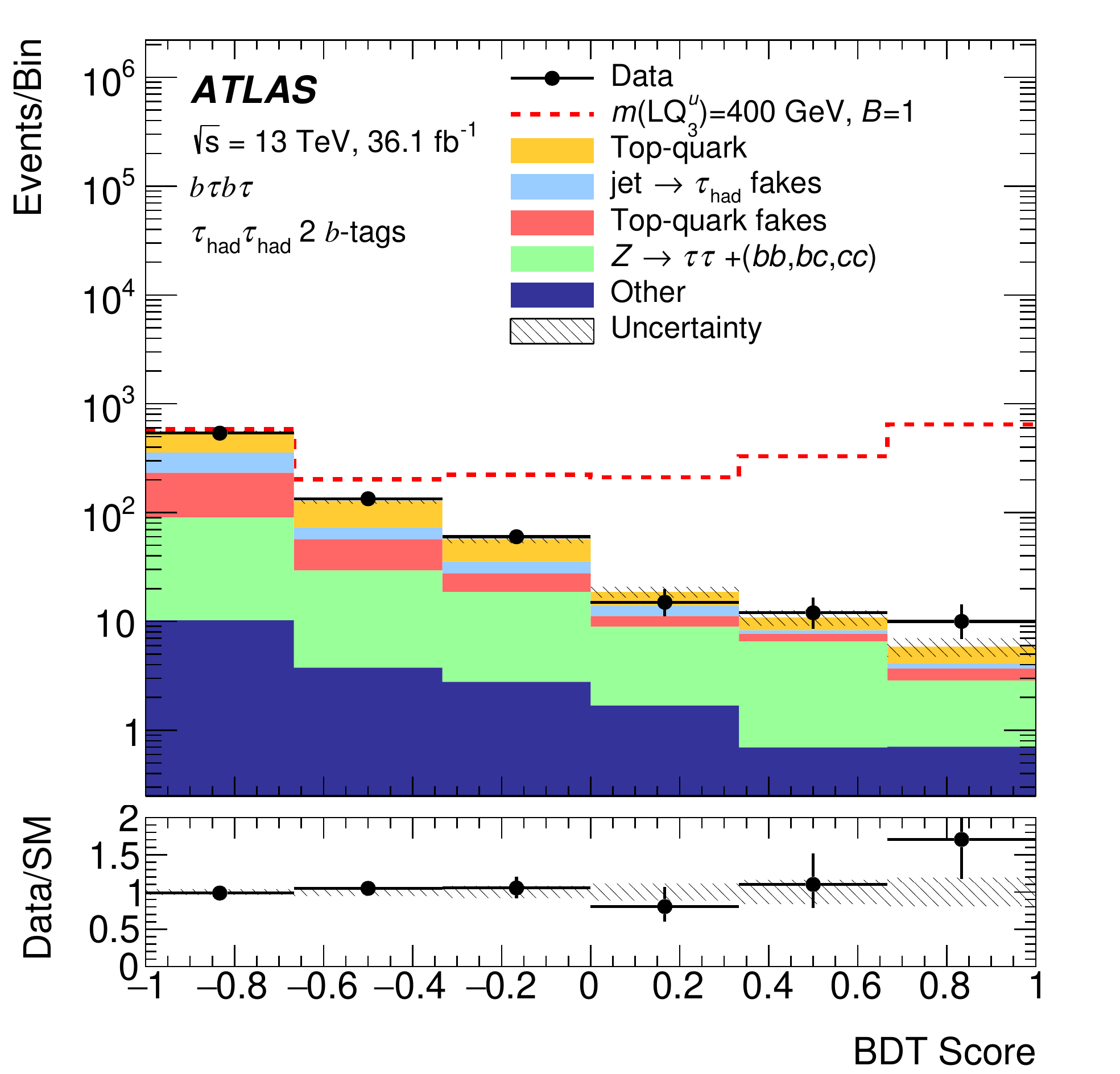}
  }
  \caption{BDT score distributions for \lephad\ (top) and \hadhad\ (bottom) channels in the 1-tag (left) and 2-tag (right) regions after performing the combined channel fit.  The stacked histograms show the various SM background contributions, which are normalized to the expected cross-section. The hatched band indicates the total statistical and systematic uncertainty in the SM background. The error bars on the black data points represent the statistical uncertainty in the data yields. The dashed histogram shows the expected additional yields from a leptoquark signal model for a up-type leptoquark sample with a mass of 400~\GeV\ ($B=1$) added on top of the SM prediction. The ratio of the data to the sum of the backgrounds is shown in the lower panel.}
  \label{fig:bdtoutput_low}
\end{figure}

\begin{figure}
  \centering
{
    \includegraphics[width=0.49\textwidth]{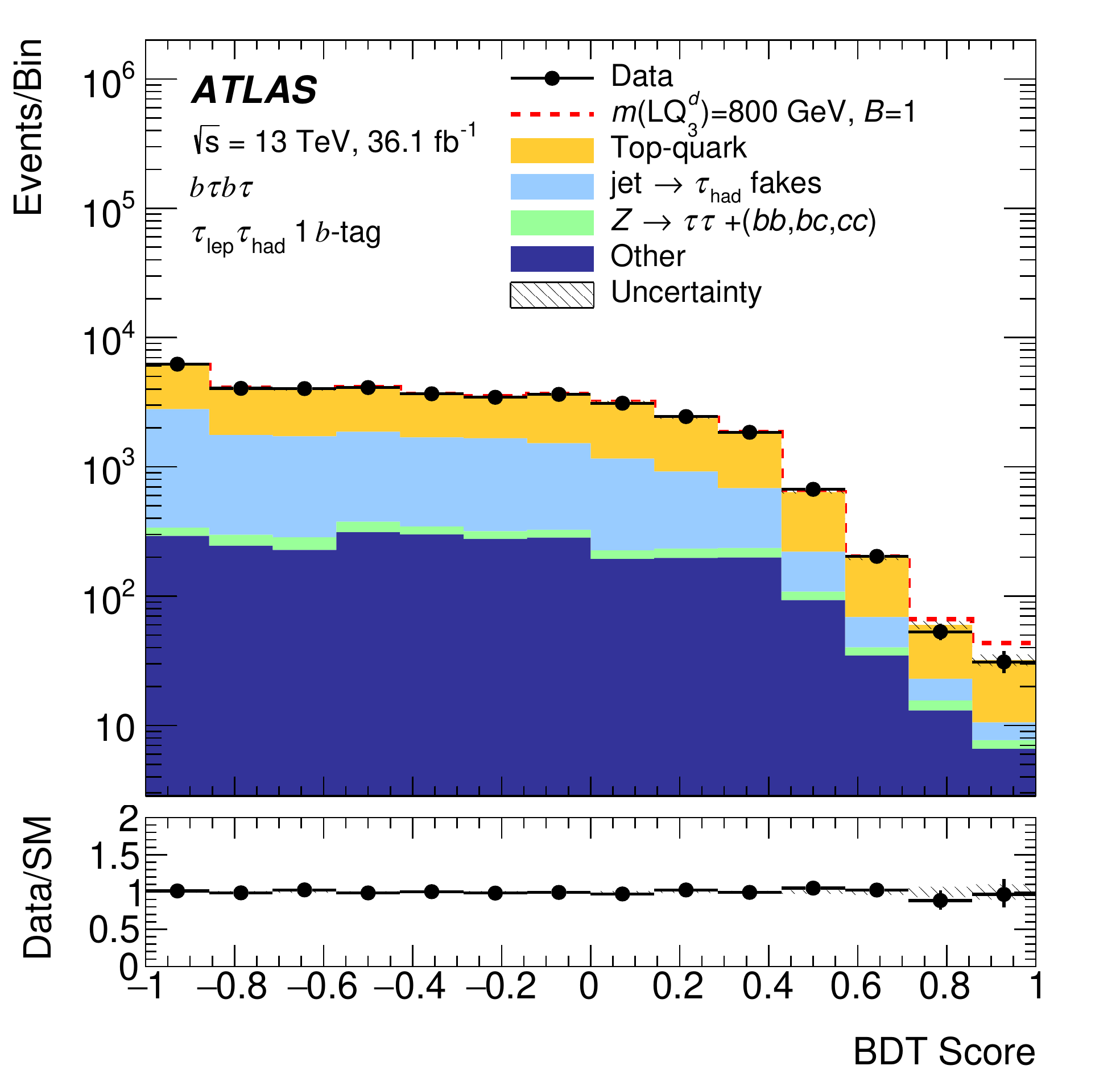}
  }
{
    \includegraphics[width=0.49\textwidth]{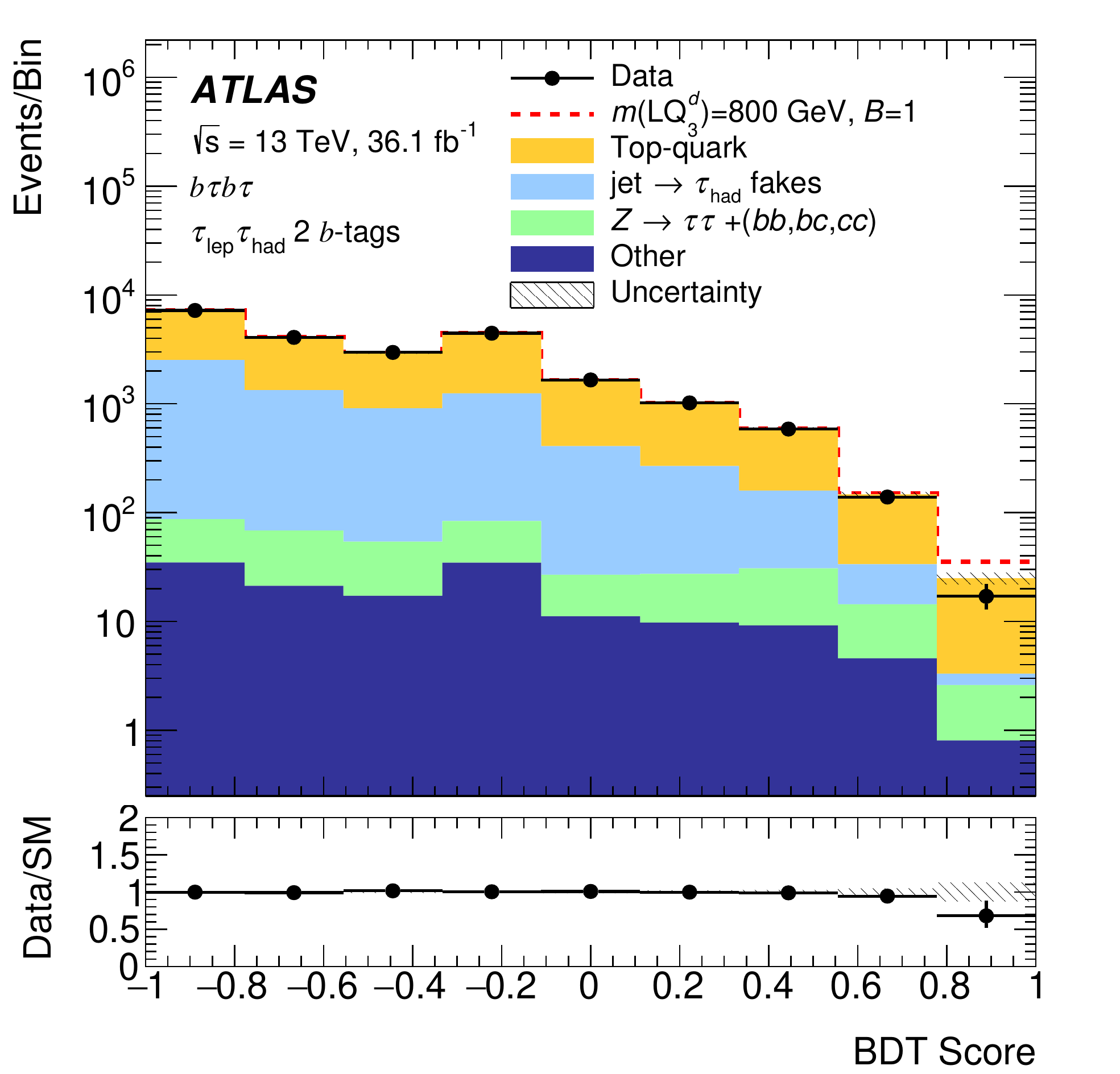}
  }\\
{
    \includegraphics[width=0.49\textwidth]{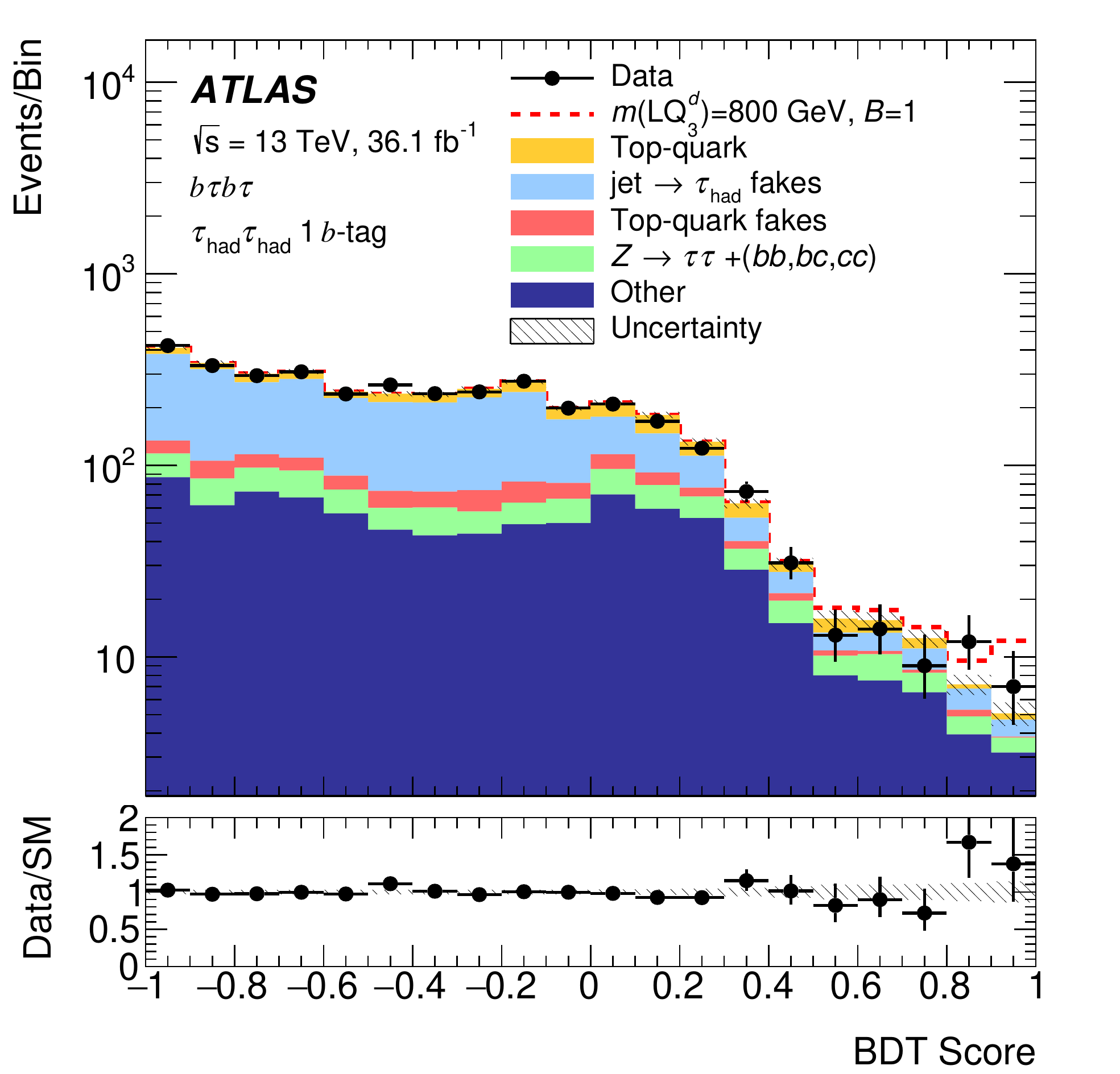}
  }
{
    \includegraphics[width=0.49\textwidth]{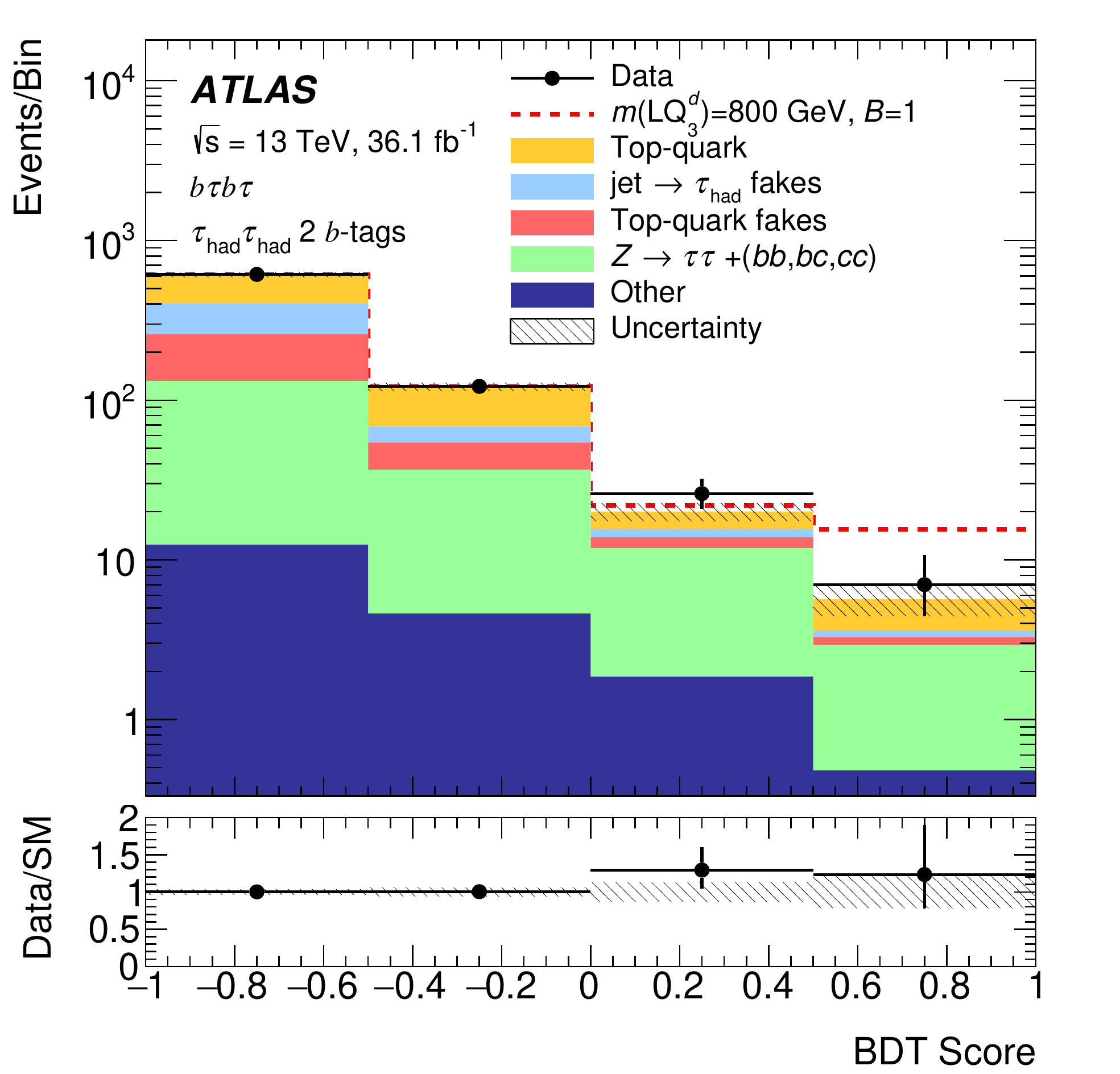}
  }
  \caption{BDT score distributions for \lephad\ (top) and \hadhad\ (bottom) channels in the 1-tag (left) and 2-tag (right) regions after performing the combined channel fit.  The stacked histograms show the various SM background contributions, which are normalized to the expected cross-section. The hatched band indicates the total statistical and systematic uncertainty in the SM background. The error bars on the black data points represent the statistical uncertainty in the data yields. The dashed histogram shows the expected additional yields from a leptoquark signal model for a down-type leptoquark sample with a mass of 800~\GeV\ ($B=1$) added on top of the SM prediction. The ratio of the data to the sum of the backgrounds is shown in the lower panel.}
  \label{fig:bdtoutput_high}

\end{figure}

Systematic uncertainties are considered and propagated through the full analysis. The uncertainties in the luminosity, background modeling, and detector modeling
are calculated as in Ref.~\cite{HIGG-2016-16}. The impact of varying the renormalization/factorization scales and choice of PDF on the 
signal acceptance was also investigated.

The fit strategy follows that in Ref.~\cite{HIGG-2016-16}. The BDT output score is the discriminating variable for all channels and signals in a single combined fit of all signal and control regions.  A CL$_\mathrm s$ method \cite{Read:2000ru} based on one-sided profile-likelihood test statistics is used to test the signal hypothesis. In the profile-likelihood function, Gaussian constraints are used for shape systematic uncertainties and log-normal constraints for normalization uncertainties. The binning in the BDT output categories at high BDT output score is modified to ensure sufficient statistics.  The Standard Model predictions are consistent with the data.  Figure~\ref{fig:1d-limits-combined} shows the expected and observed 95\% confidence level (CL) upper limits on the cross-section for scalar up-type and down-type leptoquark pair production as a function of leptoquark mass for the combined \tlhad\ + \thadhad\ channels.  The theoretical prediction for the cross-section of scalar leptoquark pair production is shown by the solid line, along with the uncertainties. These limits are used to set upper limits on the leptoquark branching ratio $B(\text{LQ}\rightarrow q \tau)$ as a function of the leptoquark mass.  From the data, masses below 1030 GeV and 930 GeV are 
excluded for \lqthreeu\ and \lqthreed\, respectively, at 95\% CL for the case 
of B equal to unity. The expected exclusion ranges are 1030 GeV and 
930 GeV, respectively

\begin{table}
\caption{Post-fit expected numbers of signal and background events, determined from a background-only fit, compared to the observed number of data events after applying the
selection criteria and requiring at least one $b$-tagged jet. Both the up-type and down-type leptoquark samples here use $B=1$. In the \tlhad\ channel, the fake-$\tau$-lepton background includes all processes in which a jet is misidentified as a $\tau$-lepton, while in the \thadhad\ case the fake background from QCD multi-jet processes and $t\bar{t}$ production are derived separately. The $t\bar{t}$ background includes events with true $\tau_\mathrm{had}$ and the very small contribution from leptons misidentified as $\tau_\mathrm{had}$. The `Other' category includes contributions from $W$+jets, $Z$+jets, and diboson processes.  The total background is not identical to the sum of the individual components since the latter are rounded for presentation, while the sum is calculated with the full precision before being rounded. The uncertainty in the total background is smaller than that in the $\ttbar$ and multi-jet backgrounds due to these being strongly anti-correlated.}
 \label{tab:postfitYields}
\begin{center}
\setlength{\tabcolsep}{0.3pc}
\begin{tabular}{ |l|rcl|rcl|rcl|rcl| }
\hline
Sample & \multicolumn{12}{|c|}{Post-fit yield} \\\cline{2-13}
& \multicolumn{6}{|c|}{\tlhad} & \multicolumn{6}{c|}{\hadhad}\\\cline{2-13}
& \multicolumn{3}{c}{1-tag} & \multicolumn{3}{c|}{2-tag} & \multicolumn{3}{c}{1-tag} & \multicolumn{3}{c|}{2-tag}\\
\hline

$t\bar{t}$ & 17800 &$\pm$& 1500 & 14460 &$\pm$& 980 & 285 &$\pm$& 83 & 238 &$\pm$& 69\\
 Single top & 2500 &$\pm$& 180 & 863 &$\pm$& 73 & 63 &$\pm$& 8 & 27 &$\pm$& 3\\
 QCD fake-$\tau$ & &-& & &-& & 1860 &$\pm$& 110 & 173 &$\pm$& 34\\
 $t\bar{t}$ fake-$\tau$ & &-& & &-& & 200 &$\pm$& 110 & 142 &$\pm$& 79\\
 Fake-$\tau$ & 13900 &$\pm$& 1700 & 6400 &$\pm$& 1000 & &-& & &-& \\
 $Z \to \tau\tau + (bb, bc, cc)$ & 520 &$\pm$& 160 & 285 &$\pm$& 83 & 258 &$\pm$& 64 & 156 &$\pm$& 36\\
 Other & 2785 &$\pm$& 270 & 158 &$\pm$& 26 & 817 &$\pm$& 95 & 21 &$\pm$& 4\\
 \hline
 Total Background & 37510 &$\pm$& 220 & 22120 &$\pm$& 160 & 3482 &$\pm$& 59 & 756 &$\pm$& 27\\
 \hline
 Data & \multicolumn{3}{c|}{37527} & \multicolumn{3}{c|}{22117} & \multicolumn{3}{c|}{3469} & \multicolumn{3}{c|}{768}\\
 \hline
 $m(\lqthreeu) = 400$~\GeV\ & 2140 &$\pm$& 140 & 1950 &$\pm$& 160 & 1430 &$\pm$& 190 & 1430 &$\pm$& 200 \\
 $m(\lqthreed) = 400$~\GeV\ & 1420 &$\pm$& 170 & 1096 &$\pm$& 82 & 850 &$\pm$& 110 & 672 &$\pm$& 88 \\
 $m(\lqthreeu) = 800$~\GeV\ & 39.1 &$\pm$& 2.8 & 25.2 &$\pm$& 2.3 & 25.6 &$\pm$& 3.9 & 16.8 &$\pm$& 2.7\\
 $m(\lqthreed) = 800$~\GeV\ & 23 &$\pm$& 2.3 & 16.6 &$\pm$& 1.4 & 17.8 &$\pm$& 2.8 & 12.4 &$\pm$& 2.2\\
 $m(\lqthreeu) = 1500$~\GeV\ & 0.25 &$\pm$& 0.02 & 0.08 &$\pm$& 0.01 & 0.16 &$\pm$& 0.03 & 0.05 &$\pm$& 0.01\\
 \hline
\end{tabular}
\end{center}
\end{table}

\begin{figure}
  \centering
  \small
\begin{minipage}[t]{0.49\textwidth}
  \includegraphics[width=1.1\textwidth]{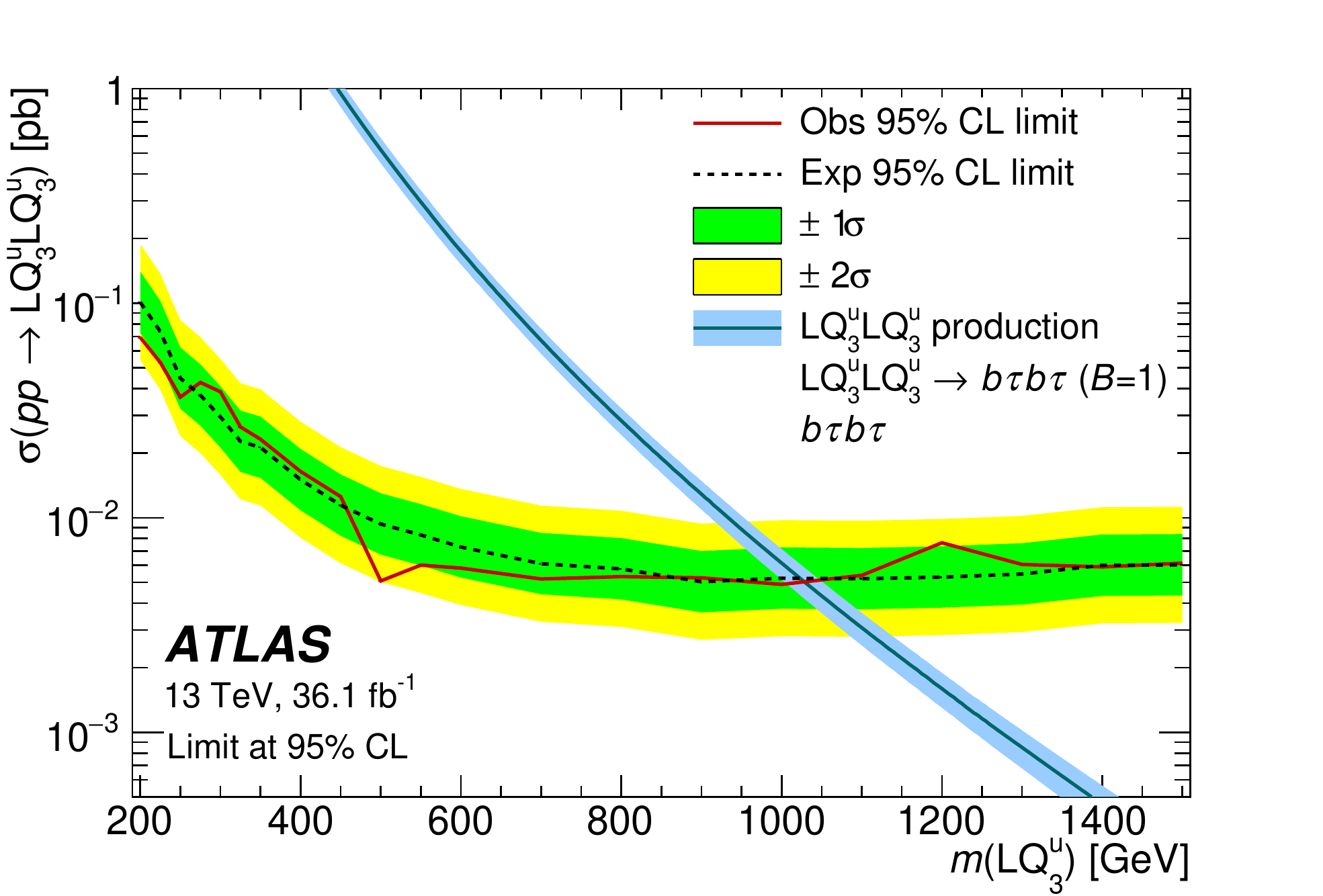}
\end{minipage}
\hfill
\begin{minipage}[t]{0.49\textwidth}
  \includegraphics[width=1.1\textwidth]{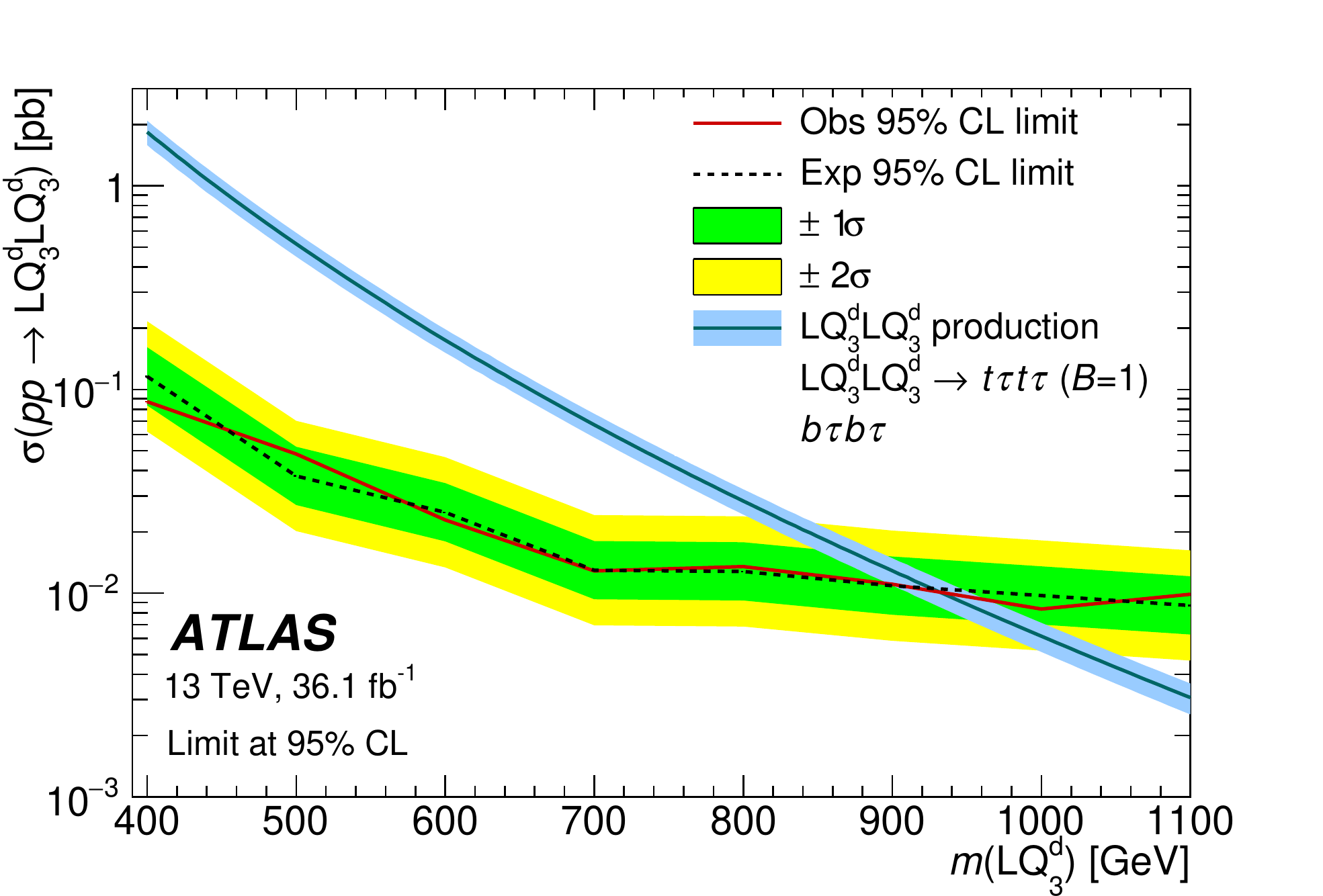}
\end{minipage}
  \caption{Expected and observed 95\% CL upper limits on the cross-section for up-type (left) and down-type (right) scalar leptoquark pair production with $B=1$ as a function of leptoquark mass for the combined \lephad\ and \hadhad\ channels. The observed limit is shown as the solid line. The thickness of the theory curve represents the theoretical uncertainty from PDFs, renormalization and factorization scales, and the strong coupling constant $\alpha_\mathrm{s}$.}
  \label{fig:1d-limits-combined}
\end{figure}

\section{The $t t$ plus \ETmiss channel with one lepton}
\label{ttmet-1l}

In this section, the LQ reinterpretation of a dedicated search for top-squark pair production~\cite{SUSY-2016-16} in final states with one lepton is described.  
Events where both LQs decay into a top quark and a neutrino are targeted, where one top quark decays hadronically and the other one semileptonically.
Events containing one isolated lepton, jets, and missing transverse momentum in the final state 
are considered. 
Two signal regions (SR) of the top-squark search~\cite{SUSY-2016-16}, tN\_med and tN\_high, have
optimal sensitivity for medium ($\sim$600~\GeV) and high ($\sim$1~\TeV) masses of the
leptoquark.
The tN\_med SR is additionally binned in \met, while the tN\_high SR is taken as a single-bin cut-and-count experiment. 
Both signal regions require at least four jets, at least one $b$-tagged jet, exactly one isolated electron or muon, and high \met. 
Variables sensitive to the direction of the \met\ are used, e.g.\ the transverse mass of the lepton and \met, denoted by \mt,\footnote{The transverse mass $\mt$ is defined as $\mt = \sqrt{2p_{\text{T}}^{\text{lep}}\met[1-\cos(\Delta\phi)]}$, where $\Delta\phi$ is the azimuthal angle between the lepton and the missing transverse momentum direction 
and $p_\text{T}^\text{lep}$ is the transverse momentum of the charged lepton.} and \amtTwo~\cite{Konar:2009qr}, which targets pair-produced heavy objects 
that each decay into a different number of measured and unmeasured particles in the detector.
The hadronically decaying top quark is reconstructed using jet reclustering,
where several jets are combined using the \antikt algorithm~\cite{Cacciari:2008gp}
with a large radius parameter that is initially set to 3.0 and then adjusted in an iterative process~\cite{SUSY-2016-16}.

The various backgrounds are estimated using simulated data. The dominant background consists of $t\bar{t}$ events, which, due to the high \mt requirements, are mainly from dileptonic $t\bar{t}$ decays in which one lepton is not reconstructed ($t\bar{t}$ 2L), even though this decay topology is strongly suppressed by 
requiring \amtTwo to be above the top-quark mass.
Other major backgrounds are due to the production of a $W$ boson in association with one or more jets ($W$+jets) and 
the production of a $t\bar{t}$ pair in association with a vector boson ($t\bar t + V$), where the latter is dominated by contributions 
from $t\bar{t}+Z(\rightarrow\nu\nu)$. 
For each SR, a set of dedicated single-bin CRs is defined in order to control the background normalization.
The $t\bar{t}$ 2L CR is defined at high \mt and with a veto on hadronically decaying top-quark candidates, while the $W$+jets CR is defined at low \mt and with a veto on hadronically decaying top-quark candidates. The top-quark candidate veto is fulfilled if either no top-quark candidate is found or if the mass is lower than the SR threshold. 
Additionally, a CR for semileptonic $t\bar t$ events ($t\bar{t}$ 1L) is defined, as these events contribute strongly to the other CRs. 
A CR for single-top events is defined similarly to the $W$+jets CR, but with at least two \bjets. 
The $t\bar t + Z$ background is estimated with a three-lepton selection.

The statistical analysis for a SR is based on a simultaneous likelihood fit to the observed events in the CRs and the SR, 
where the background samples and a signal sample are included in all regions. 
The fit determines at the same time the background normalization as well as a potential signal contribution.
For each SR, a set of validation regions is defined in addition to the CRs. The validation regions are not part of the fit 
but are used to validate the background normalization in this second set of disjunct regions.
Systematic uncertainties  
are included as nuisance parameters in the profile-likelihood estimate.
The test statistic is the profile log-likelihood ratio and its distribution is calculated using the asymptotic approximation~\cite{stat}.

The number of observed events in the data in each SR and the expected number of background events 
as calculated in 
a fit to only the CRs while neglecting a potential signal contamination are taken from Ref.~\cite{SUSY-2016-16} and shown in Tables \ref{tab:tN_high_number_events} and \ref{tab:tN_med_number_events}.
In addition, the expected number of signal events for different leptoquark masses is shown for each SR.
The contamination 
from the leptoquark signal in the CRs is below 10\% in all cases.
Figure~\ref{fig:tN_med_pull} shows the \met distribution in the tN\_med SR.

\begin{table}
\centering 
\caption{The number of observed events in the cut-and-count SR tN\_high, together with the expected number of background events including their total uncertainties, 
taken from Ref.~\cite{SUSY-2016-16}.
Additionally, the expected number of signal events 
are given for $B=0$ for up-type LQs of different masses with statistical uncertainties.
}
\label{tab:tN_high_number_events}
\setlength{\tabcolsep}{0.1pc}
	\begin{tabular}{l rcl}
		\toprule
		Observed events             &    & $8$ &              \\ 
		Total SM                      & $3.8$& $\pm$&$1.0$       \\ 
                \midrule
		$m(\lqthreeu) = 800$ \GeV    &  11.9&$\pm$&1.8   \\
		$m(\lqthreeu)  = 900$ \GeV    &    9.5&$\pm$&1.2  \\
        	$m(\lqthreeu)  = 1000$ \GeV  & 6.7&$\pm$&0.7     \\
		$m(\lqthreeu)  = 1100$ \GeV ~~~ & 3.7&$\pm$&0.3     \\ \bottomrule
	\end{tabular}
\end{table}

\begin{table}
\centering 
\caption{The number of observed events in the shape-fit SR tN\_med, together with the expected number of background events 
including their total uncertainties, taken from Ref.~\cite{SUSY-2016-16}.
Additionally, the expected number of signal events 
are given for $B=0$ for up-type LQs of different masses with statistical uncertainties.
The numbers are given for the four bins characterized by an interval of the \ETmiss variable. 
}
\label{tab:tN_med_number_events}
\small
\begin{tabular}{lcccc}
	\toprule
	\met                          & [250, 350]\,\GeV & [350, 450]\,\GeV & [450, 600]\,\GeV &  >600\,\GeV   \\ \midrule
	Observed events               &       $21$       &      $17$       &       $8$       &      $4$      \\
	Total SM                      &   $14.6\pm2.8$~   &  $11.2\pm2.2$~   &   $7.3\pm1.7$   & $3.16\pm0.74$ \\ \midrule
	$m(\lqthreeu)$ = 400 \GeV  &   $166\pm44$~   &  ~$\,58\pm32$~   &  $11\pm11$   &  $5.7\pm5.7$  \\
	$m(\lqthreeu)$ = 600 \GeV  &   $21.0\pm5.6$~   &  $49.6\pm8.8$~   &  $31.8\pm5.5\,$~   &  $1.4\pm2.1$  \\
	$m(\lqthreeu)$ = 800 \GeV  &   ~$\,5.0\pm1.5$~    &  $10.6\pm1.7$~   &  $11.2\pm2.0\,$~   &  $6.3\pm1.4$  \\
	$m(\lqthreeu)$ = 1000 \GeV &  $\,0.46\pm0.14$   &  $\,1.18\pm0.24$  &  $2.92\pm0.49$  & $4.61\pm0.64$ \\ \bottomrule
\end{tabular}
\end{table}

\begin{figure}
	\centering
	\includegraphics[width=.55\textwidth]{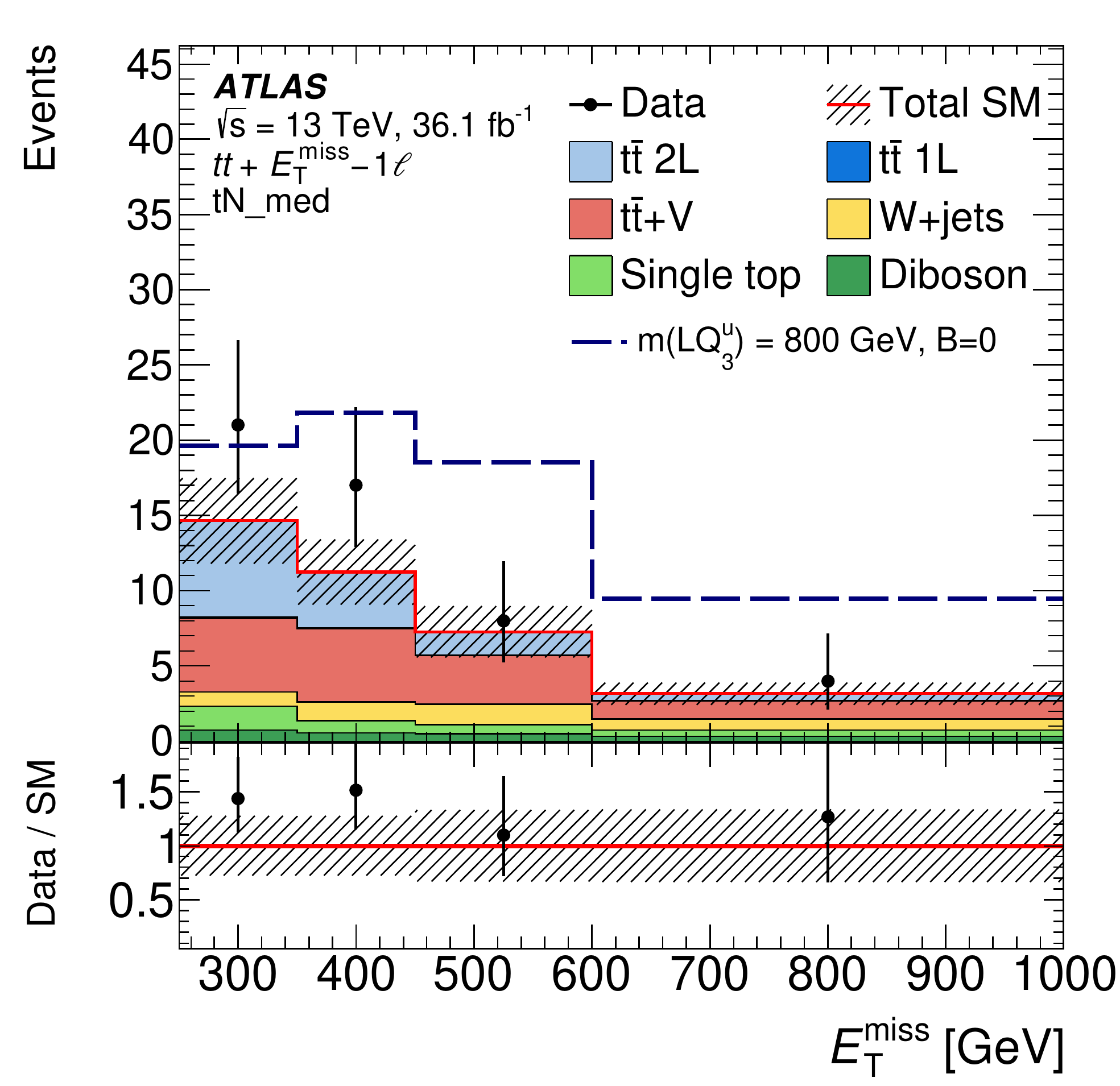}
	\caption{Observed and expected \met distributions are shown in the tN\_med signal region, as is their ratio. 
The error band includes statistical and systematic uncertainties. The expected SM backgrounds are normalized to the values determined in the fit.
The expected number of signal events 
for an up-type LQ with $m_{\textnormal{LQ}}=800$~\GeV\ and $B=0$ is
added on top of the SM prediction. The last bin contains the overflow events.
}
	\label{fig:tN_med_pull}
\end{figure}

The combination of the exclusion limits for the SRs is obtained by selecting the signal region with the better expected limit for each mass point. 
For LQ masses of 950, 1000, and 1100~\GeV, the limit of the tN\_high signal region is selected, otherwise the limit of the tN\_med signal region is selected.  
The expected and observed exclusion limits are shown in Figure \ref{fig:mass_limit_beta000}. 
The theoretical prediction for the cross-section of scalar leptoquark pair production is shown by the solid line along with the uncertainties. Pair-produced third-generation scalar leptoquarks decaying into $t \nu \bar{t} \bar{\nu}$ are excluded at 95\% CL for \mLQ < 930~\GeV. The expected exclusion range is \mLQ < 1020~\GeV.

\begin{figure}
\centering
\includegraphics[width=0.55\linewidth]{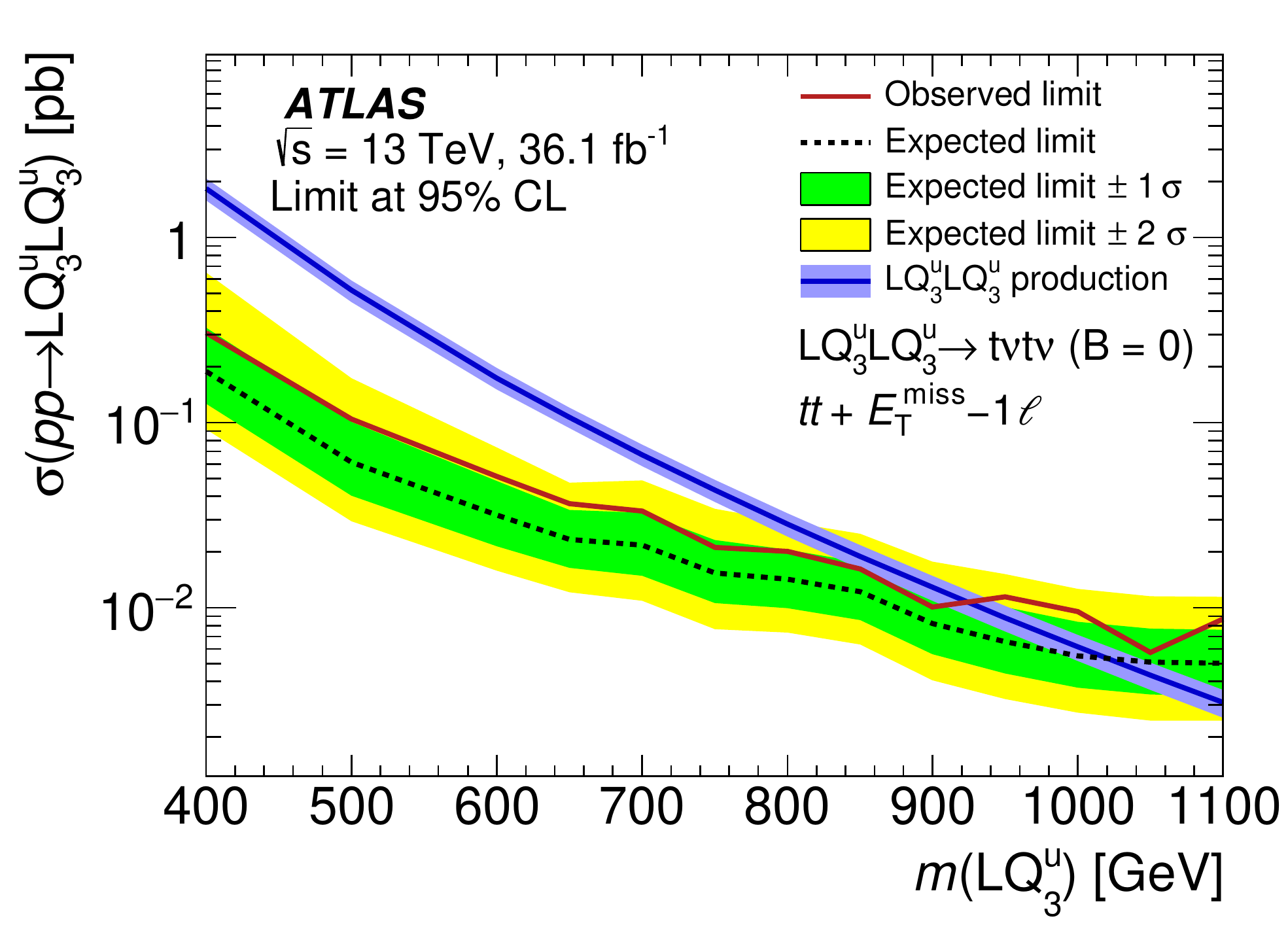}
\caption{
Observed and expected 95\% CL upper limit on the cross-section for 
up-type LQ pair production with $B=0$
as a function of the LQ mass.
The $\pm1(2)\sigma$ uncertainty bands around the expected limit represent all sources of statistical and systematic uncertainties.
The thickness of the theory curve represents the theoretical uncertainty from PDFs, renormalization and factorization scales, and the strong coupling constant $\alpha_\mathrm{s}$.
}
\label{fig:mass_limit_beta000}
\end{figure}

\section{The $t t$ plus \ETmiss channel with zero leptons}

In this section, the LQ reinterpretation of the ATLAS analysis optimized for the search of top-squark pair production~\cite{stop0LMoriond2017} 
in final states with zero leptons is discussed.  
The signature targeted is two hadronically decaying top quarks and invisible particles, closely matching the signature where both LQs decay into 
a top quark and a neutrino.
Three signal regions (SRA, SRB, SRD) of the top-squark search~\cite{stop0LMoriond2017} have the greatest sensitivity to the LQ signal models.  
SRA is optimal for high LQ masses, e.g.\ \mLQ $\approx 1$~\TeV, which typically results in high \met\ and top quarks with a significant boost. 
SRB is sensitive to medium LQ masses, which tend to have a softer \met\ spectrum and less-boosted top quarks. 
SRD targets a resonance decaying into a $b$-quark and an invisible particle, giving sensitivity to \lqthreed\ $\rightarrow b \nu$ events.

A common preselection is defined for all signal regions. At least four jets are required, of which at least one must be $b$-tagged. 
The four leading jets (ordered in \pt) must satisfy \pt\ $>$ 80, 80, 40, 40~\GeV, respectively, due to the tendency 
for signal events 
to have higher-energy jets than background events. Events containing reconstructed electrons or muons are vetoed. The $\met$ trigger threshold motivates the $\met>250$~\GeV\ requirement and rejects most of the background from multi-jet and all-hadronic $\ttbar$
events.

Similarly to the \ttmetonel\ analysis in Section~\ref{ttmet-1l}, hadronically decaying top quarks are reconstructed using jet
reclustering. 
SRA and SRB require the presence of two reclustered $R=1.2$ jets.
Both SRA and SRB are divided into three orthogonal, one-bin subregions, which are combined for maximal signal sensitivity. 
The categorization in subregions is based on the mass of the subleading (ordered in \pt) reclustered jet (\mantikttwelveone). 
In all subregions it is required that the leading reclustered $R=1.2$ jet has a mass (\mantikttwelvezero) of at least 120~\GeV. 
The subregions are denoted by TT, TW, and T0 corresponding to requirements of $\mantikttwelveone>120$~\GeV, $60<\mantikttwelveone<120$~\GeV, and $\mantikttwelveone<60$~\GeV, respectively. 
In addition to the $R=1.2$ reclustered jet mass, one of the most discriminating variables in SRA is \met, which has to be 
above 400~\GeV\ or higher depending on the subregion.
In SRB, the \met\ requirement is looser ($\met>250$~\GeV) than in SRA since the signals that SRB is targeting 
tend to have softer \met\ spectra.

Two SRD subregions, SRD-low and SRD-high, are defined for which   
at least five jets are required, two of which must be $b$-tagged. Requirements are made on the transverse momenta of the jets as well as on the scalar sum of the transverse momenta of the two $b$-tagged jets, which needs to be above 400~\GeV\ for SRD-high and above 300~\GeV\ for SRD-low. 
Tight requirements are also applied to the \mt\ calculated from \met\ and the $b$-tagged jet that has the smallest (\mtbmin) and largest (\mtbmax) 
$\Delta \phi$  relative to  
the \met\ direction.

The dominant background processes are $Z\to\nu\nu$ in association with $b$-jets, 
semileptonic \ttbar\ 
where one of the $W$ bosons decays into $\tau\nu$, and $\ttbar+Z (\to\nu\nu)$. 
To estimate the normalization of these backgrounds, control regions are designed to be as close as possible to individual signal regions 
while being strongly enhanced in the background of interest. 
For the $Z\to\nu\nu$ background, $Z\to\ell\ell$ control regions are used where the leptons are removed 
to mimic the \met\ produced by the $Z\to\nu\nu$ process. 
To estimate the \ttbar\ background, a set of one-lepton control regions is used. 
Finally, the $\ttbar+Z(\to\nu\nu)$ background is estimated using a $\ttbar+\gamma$ control region where the photon \pt\ is used 
to approximate \met. 
Several normalizations for the subdominant backgrounds, 
such as single-top and $W$+jets production, are also estimated using control regions. The normalizations are calculated using a simultaneous binned profile-likelihood fit. Signal contamination in the control regions, specifically regions used to estimate the normalization of \ttbar, single top, and $W$+jets, is negligible.

The statistical analysis is done similarly to the one described in Section~\ref{ttmet-1l}.
Here the signal yields are extracted during a simultaneous fit to all control regions plus SRD or the three subregion categories of either SRA or SRB. 
The two subregions of SRD are not orthogonal and are not statistically combined. The combined limits use the best expected limit among all signal regions.

\begin{table}
\begin{center}
\caption{Number of observed events in SRA, SRB, and SRD, together with the number of fitted background events including their total uncertainty, taken from Ref.~\cite{stop0LMoriond2017} (CR background-only fit). Additionally, the expected number of signal events for different branching ratios and LQ masses close to the exclusion limits are given with statistical uncertainties. 
}
 \label{tab:stop0LYields}
{\small
  \begin{tabular}{c||l|c|c|c|c|c} 
    \hline\hline
    {SR}      &                    & {TT}    & {TW}     & {T0}  & low & high \\ \hline \hline
    
    \multirow{4}{*}{{A}} & Observed & 11 & 9 & 18   &   \multirow{4}{*}{{-}} &   \multirow{4}{*}{{-}}    \\ \cline{2-5}
                         & SM Total & $8.6\pm2.1$ & $9.3\pm2.2$ & $18.7\pm2.7$  & & \\ \cline{2-5}
                         & $m(\lqthreeu) = 1000$~\GeV, $B = 0$ & $8.5 \pm 0.7$ & $4.8 \pm 0.6$ & ~$\,5.0 \pm 0.7$  & & \\ \cline{2-5}
                         & $m(\lqthreed) = 800$~\GeV, $B = 0$ & $3.1 \pm 1.1$ & $3.7 \pm 1.2$ & $ 15.5 \pm 2.5$   & & \\ \cline{2-5}

    \hline \hline 
    \multirow{4}{*}{{B}} & Observed & $38$              & $53$              & $206$    &\multirow{4}{*}{{-}} &   \multirow{4}{*}{{-}} \\ \cline{2-5}
                         & SM Total & $39\pm 8\,$~ & $52 \pm 7\,$~ & $179 \pm 26$ & & \\ \cline{2-5}
                         & $m(\lqthreeu) = 400$~\GeV, $B = 0.7$ & $26 \pm 7\,$~ & $18 \pm 8\,$~  & $27 \pm 9$       & & \\ \cline{2-5}
                         & $m(\lqthreed) = 400$~\GeV, $B = 0.9$ & $9 \pm 4$ & $18 \pm 9\,$~ & $ 63 \pm 9$  & & \\ \cline{2-5}
    \hline \hline 

\multirow{3}{*}{{D}} & Observed & \multicolumn{3}{c|}{-}  & 27 & 11 \\ \cline{2-7}
                         & SM Total & \multicolumn{3}{c|}{-} & $25\pm 6\,$~ & $8.5\pm1.5$ \\ \cline{2-7}
                         & $m(\lqthreed) = 800$~\GeV, $B = 0.9$ & \multicolumn{3}{c|}{-} & $2.87\pm0.35$ & $1.45\pm0.23$ \\ \cline{2-7}
    \hline\hline
  \end{tabular}
}
\end{center}
\end{table}

Figure~\ref{fig:finalSRABPlots} shows observed and expected \met\ distributions in SRA, \drbjetbjet\ distributions in SRB, 
and \mtbmax\ distributions in SRD-low and SRD-high.   
In addition, examples of signal distributions with different LQ masses and branching ratios are added on top of the SM prediction.
These examples are chosen such that these signals are close to being excluded.
The expected and observed numbers of events as well as the expected number of events for the example signals are shown in Table~\ref{tab:stop0LYields} for
each region.

\begin{figure}
  \centering
  \includegraphics[width=0.49\linewidth]{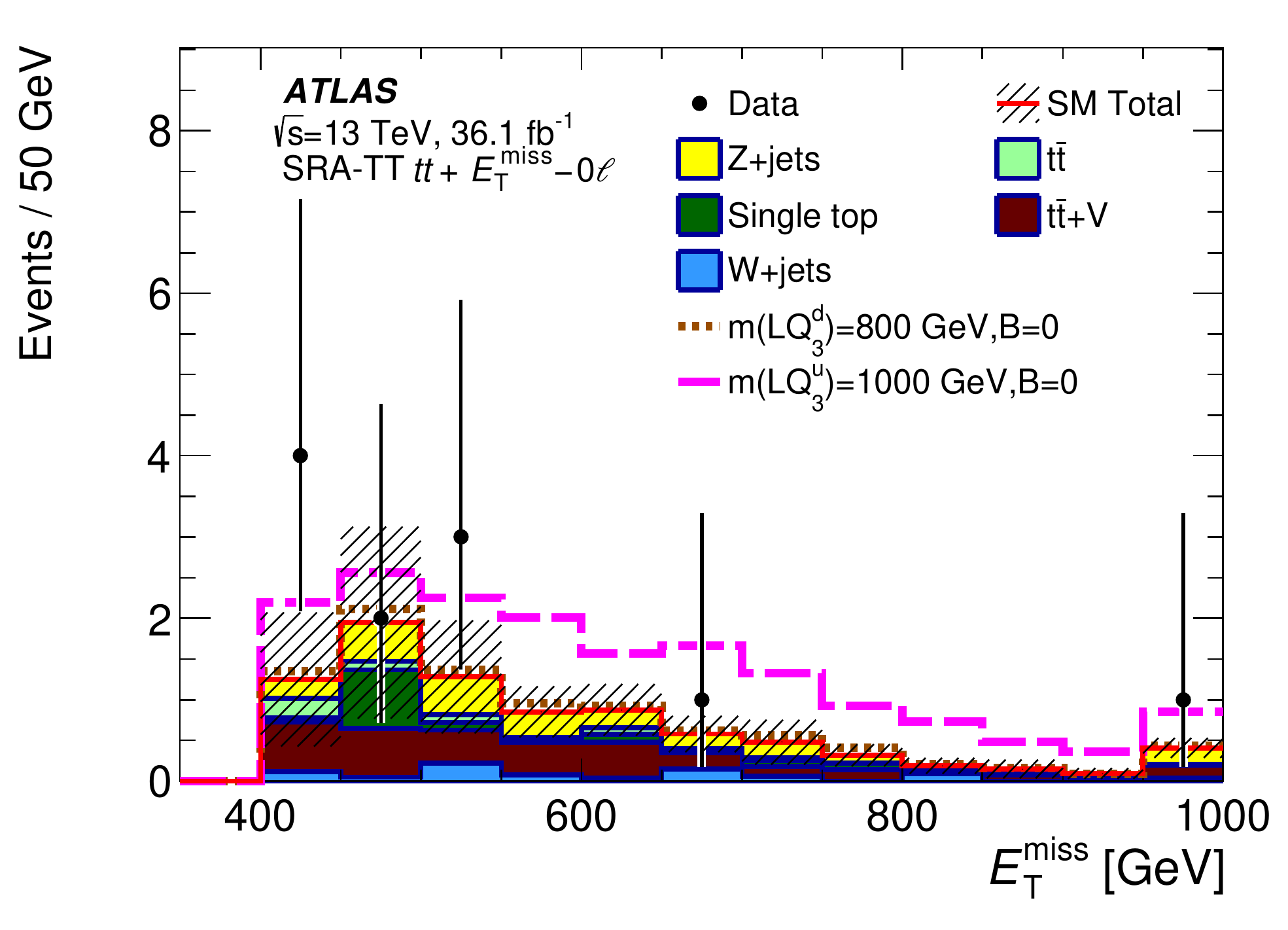}
  \includegraphics[width=0.49\linewidth]{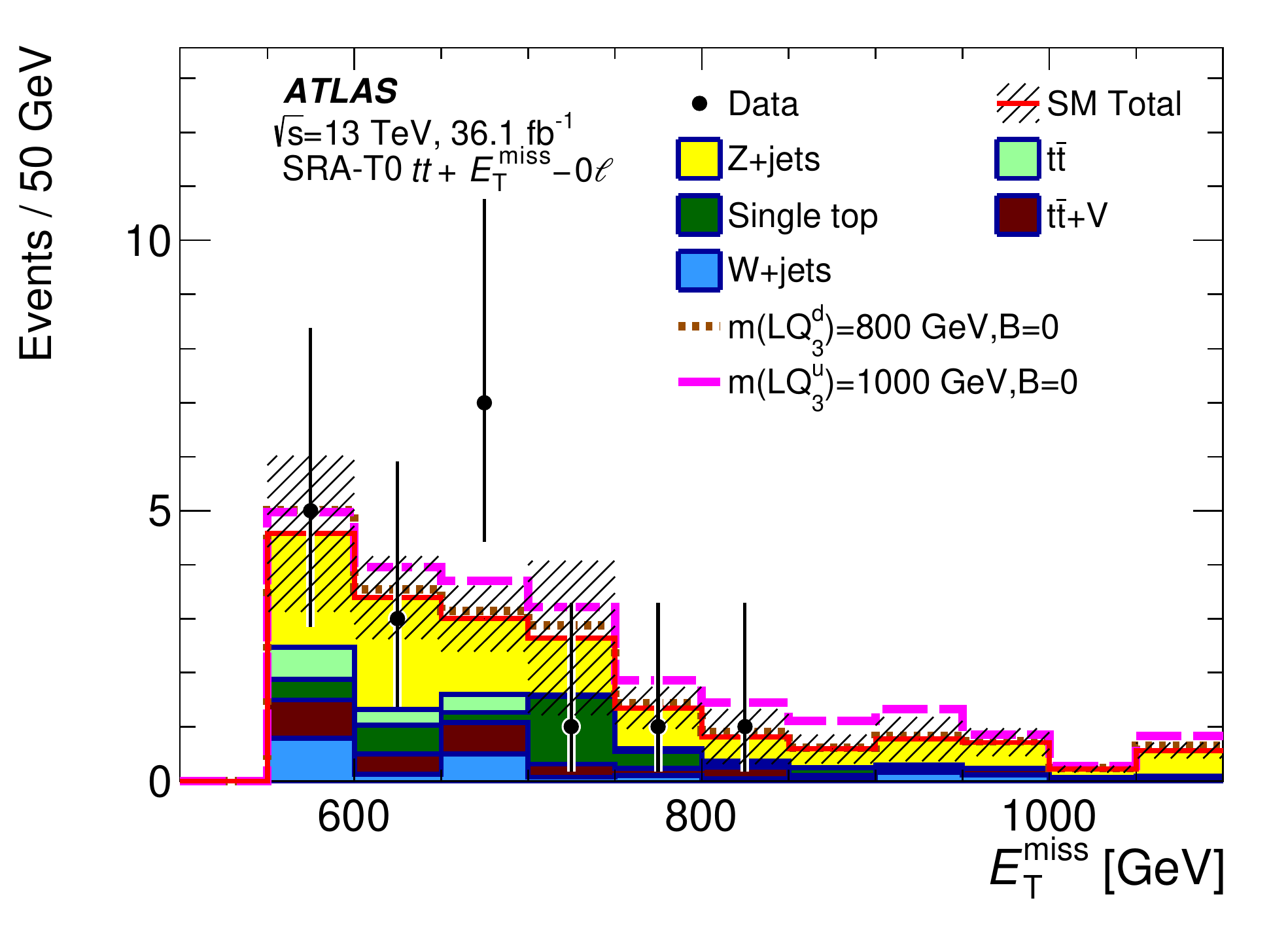}
  \includegraphics[width=0.49\linewidth]{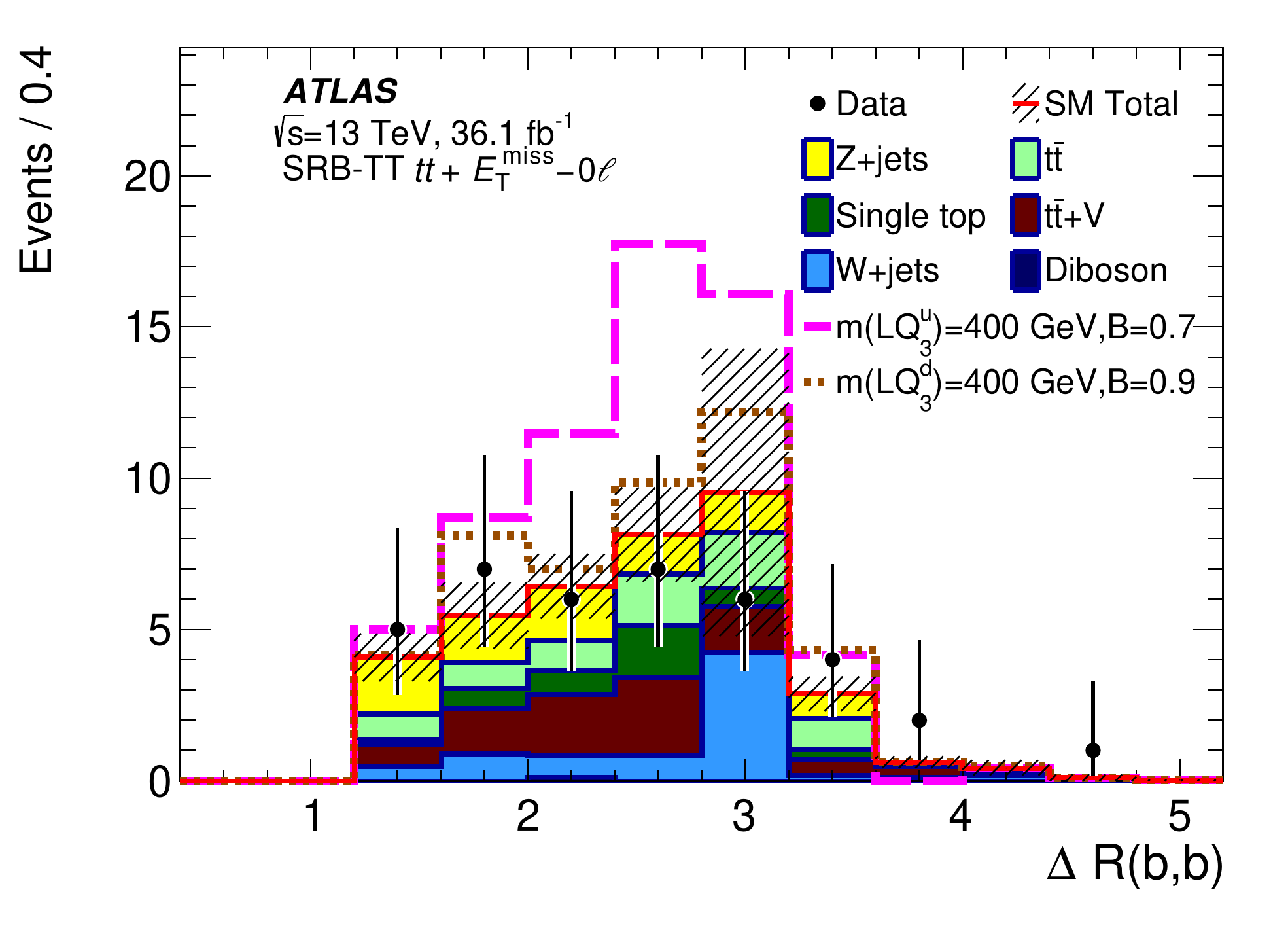}
  \includegraphics[width=0.49\linewidth]{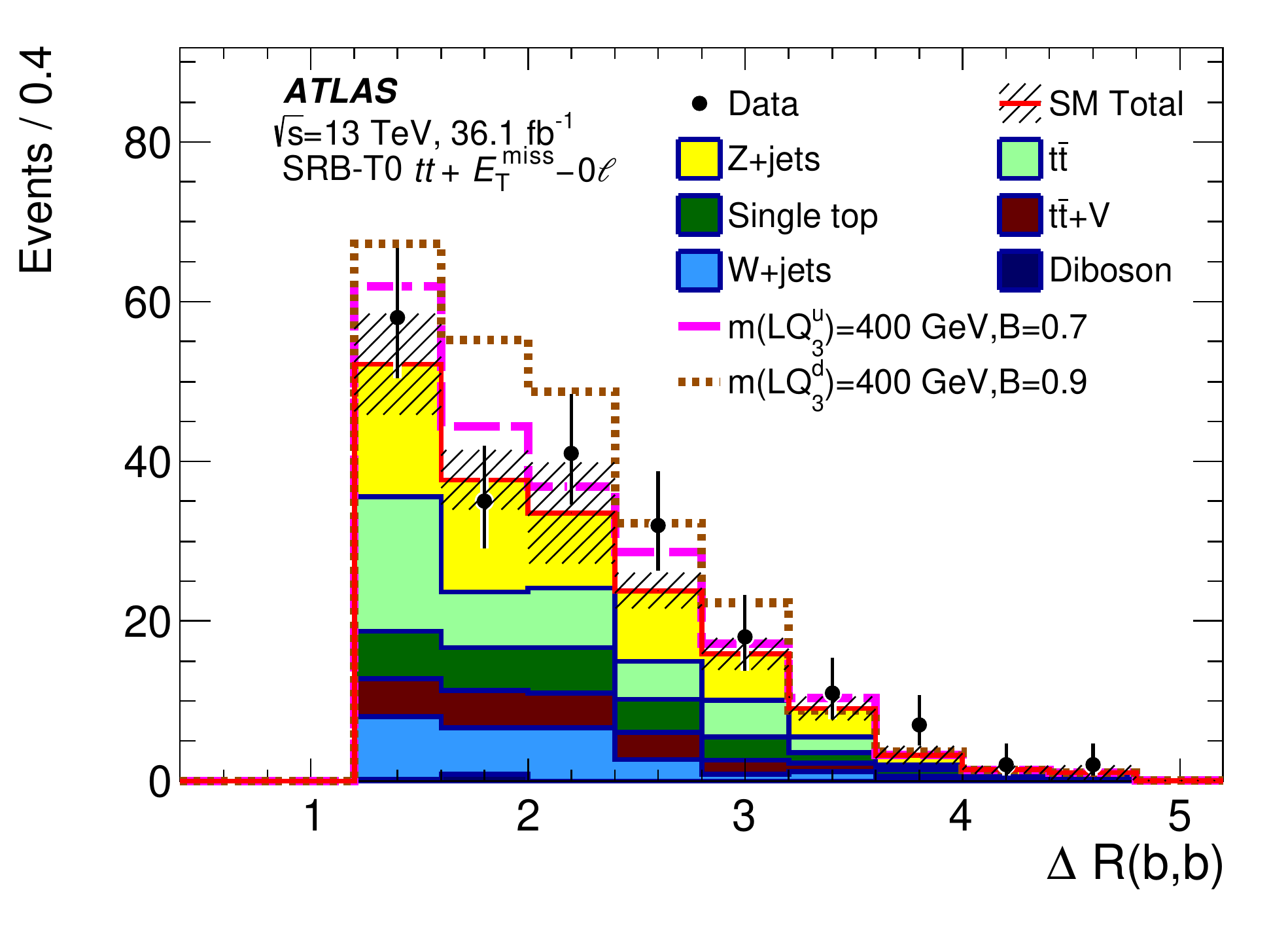}
  \includegraphics[width=0.49\linewidth]{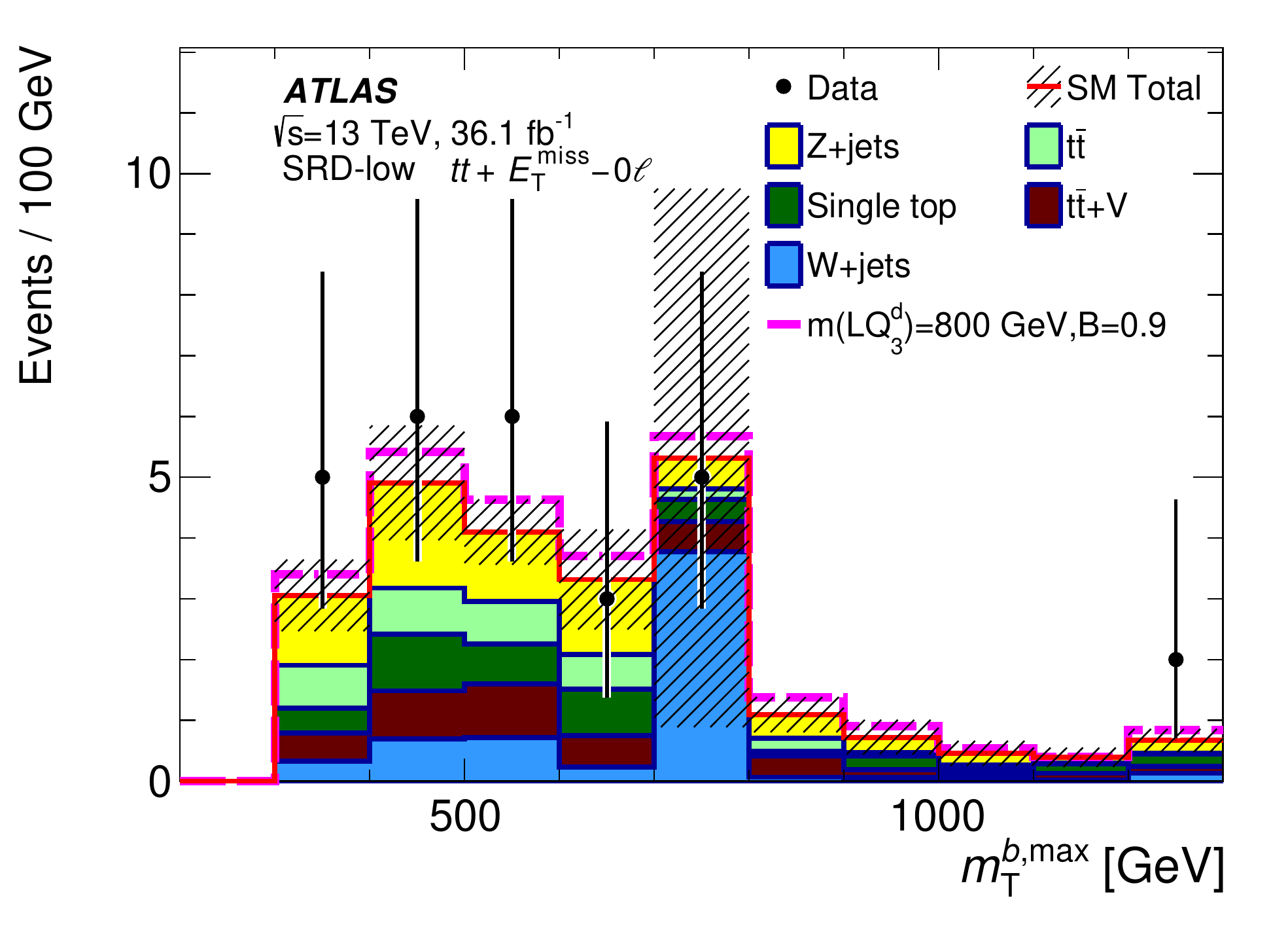}
  \includegraphics[width=0.49\linewidth]{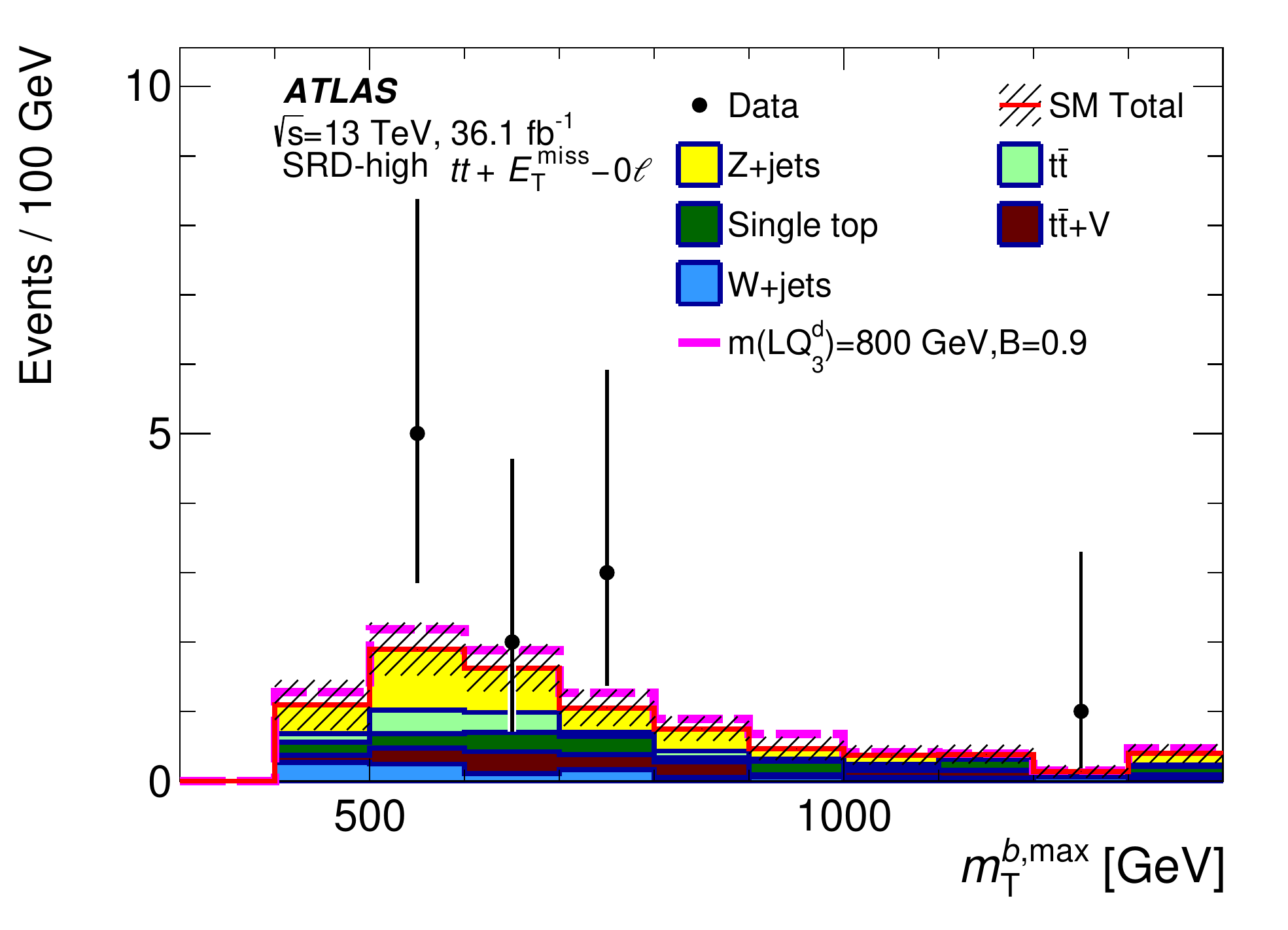}
  \caption{Distributions of \met in SRA (upper panels) and \drbjetbjet\xspace in SRB (middle panels) separately for the individual 
top reconstruction categories TT and T0 and the \mtbmax\ distribution in SRD-low and SRD-high (lower panels). 
The expected SM backgrounds are normalized to the values determined in the fit. The expected number of signal events for different LQ masses and branching ratios $B$ is
added on top of the SM prediction. The last bin contains the overflow events.
}
  \label{fig:finalSRABPlots}
\end{figure}

The expected and observed exclusion limits on the cross-section for \lqthreeu\ and \lqthreed\ for $B = 0$ are shown as a function of the 
leptoquark mass in Figure~\ref{fig:stop0LLimit}.
The theoretical prediction for the cross-section of scalar leptoquark pair-production is shown by the solid line along with the uncertainties. 
As expected, there is better sensitivity to \lqthreeu\ than to \lqthreed, excluding pair-produced \lqthreeu\ 
decaying into $t\nu\bar{t}\bar{\nu}$ for 
masses smaller than 1000~\GeV\ at 95\%~CL.
The expected limit is \mLQ < 1020~\GeV. 

\begin{figure}
  \centering
  \includegraphics[width=0.49\linewidth]{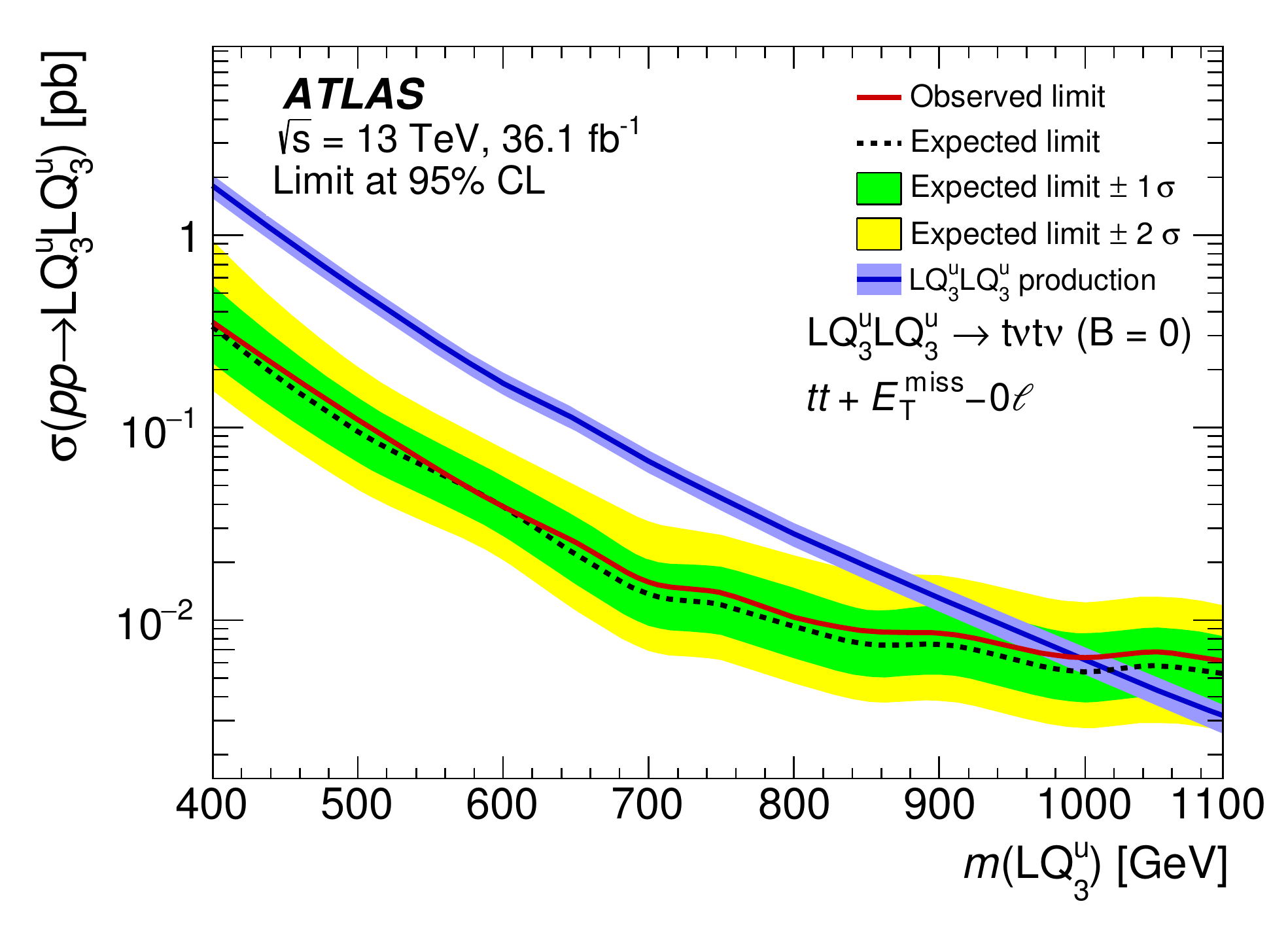}
  \includegraphics[width=0.49\linewidth]{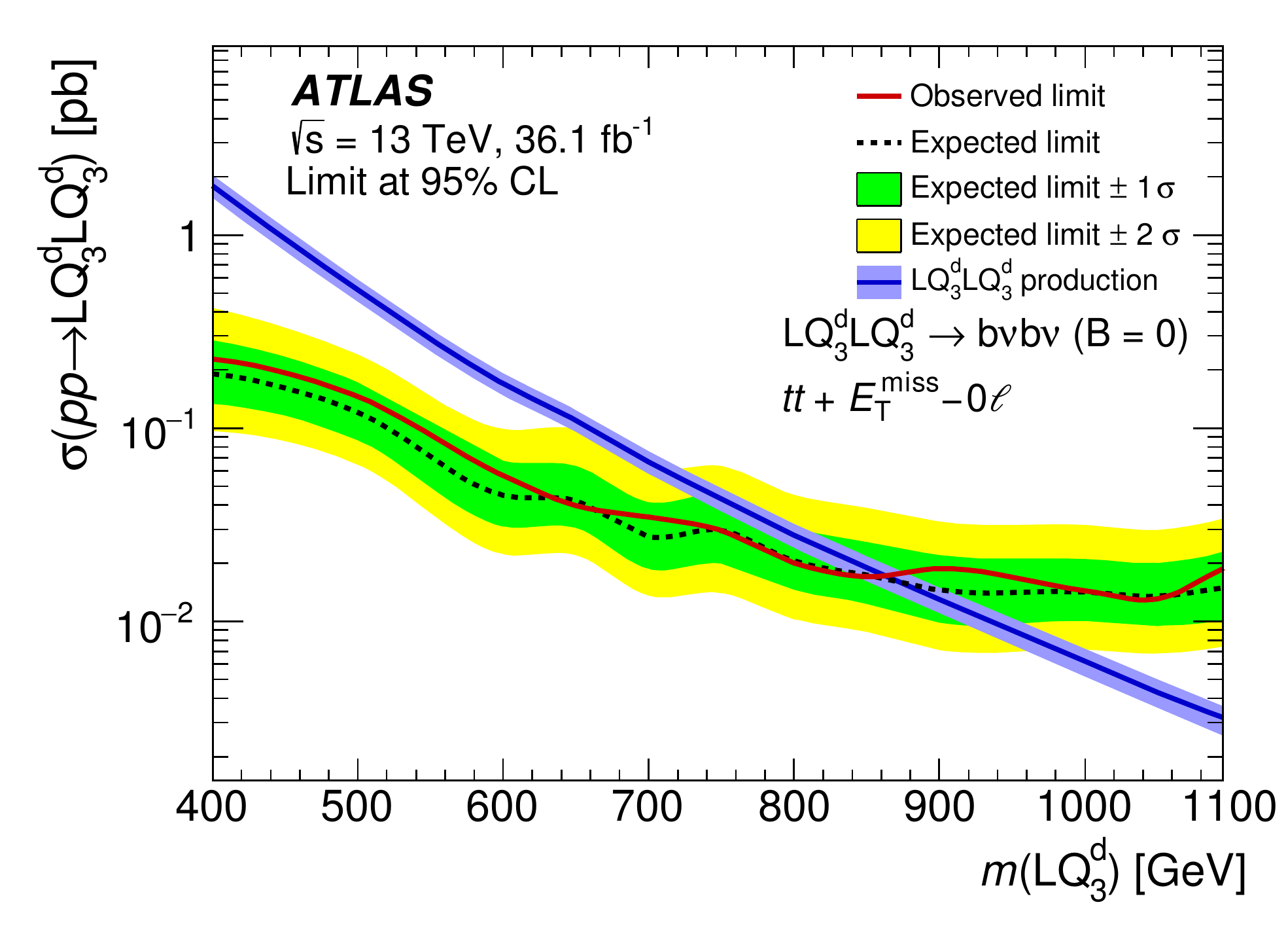}
  \caption{
Observed and expected 95\% CL upper limits on the cross-section for 
up-type (left panel) and down-type (right panel) LQ pair production with $B=0$
as a function of the LQ mass.
The $\pm1(2)\sigma$ uncertainty bands around the expected limit represent all sources of statistical and systematic uncertainties.
In the right panel 
the expected limit has an undulated behaviour
due to the use of different signal regions for 
different mass points and the statistical uncertainties of the background MC in SRD.
The thickness of the theory curve represents the theoretical uncertainty from PDFs, renormalization and factorization scales, and the strong coupling constant $\alpha_\mathrm{s}$.
}
  \label{fig:stop0LLimit}
\end{figure}

\FloatBarrier

\section{The $\tau \tau b$ plus \ETmiss channel}
\label{sec:stopstau}

The ATLAS search for top-squark pair production with decays via $\tau$-sleptons~\cite{SUSY-2016-19}
  selects events containing 
    two $\tau$-leptons, 
    at least two jets, of which at least one must be $b$-tagged, 
    and missing transverse momentum.
Its reinterpretation is expected 
  to have good sensitivity for all $B$ 
  except at very low values, with the maximum sensitivity at intermediate values.

Events are classified according to the decay of the $\tau$-leptons. 
Both the \thadhad\ and the \tlhad\ channels are considered.
Two signal selections are defined, one for the \tlhad\ channel (\SRlh) and another one for the \thadhad\ channel (\SRhh).
Both signal selections require the two leptons to have opposite electric charge. 
The main discriminating variables 
  are \met and the stransverse mass \mttwo,
  which is a generalization of the transverse mass for final states with two invisible particles~\cite{arxiv:LesterMT2, Barr:2003rg, Lester:2014yga}
  and computed from the selected lepton pair and \met.

The dominant background process with the targeted final-state signature is pair production of top quarks. 
As in the original analysis, 
  two types of contributions are discriminated,
  depending on whether the identified hadronically decaying $\tau$-lepton(s) in the selected event 
are real or candidate particles (typically jets, electrons, or, in rare cases, muons),
which are misidentified as hadronically decaying $\tau$-leptons (fake $\tau$-leptons).

In the \tlhad\ channel,
  the contribution of events with fake $\tau$-leptons is estimated using a data-driven method, 
  which is based on a measurement of the number of hadronically decaying $\tau$-leptons satisfying loose identification criteria 
  that also pass the tighter analysis selection (fake-factor method).
The contribution of events with true $\tau$-leptons is estimated from simulation,
  using a dedicated control-region selection to normalize the overall contribution to the level observed in data.
In the \thadhad\ channel,
  both the contributions with fake and with true $\tau$-leptons
  are estimated from simulation with data-driven normalization factors obtained from two dedicated control regions.
A requirement on the transverse mass computed from the transverse momentum of the leading $\tau$-lepton and the missing transverse momentum
  is used to discriminate between events with true and fake $\tau$-leptons. 
The requirement of opposite electric charge is not used for the control region targeting events with fake $\tau$-leptons, 
  since in that case the charges of the two $\tau$-leptons are not correlated. 
Two additional control regions common to both channels are used
  to obtain data-driven normalization factors for the background contributions from diboson production 
  and production of top-quark pairs in association with an additional \Wboson or \Zboson boson.
For these control regions, events that pass a single-lepton trigger and contain at least two signal leptons and two jets are used. 
The remaining definitions of these control regions and further details of the background estimation and its validation are in Ref.~\cite{SUSY-2016-19}.
Good agreement between data and predicted background yields is found in validation regions
when the normalization factors derived in the control regions are applied.

\begin{table}
\newcommand{\npmn}{$<0$&\multicolumn{3}{@{}l}{.1}}
\centering 
\caption{
  The expected number of SM background events obtained from the background fit 
    and the number of observed events in \SRhh and \SRlh, 
    together with the expected number of signal events for different mass hypotheses $m$, leptoquark types, and branching ratios $B$ into charged leptons.
  }
\label{tab:yield_stopstau_SR}      
\small
\begin{tabular}{ll r@{}l @{\,$\pm\,$} r@{}l r@{}l @{\,$\pm\,$} r@{}l}
 \toprule
                           &    &    \multicolumn{4}{c}{SR HH}      &        \multicolumn{4}{c}{SR LH}              \\ 
 \midrule
   Observed events         &    &  \multicolumn{4}{c}{$2$}  & \multicolumn{4}{c}{$3$}            \\
   Total SM                &    &  1&.9&1&.0  & 2&.2&0&.6     \\ 
 \midrule
   $ m(\lqthreeu) =  500$ \GeV& $B =0.5$&  10&.8 & 3&.4 &  27 & & 7 & \\ 
   $ m(\lqthreeu) =  750$ \GeV& $B =0  $&  \npmn        & 1& .0 & 0&.3  \\ 
   $ m(\lqthreeu) =  750$ \GeV& $B =0.5$&  2&.6 & 0&.8   &  7&.3 & 1&.5  \\
   $ m(\lqthreeu) =  750$ \GeV& $B =1  $&  2&.6 & 0&.9  &  0&.33 & 0&.1  \\
   $ m(\lqthreeu) = 1000$ \GeV& $B =0.5$&  0&.3 & 0&.09   &  1&.1 & 0&.3  \\
   $ m(\lqthreed) =  500$ \GeV& $B =0.5$&  25 & & 7 &    &  49 & & 11 &   \\
   $ m(\lqthreed) =  750$ \GeV& $B =0  $&  \npmn    &  \npmn   \\
   $ m(\lqthreed) =  750$ \GeV& $B =0.5$&  1&.9 & 0&.5  &  6&.2 & 1&.5  \\ 
   $ m(\lqthreed) =  750$ \GeV& $B =1  $&  2&.4 & 1&.1   &  2&.5 & 1&.0  \\ 
   $ m(\lqthreed) = 1000$ \GeV& $B =0.5$&  0&.53 & 0&.16   &  1&.6 & 0&.4  \\ 
\bottomrule
\end{tabular} 
\end{table}

\begin{figure}
  \centering
  \includegraphics[width=0.7\textwidth]{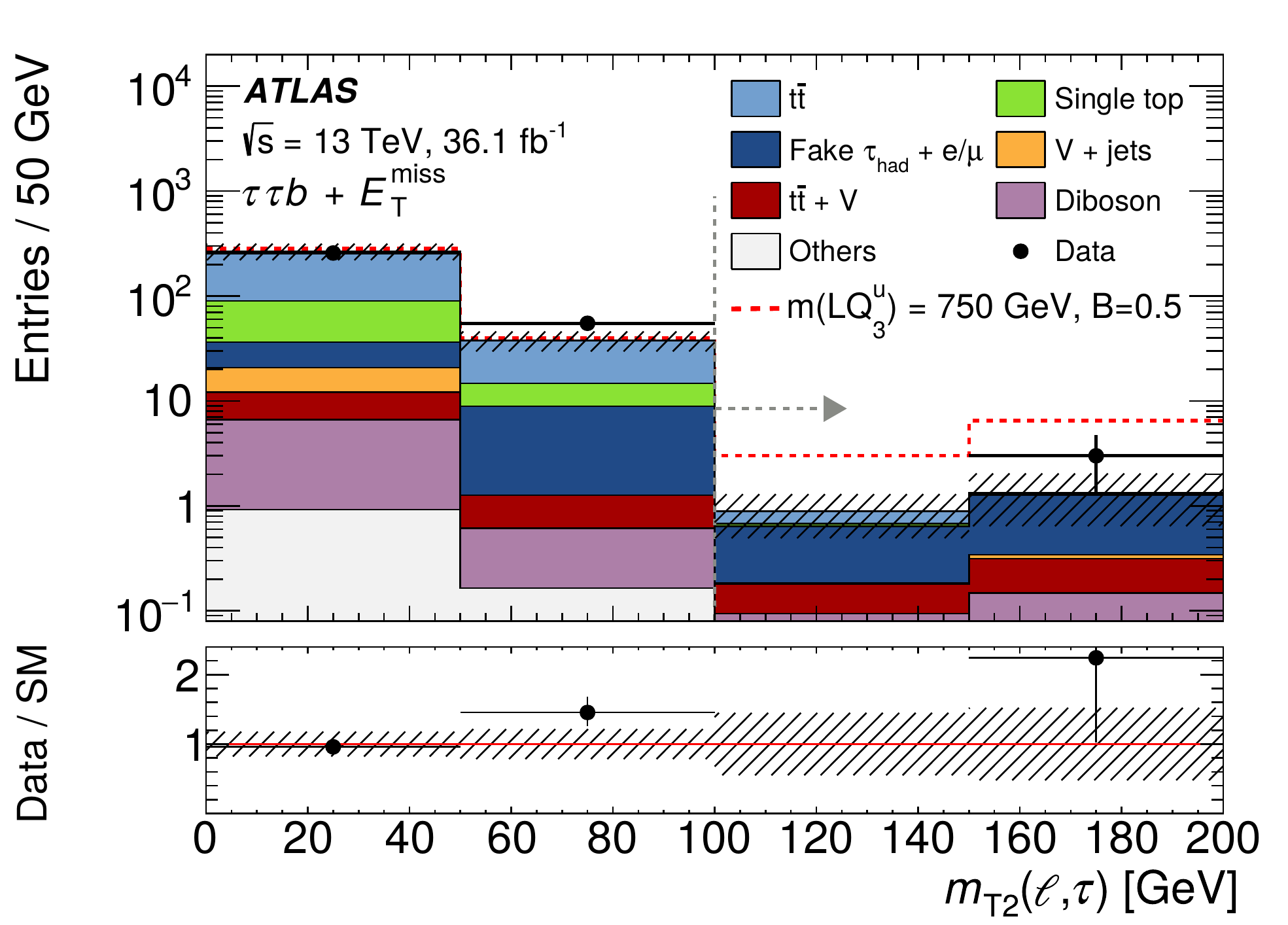}
  \caption{
    Distribution of the stransverse mass \mttwoLT~\cite{arxiv:LesterMT2, Barr:2003rg, Lester:2014yga} in the signal region of the \tlhad\ channel before applying the selection requirement on \mttwoLT,
      which is indicated by the dashed vertical line and arrow.
    The stacked histograms show the various SM background contributions, which are normalized to the values determined in the fit.
    The hatched band indicates the total statistical and systematic uncertainty in the SM background. 
    The error bars on the black data points represent the statistical uncertainty in the data yields.
    The dashed histogram shows the expected additional yields from a leptoquark signal model LQ$_3^{\textnormal{u}}$ 
      with $m_{\textnormal{LQ}}=750$~\GeV\ and $B=0.5$ added on top of the SM prediction.
    The rightmost bin includes the overflow. 
  }
  \label{fig:stopstau_SRLH_mttwo} 
\end{figure}

\begin{figure}
  \centering
  \includegraphics[width=0.495\textwidth]{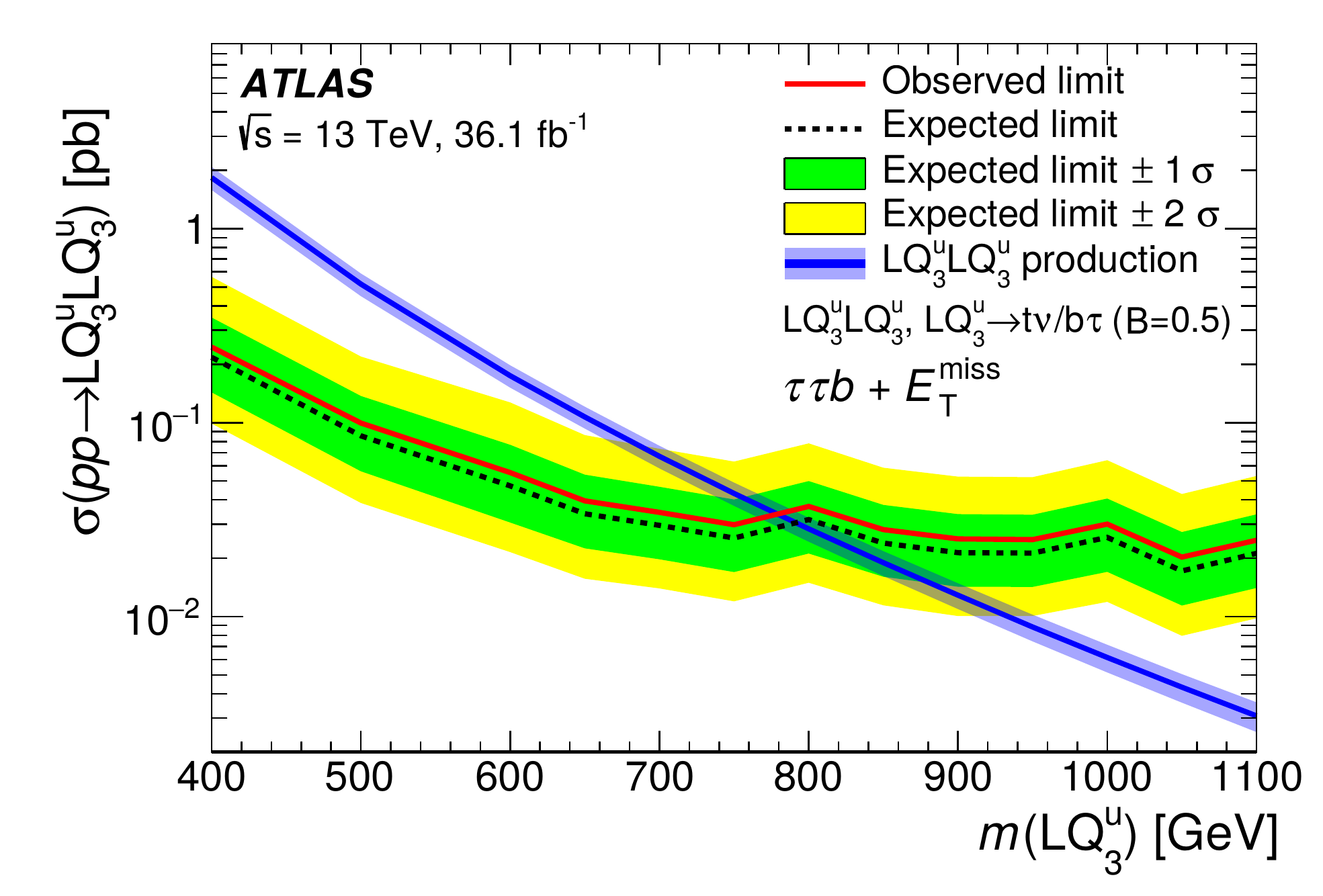}
  \includegraphics[width=0.495\textwidth]{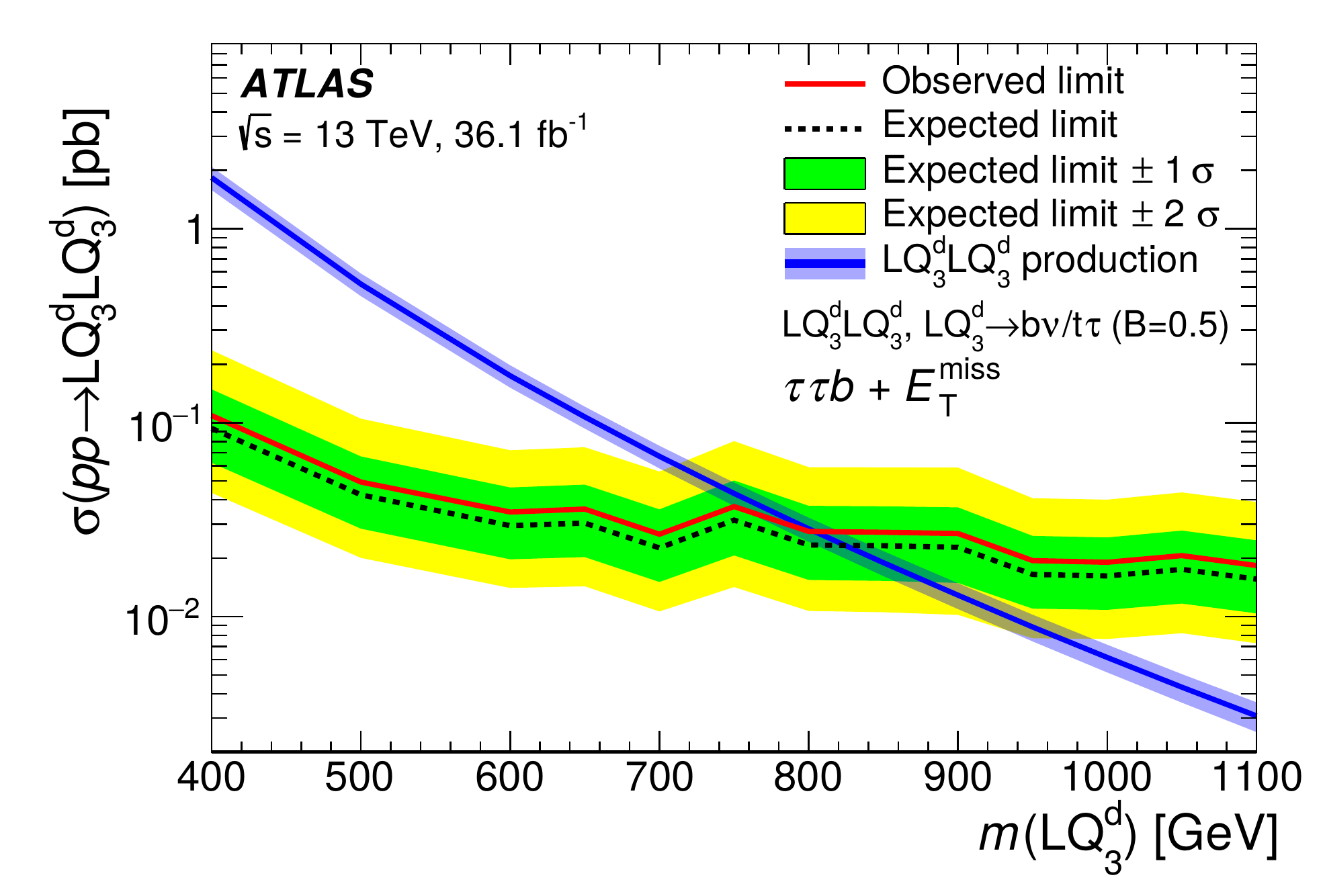}
  \caption{
    Expected cross-section limits,
      their combined statistical and systematic uncertainties excluding the theoretical cross-section uncertainty,
      and observed cross-sections limits 
      at \percent{95} CL for $B=0.5$ as a function of the leptoquark mass.
    The two channels, \thadhad\ and \tlhad, are combined for the LQ$_3^{\textnormal{u}}$ model (left) and LQ$_3^{\textnormal{d}}$ model (right).
 Statistical fluctuations are present in the expected limits
      because of the low number of simulated signal events passing the comparably tight signal-region selections.
    The thickness of the band around the theoretical cross-section curve represents the uncertainty from PDFs, 
      renormalization and factorization scales, and the strong coupling constant $\alpha_\mathrm{s}$.
  }
  \label{fig:stopstau_1D_exclusion}  
\end{figure}

The statistical analysis is done similarly to the one described in Section~\ref{ttmet-1l}.   
However, the signal regions are independent and can therefore be statistically combined in the fit.
The expected number of events from the background-only fit and the number of observed events in the signal regions 
  are shown in \tab{tab:yield_stopstau_SR} for the two analysis channels,
  together with the predicted event yields for a number of leptoquark signals.
The number of observed events agrees with the predicted Standard Model background.  Figure~\ref{fig:stopstau_SRLH_mttwo} shows the distribution of \mttwoLT,
  one of the main discriminating variables, 
  after applying all selection requirements of the signal region \SRlh except for the one on \mttwoLT,
  which is indicated by the vertical line and arrow instead.
The stacked histograms in the plot are the expected Standard Model backgrounds.
In addition, the stacked dashed histogram shows the predicted distribution of a benchmark signal model
  for up-type leptoquarks with a mass of 750~\GeV\ and branching ratio $B=0.5$.  As no significant excess in the signal regions is observed in the data,
  upper limits are set on the production cross-section.
Figure~\ref{fig:stopstau_1D_exclusion} shows the expected and observed cross-section limits 
as a function of the leptoquark mass both for up- and down-type leptoquarks. 
Leptoquark masses up to 780~\GeV\ and 800~\GeV\ are excluded for $B=0.5$ at \percent{95} confidence level
  for pair-produced up- and down-type leptoquarks, respectively.

For low leptoquark masses and high branching ratios into charged leptons,
  a non-negligible number of simulated signal events pass the control-region selections.
This leads to the background normalization factors being biased towards lower values when obtained from the exclusion fits instead of the background-only fits, 
and drives the nuisance parameters for the systematic uncertainties away from their nominal values.
Several tests were performed to check the validity of the fit procedure used to derive the exclusion.
Signal-injection tests confirmed that the fit reliably reproduces the input signal yield.
In addition, it was shown that artificially reducing the input signal yield for excluded phase-space regions 
  with a high signal contamination in the control regions,
  thereby reducing the signal contamination,
  would still allow to exclude the leptoquark signal.

\section{The $b b$ plus \ETmiss channel}

The search for direct bottom-squark pair production in either the zero- or one-lepton channel~\cite{SUSY-2016-28} is 
reinterpreted in the context of LQ production. The analysis
in the zero-lepton channel targets LQs which both decay into a bottom quark and a neutrino, i.e.\ \lqthreed\ production with $B=0$.
The one-lepton channel, also requiring two $b$-tagged jets and \ETmiss, is expected to provide sensitivity to 
\lqthreed\ production at intermediate values of $B$, as well as to small and intermediate values of $B$ for \lqthreeu\ production.

For the zero-lepton SRs (denoted by b0L), events with two high-\pt $b$-tagged jets, zero leptons, and a large amount of \met are selected. 
An exclusive jet selection (requiring 2--4 jets) is applied, preventing sizeable sensitivities to \lqthreeu\ production in this SR. 
Events with large \met arising from mismeasured jets are rejected using a selection of \dphijetimet $_{(i = 1...4)} > 0.4$. 
The two $b$-jets are required to be the two leading jets in the event and their invariant mass must be greater than 200~\GeV. 
After these baseline selections, three SRs are constructed using increasingly tighter selections on the contransverse mass \mct\ \cite{Tovey:2008ui}, 
which 
targets pair-produced heavy objects that each decay in an identical way into a visible and an invisible particle.
It is the main discriminating variable in the zero-lepton channel, with overlapping selections of $\mct >$ 350, 450, and 550~\GeV\ distinguishing the SRs.

For the one-lepton SRs (denoted by b1L), events with two $b$-tagged jets, one lepton ($e/\mu$) with $\pt > 27$\,GeV, and large \met are selected. 
Unlike the zero-lepton regions, an inclusive jet selection is used, with any two of the jets being tagged as $b$-jets. A selection of $\dphijetimet > 0.4$ for the four leading jets is used to reject events with large \met from mismeasured jets. 
Selections on the minimum invariant mass of the lepton and one of the two $b$-jets, $m_{b,\ell}^{\min}$, and on \amtTwo are used to 
reduce the \ttbar\ background, while a selection on \mt is used to reject $W$+jets events. 
After applying the previously introduced selections, two overlapping SRs are designed using \meff, the scalar sum of the \pt\ of the jets and the \ETmiss, as the main discriminating variable. The SRs are designed with selections of either $\meff >$ 600 or 750~\GeV.

The dominant backgrounds in the analysis are dependent upon the lepton multiplicity of the SR under consideration. For the zero-lepton SRs, the main SM background is $Z$-boson production ($Z \rightarrow \nu\nu$) in association with $b$-jets. Other significant sources of background arise from \ttbar\ pair production, single-top $Wt$ production, and $W$-boson production in association with $b$-jets. Control regions are defined to constrain each of the aforementioned SM backgrounds, which are designed to be kinematically close, yet orthogonal, to the SRs and also 
mutually 
orthogonal to each other. 
A two-lepton CR is used to constrain the $Z$+jets process, where the invariant mass of the leptons is required to be near the $Z$-boson mass.
The leptons are removed from the \met calculation to mimic the expected \met from the $Z \rightarrow \nu\nu$ process. The \ttbar, single-top, and $W$+jets processes are constrained in three one-lepton CRs: one CR with an inverted (relative to the SR) \amtTwo selection to create a region dominated by \ttbar; a region with an inverted \mT\ selection to constrain $W$+jets; and a final region with an inverted $m_{b,\ell}^{\min}$ selection to constrain single-top production. For the one-lepton SRs, the main SM backgrounds are \ttbar\ pair production and single-top production. A one-lepton CR is designed to constrain the \ttbar\ background by inverting the SR \amtTwo selection. The single-top background is constrained using the same CR as used for the zero-lepton single-top background. Signal contamination in these CRs is below 10\% in all regions.

\begin{table}
\caption{Number of observed events and background-only fit results in the SRs. The uncertainties contain both the statistical and
  systematic uncertainties. Two example LQ signal samples are also shown for comparison, with various assumptions about $B(\mathrm{LQ} \rightarrow q\tau)$.}
\label{tab:sbottom_SRYields}
\small
\begin{center}
\setlength{\tabcolsep}{0.0pc}
{\small
\begin{tabular*}{\textwidth}{@{\extracolsep{\fill}}lrrrrr}
\noalign{\smallskip}\hline\noalign{\smallskip}
{SR selection}           & b0L\_SRA350            & b0L\_SRA450            & b0L\_SRA550     & b1L\_SRA600            & b1L\_SRA750           \\[-0.05cm]
\noalign{\smallskip}\hline\noalign{\smallskip}
Observed events          & $81$              & $24$              & $10$     & $21$              & $13$                   \\
\noalign{\smallskip}\hline\noalign{\smallskip}
Fitted bkg events         & $70.1 \pm 13.0$          & $21.4 \pm 4.5$          & $7.2 \pm 1.5$      & $23.0 \pm 5.4$          & $14.4 \pm 3.6$          \\
\noalign{\smallskip}\hline\noalign{\smallskip}
\hline
$m_{\mathrm{LQ}}$ = 750~\GeV  \\
\hline
$B$(\lqthreed\ $\rightarrow t\tau$) = 1.0 & $ < 0.1 $  & $ < 0.1 $  & $ < 0.1 $  & $ 0.4 \pm 0.2 $  & $ 0.4 \pm 0.2  $ \\
$B$(\lqthreed\ $\rightarrow t\tau$) = 0.5 & $ 28.4 \pm 1.7 $  & $ 18.1 \pm 1.5 $  & $ 7.6 \pm 0.9 $  & $ 5.1 \pm 0.8 $  & $ 5.0 \pm 0.9  $ \\
$B$(\lqthreed\ $\rightarrow t\tau$) = 0.0 & $ 107.1 \pm 6.7 $  & $ 68.3 \pm 5.8 $  & $ 29.6 \pm 3.7 $  & $ 0.3 \pm 0.2 $  & $ 0.3 \pm 0.2  $ \\
\hline
$B$(\lqthreeu\ $\rightarrow b\tau$) = 1.0  & $ 1.3 \pm 0.6 $  & $ 0.8 \pm 0.5 $  & $ 0.2 \pm 0.2 $  & $ 0.6 \pm 0.4 $  & $ 0.6 \pm 0.3  $ \\
$B$(\lqthreeu\ $\rightarrow b\tau$) = 0.5  & $ 2.4 \pm 0.4 $  & $ 1.5 \pm 0.3 $  & $ 0.3 \pm 0.1 $  & $ 10.2 \pm 1.1 $  & $ 9.6 \pm 0.1  $ \\
$B$(\lqthreeu\ $\rightarrow b\tau$) = 0.0  & $ 2.6 \pm 1.0 $  & $ 1.7 \pm 0.6 $  & $ 0.4 \pm 0.3 $  & $ 16.7 \pm 3.3 $  & $ 14.7 \pm 0.3  $ \\
\noalign{\smallskip}\hline\noalign{\smallskip}
\hline
\end{tabular*}
}
\end{center}
\end{table}

The statistical analysis is done in full analogy to the one described in Section~\ref{ttmet-1l}. 
The number of observed events in data for each SR and the expected post-fit SM background yields are presented in Table \ref{tab:sbottom_SRYields}. 
Two example LQ signal sample yields with \mlq\ = 750~\GeV\ are presented for comparison, 
with various assumptions about the branching ratio.
Figure~\ref{fig:sbottom_SRPlots} presents the distributions of four key kinematic variables in the SRs, with two signal samples added on top of the SM background. The top row shows the \mct and \met in the zero-lepton b0L\_SRA350 region. The bottom row shows the \amtTwo and \meff\ in the one-lepton b1L\_SRA600 region.

The combination of the exclusion limits for the SRs is obtained by selecting the signal region with the
best expected limit for each mass point.
The expected and observed exclusion limits on the cross-section for \lqthreed\ and $B = 0$ are shown in Figure~\ref{fig:sbottom_BR_limit}
as a function of the leptoquark mass.
Also shown is the theoretical prediction for the cross-section of scalar leptoquark pair production including the uncertainties. 
The expected and observed 95\%~CL lower limits on the mass of a down-type LQ decaying into $b\nu\bar{b}\bar{\nu}$ are 980~\GeV\ and 970~\GeV, respectively.

\begin{figure}
  \centering
  \includegraphics[width=0.49\linewidth]{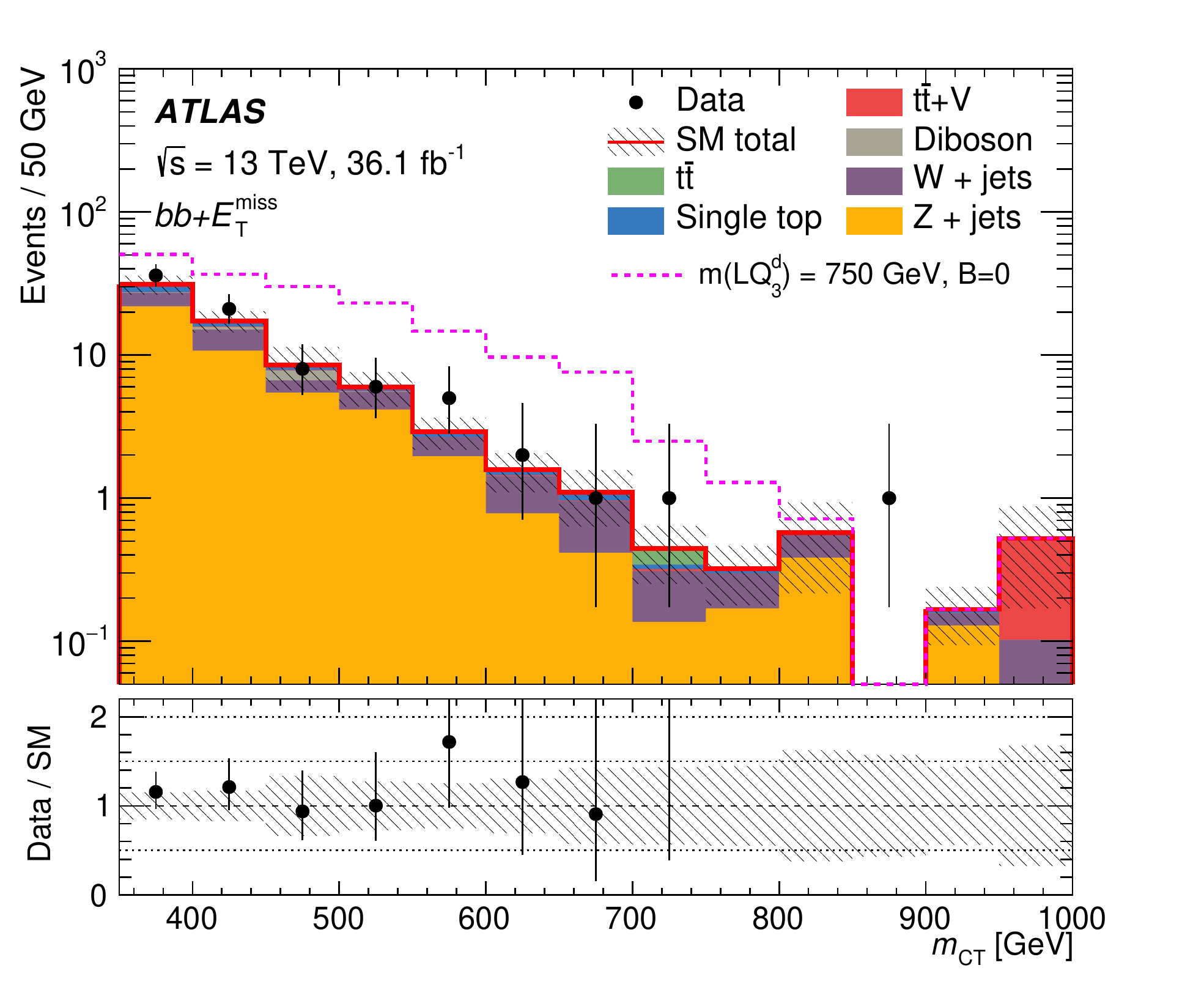}
  \includegraphics[width=0.49\linewidth]{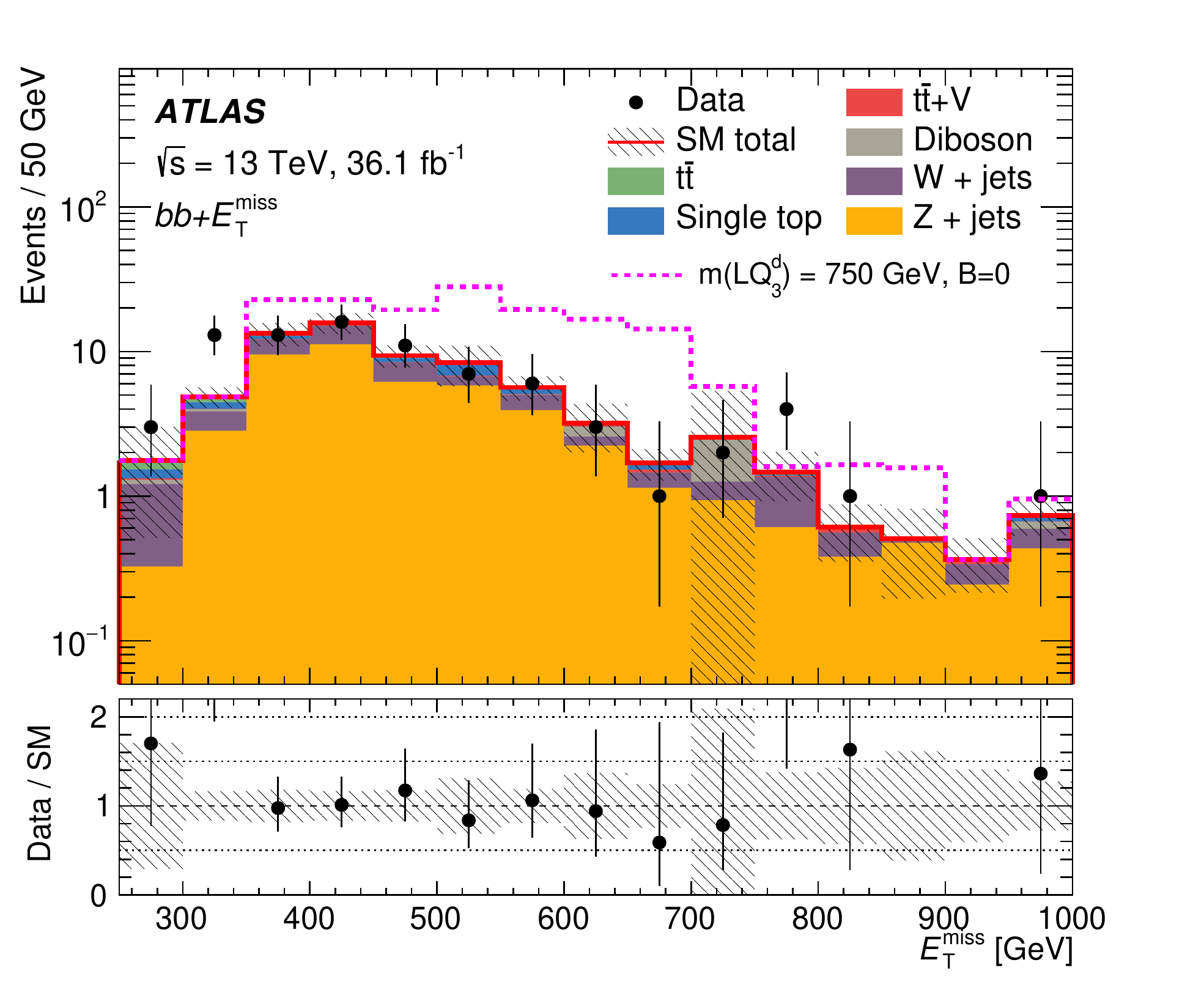}
  \includegraphics[width=0.49\linewidth]{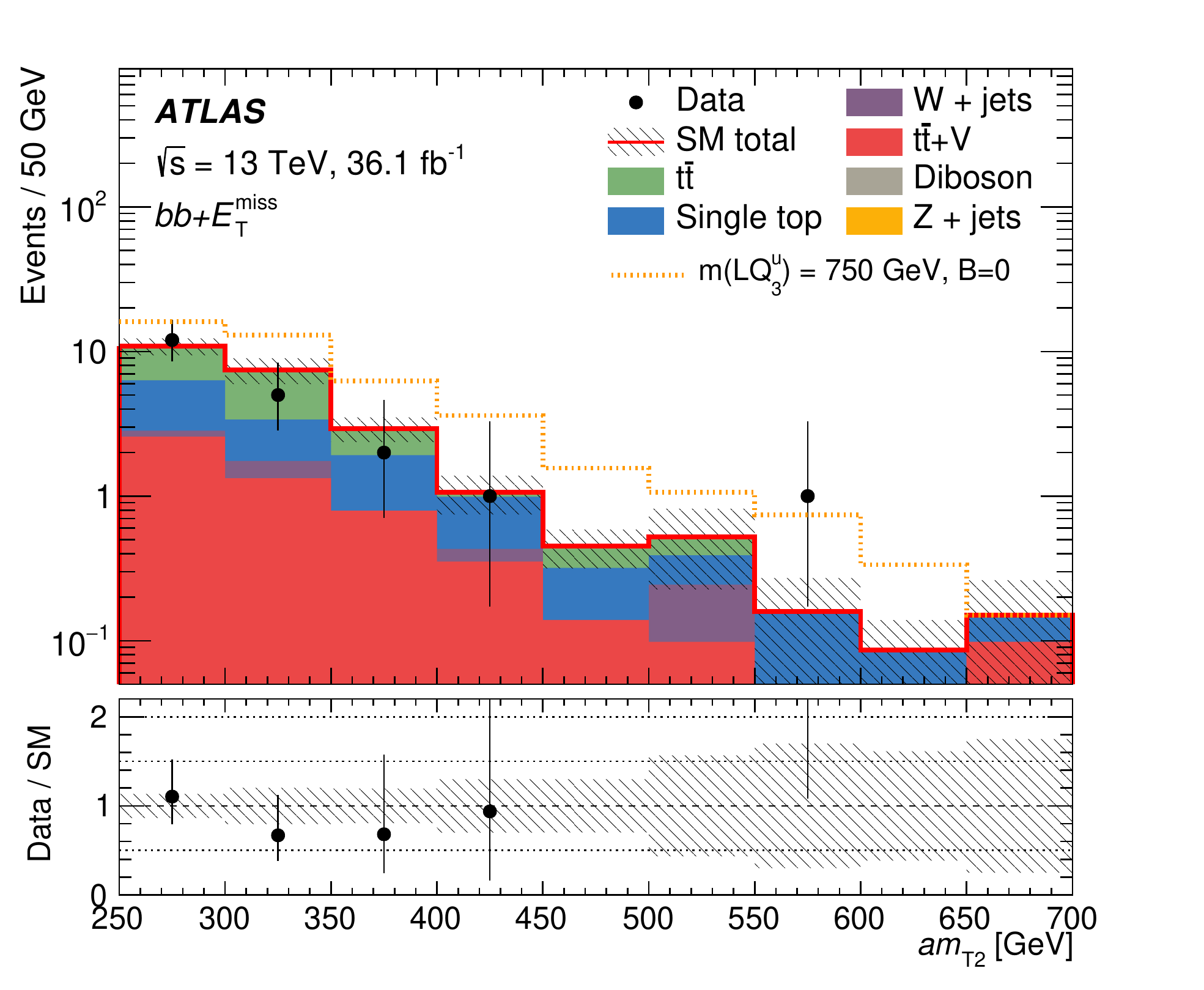}
  \includegraphics[width=0.49\linewidth]{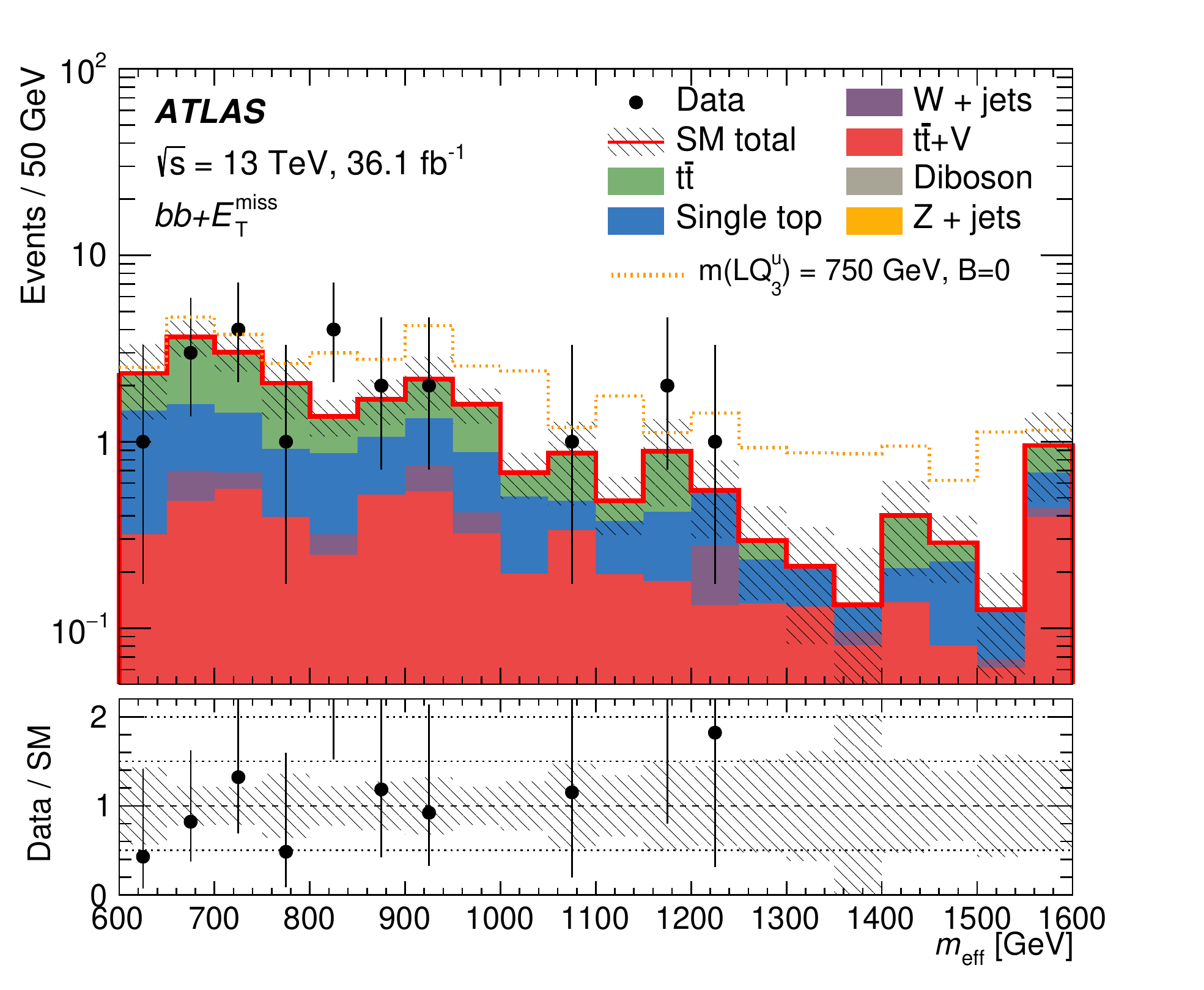}
  \caption{Distributions for key kinematic variables in the zero- and one-lepton SR selections for
the contransverse mass, $m_{\mathrm{CT}}$~\cite{Tovey:2008ui} and \met\ in the 
zero-lepton SR (top) and $am_{\mathrm{T2}}$~\cite{Konar:2009qr} and $m_{\mathrm{eff}}$ in the one-lepton SRs (bottom).
Two example LQ signal samples are added on top of the SM background, \lqthreed\ (dashed lines top plots) and \lqthreeu\ (dashed lines bottom plots). The assumed $B(\mathrm{LQ} \rightarrow q\tau)$ for both signal samples is zero, and $m_{\mathrm{LQ}}$ is 750~\GeV. The expected SM backgrounds are normalized to the values determined in the fit. The last bin contains the overflow events.}
  \label{fig:sbottom_SRPlots}
\end{figure}

\begin{figure}
  \centering
  \includegraphics[width=0.6\linewidth]{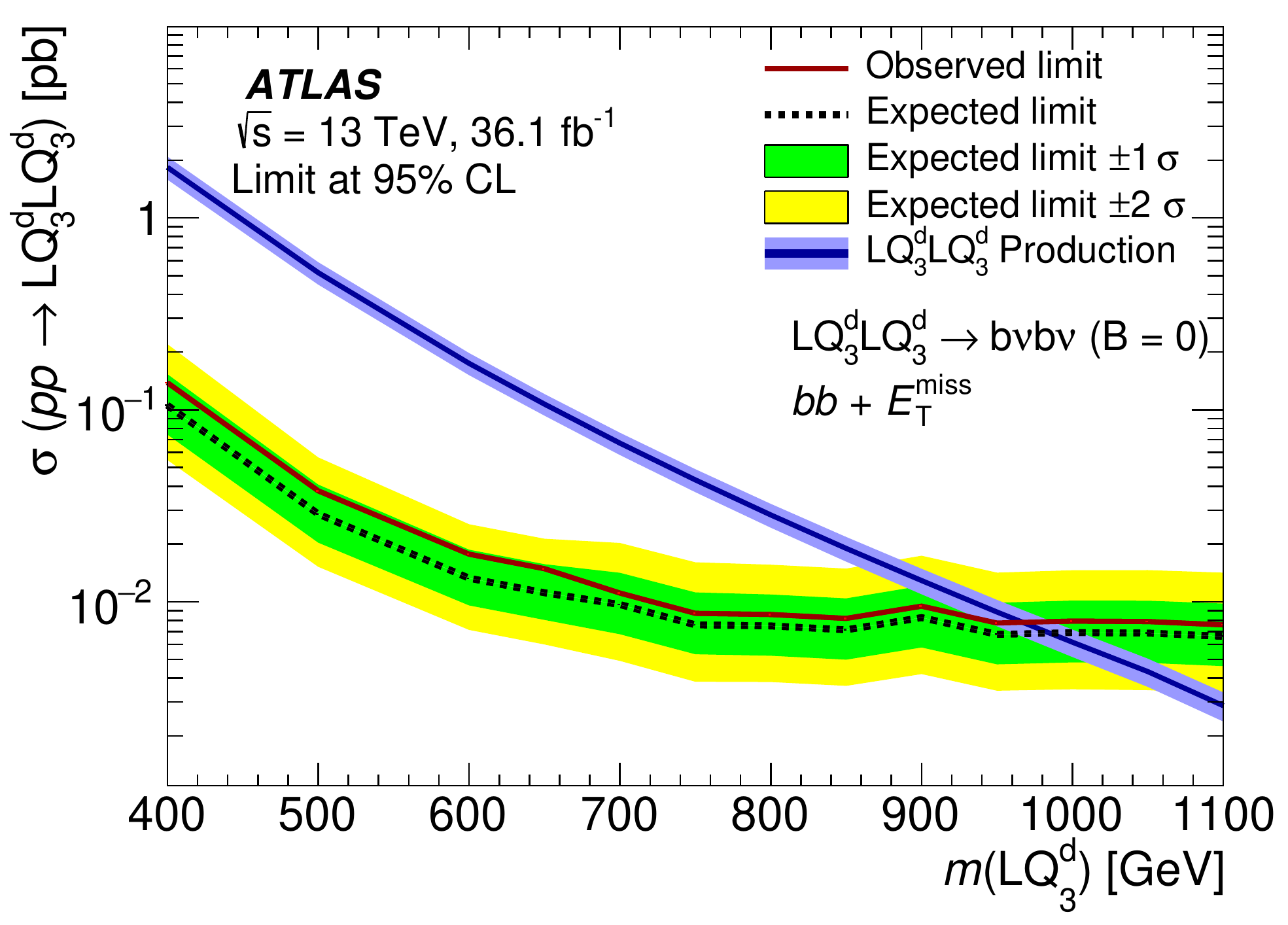}
  \caption{
Observed and expected 95\% CL upper limits on the cross-section for 
down-type LQ pair production with $B=0$
as a function of the LQ mass.
The $\pm1(2)\sigma$ uncertainty bands around the expected limit represent all sources of statistical and systematic uncertainties.
The thickness of the theory curve represents the theoretical uncertainty from PDFs, renormalization and factorization scales, and the strong coupling constant $\alpha_\mathrm{s}$.
}
  \label{fig:sbottom_BR_limit}
\end{figure}

\FloatBarrier

\section{    
Limits on the LQ mass as a function of $B$}
\label{sec:results}

The limits on cross-section and mass 
for a fixed value of $B$ that is expected to have the
highest sensitivity for the respective analysis, are presented
in the previous five sections. 
Here, Figure~\ref{fig:br_vs_massU} 
shows the limits on the LQ mass as a function of $B$ for 
all five analyses for \lqthreeu\ and \lqthreed\ pair production.
The region to the left of the contour lines is excluded at \percent{95} confidence level. 

The strongest limits in terms of mass exclusion are for \lqthreeu\ for 
$B=1$ and $B = 0$ in the $b \tau b \tau$ and $t t +$\ETmiss channel, respectively, and for \lqthreed\ for $B= 0$ 
in the $b b +$\ETmiss channel. These are the cases where the channels are optimized ($b \tau b \tau$) 
or optimal ($t t +$\ETmiss, $b b +$\ETmiss), as discussed in the introduction.
However, as can be seen from Figure~\ref{fig:br_vs_massU}, 
all channels exhibit good sensitivities to both types of LQs and to a larger range of $B$ values,
except for the \ttmetonel\ channel, which
is not sensitive to \lqthreed\ mainly due to the requirement of exactly one lepton.

Therefore, good sensitivity to both types of LQs at all values of $B$ is obtained,
excluding 
masses below 800~\GeV\ for both
\lqthreeu\ and \lqthreed\ independently of the branching ratio, with masses below 1000~\GeV\ and 1030~\GeV\ (970~\GeV\ and 920~\GeV) 
being excluded for the limiting cases of $B$ equal to zero and unity for \lqthreeu\ (\lqthreed). 

\begin{figure}
  \centering
    \includegraphics[width=0.8\linewidth]{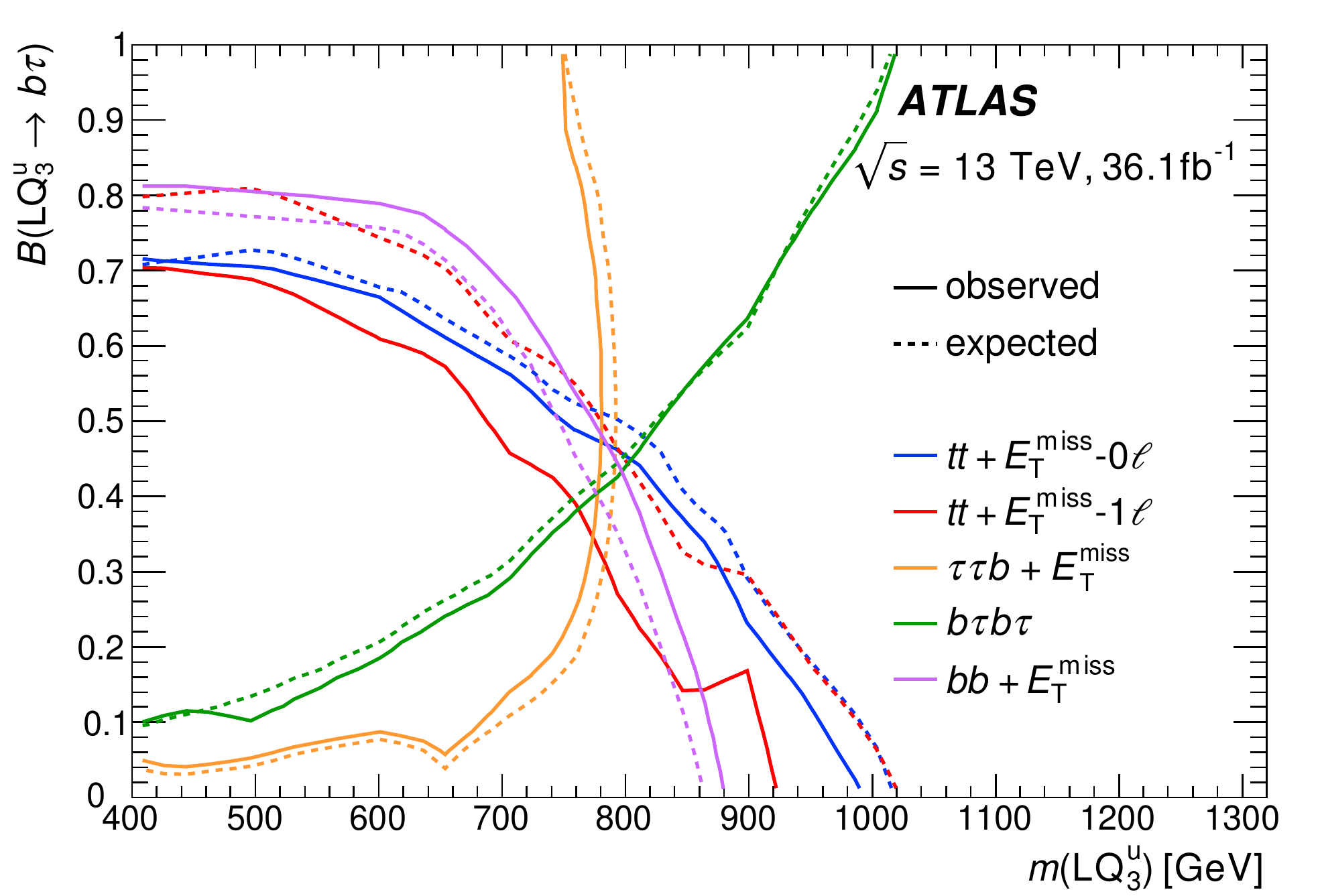}
    \includegraphics[width=0.8\linewidth]{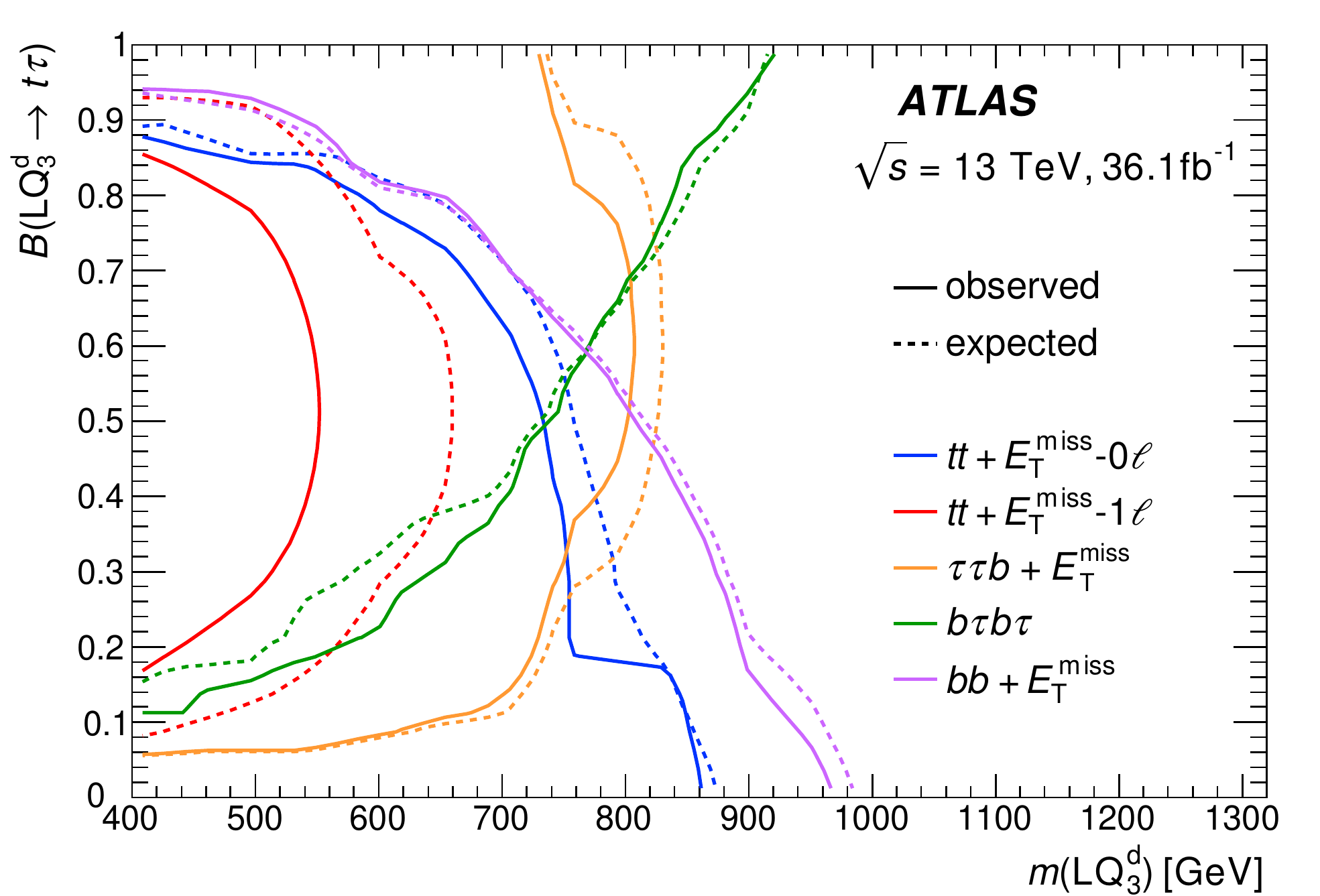}
  \caption{
    Limits on the branching ratio into charged leptons
for scalar third-generation up-type (upper panel) leptoquark 
pair production (\lqthreeu $\rightarrow b\tau / t\nu $) and down-type (lower panel) leptoquark 
pair production (\lqthreed $\rightarrow t\tau / b\nu$) as a function of the leptoquark mass.
The limits are based on a dedicated LQ search for two $b$-jets and two $\tau$-leptons 
($b \tau b \tau$),   
and reinterpretations of the search for bottom-squark 
pair production ($b b +$\ETmiss)~\cite{SUSY-2016-28}, for top-squark pair production with one 
($t t +$\ETmiss-$1\ell$)~\cite{SUSY-2016-16} 
or zero leptons ($t t +$\ETmiss-$0\ell$)~\cite{stop0LMoriond2017} in the final state, and for top squarks decaying 
via $\tau$-sleptons ($\tau \tau b +$\ETmiss)~\cite{SUSY-2016-19}.
The region to the left of 
the contour lines is excluded at \percent{95} confidence level.
  }
  \label{fig:br_vs_massU}
\end{figure}

\FloatBarrier

\section{Conclusion}
\label{sec:conclusion}

Pair production of scalar third-generation 
leptoquarks is investigated for all possible decays of the leptoquarks into a quark ($t$, $b$)
and a lepton ($\tau$, $\nu$) of the third generation.
LHC proton--proton collision 
data recorded by the ATLAS detector in 2015 and 2016 at a centre-of-mass energy of \tttev\ are used,
with  an integrated luminosity of \ilumi.
The results are based on reinterpretations of previously published ATLAS results as well as a dedicated search, 
where no significant excess above the SM background expectation is observed.

Upper limits on the LQ pair-production cross-section as a function of the LQ mass
are reported for both the up-type (\lqthreeu\ $\rightarrow t \nu / b \tau$)
and down-type (\lqthreed\ $\rightarrow b \nu / t \tau$) leptoquarks for 
branching ratios into charged leptons equal to zero, 0.5, or unity.
Based on the theoretical prediction for the LQ pair-production cross-section,
these upper limits on the cross-section can be converted to lower limits on the mass,
excluding leptoquarks with masses below about 1~\TeV\ for both LQ types and for both 
limiting cases of branching ratios into charged leptons of zero or unity. 

In addition, mass limits are shown 
as a function of the branching ratio into charged leptons for both the up- and down-type leptoquarks.
Even for intermediate values of the branching ratio, masses below at least 800~\GeV\ are 
excluded. These mass limits quickly increase to about 1~\TeV\ for small and large branching ratios.

\section*{Acknowledgements}


We thank CERN for the very successful operation of the LHC, as well as the
support staff from our institutions without whom ATLAS could not be
operated efficiently.

We acknowledge the support of ANPCyT, Argentina; YerPhI, Armenia; ARC, Australia; BMWFW and FWF, Austria; ANAS, Azerbaijan; SSTC, Belarus; CNPq and FAPESP, Brazil; NSERC, NRC and CFI, Canada; CERN; CONICYT, Chile; CAS, MOST and NSFC, China; COLCIENCIAS, Colombia; MSMT CR, MPO CR and VSC CR, Czech Republic; DNRF and DNSRC, Denmark; IN2P3-CNRS, CEA-DRF/IRFU, France; SRNSFG, Georgia; BMBF, HGF, and MPG, Germany; GSRT, Greece; RGC, Hong Kong SAR, China; ISF and Benoziyo Center, Israel; INFN, Italy; MEXT and JSPS, Japan; CNRST, Morocco; NWO, Netherlands; RCN, Norway; MNiSW and NCN, Poland; FCT, Portugal; MNE/IFA, Romania; MES of Russia and NRC KI, Russian Federation; JINR; MESTD, Serbia; MSSR, Slovakia; ARRS and MIZ\v{S}, Slovenia; DST/NRF, South Africa; MINECO, Spain; SRC and Wallenberg Foundation, Sweden; SERI, SNSF and Cantons of Bern and Geneva, Switzerland; MOST, Taiwan; TAEK, Turkey; STFC, United Kingdom; DOE and NSF, United States of America. In addition, individual groups and members have received support from BCKDF, CANARIE, CRC and Compute Canada, Canada; COST, ERC, ERDF, Horizon 2020, and Marie Sk{\l}odowska-Curie Actions, European Union; Investissements d' Avenir Labex and Idex, ANR, France; DFG and AvH Foundation, Germany; Herakleitos, Thales and Aristeia programmes co-financed by EU-ESF and the Greek NSRF, Greece; BSF-NSF and GIF, Israel; CERCA Programme Generalitat de Catalunya, Spain; The Royal Society and Leverhulme Trust, United Kingdom. 

The crucial computing support from all WLCG partners is acknowledged gratefully, in particular from CERN, the ATLAS Tier-1 facilities at TRIUMF (Canada), NDGF (Denmark, Norway, Sweden), CC-IN2P3 (France), KIT/GridKA (Germany), INFN-CNAF (Italy), NL-T1 (Netherlands), PIC (Spain), ASGC (Taiwan), RAL (UK) and BNL (USA), the Tier-2 facilities worldwide and large non-WLCG resource providers. Major contributors of computing resources are listed in Ref.~\cite{ATL-GEN-PUB-2016-002}.

\printbibliography

@ARTICLE{techni2,
  author = {Dimopoulos, S.},
  title = {Technicoloured signatures},
  journal = {Nucl. Phys. B},
  year = {1980},
  volume = {168},
  pages = {69-92},
  doi = {10.1016/0550-3213(80)90277-1},
}

@ARTICLE{dimop_suss,
  author = {Dimopoulos, S. K. and Susskind, L.},
  title = {Mass without scalars},
  journal = {Nucl. Phys. B},
  year = {1979},
  volume = {155},
  pages = {237-252},
  number = {1},
  doi = {10.1016/0550-3213(79)90364-X},
}

@ARTICLE{techni3,
  author = {Eichten, E. J. and Lane, K.},
  title = {Dynamical breaking of weak interaction symmetries},
  journal = {Phys. Lett. B},
  year = {1980},
  volume = {90},
  pages = {125-130},
  number = {1,2},
  doi = {10.1016/0370-2693(80)90065-9},
}

@ARTICLE{georgi_glashow_unification,
  author = {Georgi, H. and Glashow, S.L.},
  title = {{Unity of All Elementary Particle Forces}},
  journal = {Phys. Rev. Lett.},
  year = {1974},
  volume = {32},
  pages = {438-441},
  doi = {10.1103/PhysRevLett.32.438}
}

@ARTICLE{Buchmuller:1986zs,
  author = {Buchm\"uller, W. and R\"uckl, R. and Wyler, D.},
  title = {{Leptoquarks in lepton - quark collisions}},
  journal = {Phys. Lett. B},
  year = {1987},
  volume = {191},
  pages = {442-448},
  doi = {10.1016/0370-2693(87)90637-X},
  reportnumber = {DESY-86-150, ITP-UH-14-86},
  slaccitation = {%%CITATION = PHLTA,B191,442;%%},
  related = {Buchmuller:1986zs_erratum},
  relatedstring = {Erratum:},
}

@ARTICLE{comp,
  author = {Buchm{\"u}ller, W. and Wyler, D.},
  title = {{Constraints on SU(5)-type leptoquarks}},
  journal = {Phys. Lett. B},
  year = {1986},
  volume = {177},
  pages = {377-382},
  doi = {10.1016/0370-2693(86)90771-9}
}

@ARTICLE{pati_salam_colour,
  author = {Pati, J. C. and Salam, A.},
  title = {{Lepton number as the fourth ``colour''}},
  journal = {Phys. Rev. D},
  year = {1974},
  volume = {10},
  pages = {275-289},
  doi = {10.1103/PhysRevD.10.275},
  related = {pati_salam_colour_erratum},
  relatedstring = {Erratum:},
}

@ARTICLE{string,
  author  = {Angelopoulos, V. D. and Ellis, J. R. and Kowalski, H. and others},
  title   = {Search for new quarks suggested by the superstring},
  journal = {Nucl. Phys. B},
  year    = {1987},
  volume  = {292},
  pages   = {59-92},
  doi     = {doi:10.1016/0550-3213(87)90637-7},
}

@article{xsref,
      author         = "Borschensky, Christoph and Krämer, Michael and Kulesza,
                        Anna and Mangano, Michelangelo and Padhi, Sanjay and
                        Plehn, Tilman and Portell, Xavier",
      title          = "{Squark and gluino production cross sections in pp
                        collisions at $\sqrt{s}$ = 13, 14, 33 and 100 TeV}",
      journal        = "Eur. Phys. J. C",
      volume         = "74",
      year           = "2014",
      number         = "12",
      pages          = "3174",
      doi            = "10.1140/epjc/s10052-014-3174-y",
      eprint         = "1407.5066",
      archivePrefix  = "arXiv",
      primaryClass   = "hep-ph",
      reportNumber   = "MS-TP-14-25, CERN-PH-TH-2014-137, TTK-14-13",
      SLACcitation   = "%%CITATION = ARXIV:1407.5066;%%"
}

@Article{stop0LMoriond2017,
    author         = "{ATLAS Collaboration}",
    title          = "{Search for a scalar partner of the top quark in the jets plus missing transverse momentum final state at $\sqrt{s}$=13 TeV with the ATLAS detector}",
    journal        = "JHEP",
      volume         = "12",
      year           = "2017",
      pages          = "085",
      doi            = "10.1007/JHEP12(2017)085",
      eprint         = "1709.04183",
      archivePrefix  = "arXiv",
      primaryClass   = "hep-ex",
      reportNumber   = "CERN-EP-2017-162",
      SLACcitation   = "%%CITATION = ARXIV:1709.04183;%%"
}

@article{arxiv:LesterMT2,
  author         = "Lester, C. G. and Summers, D. J.",
  Title          = {{Measuring masses of semi-invisibly decaying particle pairs produced at hadron colliders}},
  journal        = "Phys. Lett. B",
  volume         = "463",
  year           = "1999",
  pages          = "99",
  doi            = "10.1016/S0370-2693(99)00945-4",
  archivePrefix  = "arXiv",
  eprint         = "hep-ph/9906349",
  primaryClass   = "hep-ph",
  reportNumber   = "CAVENDISH-HEP-99-07",
  SLACcitation   = "%%CITATION = HEP-PH/9906349;%%"
}

@Article{Barr:2003rg,
  Title                    = {A variable for measuring masses at hadron colliders when missing energy is expected; $m_{T2}$: the truth behind the glamour},
  Author                   = {Barr, Alan and Lester, Christopher and Stephens, P.},
  Journal                  = {J. Phys. G},
  Year                     = {2003},
  Pages                    = {2343-2363},
  Volume                   = {29},
  Doi                      = {10.1088/0954-3899/29/10/304},
  Archiveprefix            = {arXiv},
  Eprint                   = {hep-ph/0304226},
  Slaccitation             = {%%CITATION = HEP-PH/0304226;%%}
}

@article{Lester:2014yga,
      author         = "Lester, Chrisopher G. and Nachman, Benjamin",
      title          = "{Bisection-based asymmetric $M_{T2}$ computation: a
                        higher precision calculator than existing symmetric
                        methods}",
      journal        = "JHEP",
      volume         = "03",
      year           = "2015",
      pages          = "100",
      doi            = "10.1007/JHEP03(2015)100",
      eprint         = "1411.4312",
      archivePrefix  = "arXiv",
      primaryClass   = "hep-ph",
      reportNumber   = "CAV-HEP-14-13",
      SLACcitation   = "%%CITATION = ARXIV:1411.4312;%%"
}

@Booklet{ATLAS-TDR-19,
     author       = "{ATLAS Collaboration}",
     title        = "ATLAS Insertable B-Layer Technical Design Report",
     howpublished = "ATLAS-TDR-19",
     year         = "2010",
     url          = "https://cds.cern.ch/record/1291633",
     addendum     = "{\textit{ATLAS Insertable B-Layer Technical Design Report Addendum}}, ATLAS-TDR-19-ADD-1, 2012, {\scriptsize{URL}}: \url{https://cds.cern.ch/record/1451888}",
}

@article{Alwall:2014hca,
      author         = "Alwall, J. and Frederix, R. and Frixione, S. and Hirschi,
                        V. and Maltoni, F. and Mattelaer, O. and Shao, H. -S. and
                        Stelzer, T. and Torrielli, P. and Zaro, M.",
      title          = "{The automated computation of tree-level and
                        next-to-leading order differential cross sections, and
                        their matching to parton shower simulations}",
      journal        = "JHEP",
      volume         = "07",
      year           = "2014",
      pages          = "079",
      doi            = "10.1007/JHEP07(2014)079",
      eprint         = "1405.0301",
      archivePrefix  = "arXiv",
      primaryClass   = "hep-ph",
      reportNumber   = "CERN-PH-TH-2014-064, CP3-14-18, LPN14-066, MCNET-14-09,
                        ZU-TH-14-14",
      SLACcitation   = "%%CITATION = ARXIV:1405.0301;%%"
}

@article{Mandal:2015lca,
      author         = "Mandal, Tanumoy and Mitra, Subhadip and Seth, Satyajit",
      title          = "{Pair production of scalar leptoquarks at the LHC to NLO parton shower accuracy}",
      journal        = "Phys. Rev. D",
      volume         = "93",
      year           = "2016",
      number         = "3",
      pages          = "035018",
      doi            = "10.1103/PhysRevD.93.035018",
      eprint         = "1506.07369",
      archivePrefix  = "arXiv",
      primaryClass   = "hep-ph",
      reportNumber   = "HRI-RECAPP-2015-013, MITP-15-043",
      SLACcitation   = "%%CITATION = ARXIV:1506.07369;%%"
}

@article{Sjostrand:2014zea,
      author         = "Sjöstrand, Torbjörn and Ask, Stefan and Christiansen,
                        Jesper R. and Corke, Richard and Desai, Nishita and Ilten,
                        Philip and Mrenna, Stephen and Prestel, Stefan and
                        Rasmussen, Christine O. and Skands, Peter Z.",
      title          = "{An introduction to PYTHIA 8.2}",
      journal        = "Comput. Phys. Commun.",
      volume         = "191",
      year           = "2015",
      pages          = "159-177",
      doi            = "10.1016/j.cpc.2015.01.024",
      eprint         = "1410.3012",
      archivePrefix  = "arXiv",
      primaryClass   = "hep-ph",
      reportNumber   = "LU-TP-14-36, MCNET-14-22, CERN-PH-TH-2014-190,
                        FERMILAB-PUB-14-316-CD, DESY-14-178, SLAC-PUB-16122",
      SLACcitation   = "%%CITATION = ARXIV:1410.3012;%%"
}

@article{Ball:2014uwa,
      author         = "Ball, Richard D. and others",
      title          = "{Parton distributions for the LHC Run II}",
      collaboration  = "NNPDF",
      journal        = "JHEP",
      volume         = "04",
      year           = "2015",
      pages          = "040",
      doi            = "10.1007/JHEP04(2015)040",
      eprint         = "1410.8849",
      archivePrefix  = "arXiv",
      primaryClass   = "hep-ph",
      reportNumber   = "EDINBURGH-2014-15, IFUM-1034-FT, CERN-PH-TH-2013-253,
                        OUTP-14-11P, CAVENDISH-HEP-14-11",
      SLACcitation   = "%%CITATION = ARXIV:1410.8849;%%"
}

@article{Artoisenet:2012st,
      author         = "Artoisenet, Pierre and Frederix, Rikkert and Mattelaer,
                        Olivier and Rietkerk, Robbert",
      title          = "{Automatic spin-entangled decays of heavy resonances in
                        Monte Carlo simulations}",
      journal        = "JHEP",
      volume         = "03",
      year           = "2013",
      pages          = "015",
      doi            = "10.1007/JHEP03(2013)015",
      eprint         = "1212.3460",
      archivePrefix  = "arXiv",
      primaryClass   = "hep-ph",
      reportNumber   = "NIKHEF-2012-021, CERN-PH-TH-2012-329",
      SLACcitation   = "%%CITATION = ARXIV:1212.3460;%%"
}

@article{Sirunyan:2018nkj,
      author         = "{CMS Collaboration}",
      title          = "{Search for third-generation scalar leptoquarks decaying
                        to a top quark and a $\tau$ lepton at $\sqrt{s}=$ 13 TeV}",
      collaboration  = "CMS",
      journal        = "Eur. Phys. J. C",
      volume         = "78",
      year           = "2018",
      pages          = "707",
      doi            = "10.1140/epjc/s10052-018-6143-z",
      eprint         = "1803.02864",
      archivePrefix  = "arXiv",
      primaryClass   = "hep-ex",
      reportNumber   = "CMS-B2G-16-028, CERN-EP-2018-019, CMS-B2G-16-028,
                        CERN-EP-2018-019",
      SLACcitation   = "%%CITATION = ARXIV:1803.02864;%%"
}

@article{stat,
    author         = "Cowan, Glen and Cranmer, Kyle and Gross, Eilam and
    Vitells, Ofer",
    title          = "{Asymptotic formulae for likelihood-based tests of new
    physics}",
    journal        = "Eur. Phys. J. C",
    volume         = "71",
    year           = "2011",
    pages          = "1554",
    doi            = "10.1140/epjc/s10052-011-1554-0",
    eprint         = "1007.1727",
    archivePrefix  = "arXiv",
    primaryClass   = "physics.data-an",
    SLACcitation   = "%%CITATION = ARXIV:1007.1727;%%",
    related        = "stat_erratum",
    relatedstring  = "Erratum:"
}

@inproceedings{Read:2000ru,
      author         = "Read, Alexander L.",
      title          = "{Modified frequentist analysis of search results (The
                        CL(s) method)}",
      booktitle      = "{Workshop on confidence limits, CERN, Geneva,
                        Switzerland, 17-18 Jan 2000: Proceedings}",
      url            = "http://cds.cern.ch/record/451614",
      year           = "2000",
      pages          = "81-101",
      reportNumber   = "CERN-OPEN-2000-205",
      SLACcitation   = "%%CITATION = CERN-OPEN-2000-205;%%"
}

@article{Tovey:2008ui,
      author         = "Tovey, Daniel R.",
      title          = "{On measuring the masses of pair-produced semi-invisibly
                        decaying particles at hadron colliders}",
      journal        = "JHEP",
      volume         = "04",
      year           = "2008",
      pages          = "034",
      doi            = "10.1088/1126-6708/2008/04/034",
      eprint         = "0802.2879",
      archivePrefix  = "arXiv",
      primaryClass   = "hep-ph",
      SLACcitation   = "%%CITATION = ARXIV:0802.2879;%%"
}

@article{Konar:2009qr,
      author         = "Konar, Partha and Kong, Kyoungchul and Matchev,
                        Konstantin T. and Park, Myeonghun",
      title          = "{Dark matter particle spectroscopy at the LHC: generalizing M$_{T_2}$ to asymmetric event topologies}",
      journal        = "JHEP",
      volume         = "04",
      year           = "2010",
      pages          = "086",
      doi            = "10.1007/JHEP04(2010)086",
      eprint         = "0911.4126",
      archivePrefix  = "arXiv",
      primaryClass   = "hep-ph",
      reportNumber   = "SLAC-PUB-14918",
      SLACcitation   = "%%CITATION = ARXIV:0911.4126;%%"
}

@article{Cacciari:2008gp,
      author         = "Cacciari, Matteo and Salam, Gavin P. and Soyez, Gregory",
      title          = "{The anti-k(t) jet clustering algorithm}",
      journal        = "JHEP",
      volume         = "04",
      year           = "2008",
      pages          = "063",
      doi            = "10.1088/1126-6708/2008/04/063",
      eprint         = "0802.1189",
      archivePrefix  = "arXiv",
      primaryClass   = "hep-ph",
      reportNumber   = "LPTHE-07-03",
      SLACcitation   = "%%CITATION = ARXIV:0802.1189;%%"
}

@article{Belyaev:2005ew,
      author         = "Belyaev, Alexander and Leroy, Claude and Mehdiyev, Rashid
                        and Pukhov, Alexander",
      title          = "{Leptoquark single and pair production at LHC with
                        CalcHEP/CompHEP in the complete model}",
      journal        = "JHEP",
      volume         = "09",
      year           = "2005",
      pages          = "005",
      doi            = "10.1088/1126-6708/2005/09/005",
      eprint         = "hep-ph/0502067",
      archivePrefix  = "arXiv",
      primaryClass   = "hep-ph",
      reportNumber   = "MSU-HEP-070204",
      SLACcitation   = "%%CITATION = HEP-PH/0502067;%%"
}

@article{Sirunyan:2018vhk,
      author         = "{CMS Collaboration}",
      title          = "{Search for heavy neutrinos and third-generation
                        leptoquarks in hadronic states of two $\tau$ leptons and
                        two jets in proton-proton collisions at $\sqrt{s} =$ 13
                        TeV}",
      collaboration  = "CMS",
      year           = "2018",
      eprint         = "1811.00806",
      archivePrefix  = "arXiv",
      primaryClass   = "hep-ex",
      reportNumber   = "CMS-EXO-17-016, CERN-EP-2018-272",
      SLACcitation   = "%%CITATION = ARXIV:1811.00806;%%"
}

@article{Hiller:2014yaa,
      author         = "Hiller, Gudrun and Schmaltz, Martin",
      title          = "{$R_K$ and future $b \to s \ell \ell$ physics beyond the
                        standard model opportunities}",
      journal        = "Phys. Rev. D",
      volume         = "90",
      year           = "2014",
      pages          = "054014",
      doi            = "10.1103/PhysRevD.90.054014",
      eprint         = "1408.1627",
      archivePrefix  = "arXiv",
      primaryClass   = "hep-ph",
      reportNumber   = "DO-TH-14-17",
      SLACcitation   = "%%CITATION = ARXIV:1408.1627;%%"
}

@article{Gripaios:2014tna,
      author         = "Gripaios, Ben and Nardecchia, Marco and Renner, S. A.",
      title          = "{Composite leptoquarks and anomalies in $B$-meson
                        decays}",
      journal        = "JHEP",
      volume         = "05",
      year           = "2015",
      pages          = "006",
      doi            = "10.1007/JHEP05(2015)006",
      eprint         = "1412.1791",
      archivePrefix  = "arXiv",
      primaryClass   = "hep-ph",
      SLACcitation   = "%%CITATION = ARXIV:1412.1791;%%"
}

@article{Cline:2017aed,
      author         = "Cline, James M.",
      title          = "{$B$ decay anomalies and dark matter from vectorlike
                        confinement}",
      journal        = "Phys. Rev. D",
      volume         = "97",
      year           = "2018",
      number         = "1",
      pages          = "015013",
      doi            = "10.1103/PhysRevD.97.015013",
      eprint         = "1710.02140",
      archivePrefix  = "arXiv",
      primaryClass   = "hep-ph",
      SLACcitation   = "%%CITATION = ARXIV:1710.02140;%%"
}

@article{DiLuzio:2017chi,
      author         = "Di Luzio, Luca and Nardecchia, Marco",
      title          = "{What is the scale of new physics behind the $B$-flavour
                        anomalies?}",
      journal        = "Eur. Phys. J. C",
      volume         = "77",
      year           = "2017",
      number         = "8",
      pages          = "536",
      doi            = "10.1140/epjc/s10052-017-5118-9",
      eprint         = "1706.01868",
      archivePrefix  = "arXiv",
      primaryClass   = "hep-ph",
      reportNumber   = "IPPP-17-50, CERN-TH-2017-126",
      SLACcitation   = "%%CITATION = ARXIV:1706.01868;%%"
}

@article{Freytsis:2015qca,
      author         = "Freytsis, Marat and Ligeti, Zoltan and Ruderman, Joshua
                        T.",
      title          = "{Flavor models for $\bar{B} \to D^{(*)} \tau \bar{\nu}$}",
      journal        = "Phys. Rev. D",
      volume         = "92",
      year           = "2015",
      number         = "5",
      pages          = "054018",
      doi            = "10.1103/PhysRevD.92.054018",
      eprint         = "1506.08896",
      archivePrefix  = "arXiv",
      primaryClass   = "hep-ph",
      SLACcitation   = "%%CITATION = ARXIV:1506.08896;%%"
}

@article{Bauer:2015knc,
      author         = "Bauer, Martin and Neubert, Matthias",
      title          = "{Minimal Leptoquark Explanation for the R$_{D^{(*)}}$ ,
                        R$_K$ , and $(g-2)_g$ Anomalies}",
      journal        = "Phys. Rev. Lett.",
      volume         = "116",
      year           = "2016",
      number         = "14",
      pages          = "141802",
      doi            = "10.1103/PhysRevLett.116.141802",
      eprint         = "1511.01900",
      archivePrefix  = "arXiv",
      primaryClass   = "hep-ph",
      reportNumber   = "MITP-15-100",
      SLACcitation   = "%%CITATION = ARXIV:1511.01900;%%"
}

@article{Buttazzo:2017ixm,
      author         = "Buttazzo, Dario and Greljo, Admir and Isidori, Gino and
                        Marzocca, David",
      title          = "{B-physics anomalies: a guide to combined explanations}",
      journal        = "JHEP",
      volume         = "11",
      year           = "2017",
      pages          = "044",
      doi            = "10.1007/JHEP11(2017)044",
      eprint         = "1706.07808",
      archivePrefix  = "arXiv",
      primaryClass   = "hep-ph",
      reportNumber   = "ZU-TH-18-17",
      SLACcitation   = "%%CITATION = ARXIV:1706.07808;%%"
}

@article{Aaboud:2019jcc,
      author         = "{ATLAS Collaboration}",
      title          = "{Searches for scalar leptoquarks and differential
                        cross-section measurements in dilepton-dijet events in
                        proton-proton collisions at a centre-of-mass energy of
                        $\sqrt{s}$ = 13 TeV with the ATLAS experiment}",
      collaboration  = "ATLAS",
      year           = "2019",
      eprint         = "1902.00377",
      archivePrefix  = "arXiv",
      primaryClass   = "hep-ex",
      reportNumber   = "CERN-EP-2018-262",
      SLACcitation   = "%%CITATION = ARXIV:1902.00377;%%"
}

@article{Kramer:2004df,
      author         = "Kramer, M. and Plehn, T. and Spira, M. and Zerwas, P. M.",
      title          = "{Pair production of scalar leptoquarks at the CERN LHC}",
      journal        = "Phys. Rev.",
      volume         = "D71",
      year           = "2005",
      pages          = "057503",
      doi            = "10.1103/PhysRevD.71.057503",
      eprint         = "hep-ph/0411038",
      archivePrefix  = "arXiv",
      primaryClass   = "hep-ph",
      reportNumber   = "CERN-PH-TH-2004-207, DESY-04-200, EDINBURGH-2004-28,
                        FERMILAB-PUB-04-298-T, PSI-PR-04-12",
      SLACcitation   = "%%CITATION = HEP-PH/0411038;%%"
}

\clearpage 

 
\begin{flushleft}
{\Large The ATLAS Collaboration}

\bigskip

M.~Aaboud$^\textrm{\scriptsize 34d}$,    
G.~Aad$^\textrm{\scriptsize 100}$,    
B.~Abbott$^\textrm{\scriptsize 126}$,    
D.C.~Abbott$^\textrm{\scriptsize 101}$,    
O.~Abdinov$^\textrm{\scriptsize 13,*}$,    
D.K.~Abhayasinghe$^\textrm{\scriptsize 92}$,    
S.H.~Abidi$^\textrm{\scriptsize 165}$,    
O.S.~AbouZeid$^\textrm{\scriptsize 39}$,    
N.L.~Abraham$^\textrm{\scriptsize 154}$,    
H.~Abramowicz$^\textrm{\scriptsize 159}$
% ...
\bigskip
\\

% ...
$^{13}$Institute of Physics, Azerbaijan Academy of Sciences, Baku; Azerbaijan.\\
% ...
$^{34}$$^{(a)}$Facult\'e des Sciences Ain Chock, R\'eseau Universitaire de Physique des Hautes Energies - Universit\'e Hassan II, Casablanca;$^{(b)}$Centre National de l'Energie des Sciences Techniques Nucleaires (CNESTEN), Rabat;$^{(c)}$Facult\'e des Sciences Semlalia, Universit\'e Cadi Ayyad, LPHEA-Marrakech;$^{(d)}$Facult\'e des Sciences, Universit\'e Mohamed Premier and LPTPM, Oujda;$^{(e)}$Facult\'e des sciences, Universit\'e Mohammed V, Rabat; Morocco.\\
% ...
$^{39}$Niels Bohr Institute, University of Copenhagen, Copenhagen; Denmark.\\
% ...
$^{92}$Department of Physics, Royal Holloway University of London, Egham; United Kingdom.\\
% ...
$^{100}$CPPM, Aix-Marseille Universit\'e, CNRS/IN2P3, Marseille; France.\\
$^{101}$Department of Physics, University of Massachusetts, Amherst MA; United States of America.\\
% ...
$^{126}$Homer L. Dodge Department of Physics and Astronomy, University of Oklahoma, Norman OK; United States of America.\\
% ...
$^{154}$Department of Physics and Astronomy, University of Sussex, Brighton; United Kingdom.\\
% ...
$^{159}$Raymond and Beverly Sackler School of Physics and Astronomy, Tel Aviv University, Tel Aviv; Israel.\\
% ...
$^{165}$Department of Physics, University of Toronto, Toronto ON; Canada.\\
% ...
$^{a}$ Also at Borough of Manhattan Community College, City University of New York, NY; United States of America.\\
$^{b}$ Also at California State University, East Bay; United States of America.\\
$^{c}$ Also at Centre for High Performance Computing, CSIR Campus, Rosebank, Cape Town; South Africa.\\
$^{d}$ Also at CERN, Geneva; Switzerland.\\
$^{e}$ Also at CPPM, Aix-Marseille Universit\'e, CNRS/IN2P3, Marseille; France.\\
% ...
$^{*}$ Deceased

\end{flushleft}


\end{document}